\documentclass{article}
\title{\bf
Tropical Robinson-Schensted-Knuth \\
correspondence and birational \\
Weyl group actions
}
\author{
Masatoshi NOUMI and Yasuhiko YAMADA\\
{\normalsize Department of Mathematics, Kobe University}
}
\date{}
\newcommand{\comment}[1]{}
\comment{
By using an elementary matrix approach, based on the technique
of discrete Toda equation, we construct subtraction-free rational 
and piecewise linear transformations associated with various 
combinatorial algorithms, including the RSK correspondence.  
We also investigate birational Weyl group actions
related to those algorithms. 
}
\usepackage{amssymb}
\newtheorem{theorem}{Theorem}[section]
\newtheorem{lemma}[theorem]{Lemma}
\newtheorem{proposition}[theorem]{Proposition}
\newtheorem{corollary}[theorem]{Corollary}

\newtheorem{example}[theorem]{Example}
\newtheorem{remark}[theorem]{Remark}
\newcommand{\eqref}[1]{$(\ref{#1})$}
\newcommand{\proof}{\par\medskip\noindent{\it Proof\/}:\ \ }

\newcommand{\qed}{\hfill Q.E.D.\par\medskip}
\newcommand{\Mat}{\mbox{\rm Mat}}

\newcommand{\CB}{\mathcal{B}}

\newcommand{\CF}{\mathcal{F}}
\newcommand{\BC}{\mathbb{C}}
\newcommand{\BR}{\mathbb{R}}
\newcommand{\BQ}{\mathbb{Q}}
\newcommand{\BN}{\mathbb{N}}
\newcommand{\BZ}{\mathbb{Z}}
\newcommand{\BK}{\mathbb{K}}

\newcommand{\BS}{\mbox{\boldmath$S$}}
\newcommand{\ba}{\mbox{\boldmath$a$}}
\newcommand{\bb}{\mbox{\boldmath$b$}}
\newcommand{\be}{\mbox{\boldmath$e$}}
\newcommand{\bt}{\mbox{\boldmath$t$}}
\newcommand{\bu}{\mbox{\boldmath$u$}}
\newcommand{\bv}{\mbox{\boldmath$v$}}
\newcommand{\bx}{\mbox{\boldmath$x$}}
\newcommand{\by}{\mbox{\boldmath$y$}}

\newcommand{\bp}{\mbox{\boldmath$p$}}
\newcommand{\bq}{\mbox{\boldmath$q$}}

\newcommand{\btau}{\mbox{\boldmath$\tau$}}
\newcommand{\bsigma}{\mbox{\boldmath$\sigma$}}

\newcommand{\bmu}{\mbox{\boldmath$\mu$}}
\newcommand{\bnu}{\mbox{\boldmath$\nu$}}
\newcommand{\br}[1]{\{#1\}}
\newcommand{\pr}[1]{\langle{#1}\rangle}
\newcommand{\diag}[1]{\mbox{\rm diag}(#1)}
\newcommand{\dsum}[2]{{\displaystyle\sum_{#1}^{#2}}}
\newcommand{\dfrac}[2]{{\displaystyle\frac{#1}{#2}}}
\newcommand{\dmax}[1]{{\displaystyle\max_{#1}}}
\newcommand{\iso}{\stackrel{\sim}{\to}}
\newcommand{\tbl}[1]{
\arraycolsep=1pt\renewcommand{\arraystretch}{0.8}
\begin{array}{llllllllll}#1\end{array}}

\newcommand{\cross}{\begin{picture}(12,10)(-2,0)
\put(-8,0){\Large$\longrightarrow$}
\put(0,0){\Large$\downarrow$}
\end{picture}}
\newcommand{\trp}{{\mbox{\scriptsize\rm t}}}
\newcommand{\RSKs}{{RSK$^\ast$\,}}

\newcommand{\SI}[1]{{#1}^{\mbox{\scriptsize\sf\,s}}}
\newcommand{\sSI}[1]{\SI{#1}}
\newcommand{\wt}{\mbox{\rm wt}}
\newcommand{\mod}{\mbox{\rm\ mod\ }}
\newcommand{\Res}{\mbox{\rm Res}}
\begin{document}
\maketitle
\section*{Introduction }
This paper is an outcome of our attempt to 
understand internal connections 
among several appearances of the subtraction-free birational 
transformations. 

There is a well-known procedure for passing 
from subtraction-free rational functions 
to piecewise linear functions.  
Roughly, this is the procedure of replacing the operations 
$$
a\,b \to a+b,\quad 
a/b \to a-b,\quad a+b\,\to \max\br{a,b}\quad(\mbox{or}\ \min\br{a,b}). 
$$
It can be applied consistently 
to an arbitrary rational function expressed as a ratio of 
two polynomials with positive real coefficients, in order 
to produce a combination of $+$, $-$ and $\max$ (or $\min$), 
representing a piecewise linear function.  
In combinatorics, this procedure has been employed 
for the {\em algebraization} of combinatorial algorithms.  
A large class of combinatorial algorithms can be 
described as piecewise linear transformations 
among discrete variables which take integer values. 
For such a piecewise  linear transformation, it is 
meaningful in many cases to find a good 
subtraction-free rational counterpart; 
algebraic computation of subtraction-free 
rational functions may possibly bring out 
unexpected solutions to combinatorial problems. 
For this {\em tropical approach} to combinatorics, 
we refer the reader to \cite{BFZ}, \cite{K} and the 
references therein. 

In the context of discrete integrable systems, 
the same procedure 
is known as {\em ultra-discretization} \cite{TTMS}. 
A remarkable example is the ultra-discretization of 
discrete Toda equation which provides with soliton cellular 
automata, called the box-ball systems \cite{TS}.  
It is already recognized that 
the theory of box-ball systems is precisely 
the dynamics of 
{\em crystal bases} which arise as 
the $q\to 0$ limit of representations of 
quantum groups (see \cite{HHIKTT}, for example).  
The ultra-discretization of certain $q$-Painlev\'e systems 
can also be understood as a non-autonomous deformation 
of box-ball systems \cite{KNY1,KNY2,KNY3}; 
the time evolution of such 
(ultra-)discrete systems arises from the translation 
lattice of affine Weyl groups. 

Another important aspect is the connection with the theory of totally 
positive matrices.  
Totally positive matrices have been studied extensively 
from the viewpoint of geometric approach to canonical 
bases \cite{L}, \cite{BFZ}, \cite{FZ}, \cite{BKa};
they provide a basic tool for 
producing nice subtraction-free rational transformations. 
\par\medskip 
The purpose of this paper is to develop 
a new, elementary approach to 
the application of subtraction-free birational transformations 
to combinatorial problems. 
Our method is based on the decomposition 
and exchange of matrices, and the path representation  
of minor determinants. 
We employ such techniques 
to construct both subtraction-free rational and 
piecewise linear transformations
for typical combinatorial algorithms, such as 
the bumping procedure, 
the Sch\"utzenberger involution and 
the Robinson-Schensted-Knuth correspondence
(RSK correspondence, for short).  
Our {\em matrix approach} can be regarded as an integration 
of the idea of totally positive matrices and 
the technique of discrete Toda equations. 
We also investigate certain birational and piecewise linear
actions of (affine) Weyl groups 
on matrices and tableaux. 

This work was motivated by the impressive paper 
\cite{K} by A.N.~Kirillov.   
It was a great surprise for the authors to find that  
many formulas in \cite{K}, arising from combinatorics,  
were essentially the same as what we had encountered with 
in the context of discrete Painlev\'e systems. 
The matrix approach, as we will develop below, was a 
natural consequence of our attempt to clarify the theoretical 
background of this remarkable coincidence.

In view of the elementary nature of our approach, 
we have tried to make this paper as self-contained as possible.
Many of the explicit formulas discussed in this paper 
can already be found in the literature
(\cite{BKi}, \cite{K}, \cite{KB}).  
Also, many of the statements on decomposition of matrices 
are essentially contained in a series of works \cite{BFZ}, \cite{FZ}, 
\cite{L} on totally positive matrices.  
We expect however that the results and techniques developed 
in this paper would be applicable to various problems 
both in combinatorics and 
in discrete integrable systems.

\par\medskip
The authors would like to express their thanks 
to Professors 
S.~Fomin, A.N.~Kirillov, 
and  A.~Zelevinsky for valuable discussions. 

\par\bigskip
\noindent
{\sl Notes:}\ \ 
The use of the phrase ``tropical'' comes originally 
from computer science; 
as in ``tropical semirings'',  this word has been used 
in a restrictive way to refer to the semiring structure 
on various set of numbers 
with respect to the pair of operations $(\mbox{\rm min},+)$.  
We thank Prof.\,Fomin for directing 
our attention to this point. 
In the combinatorial literature, 
the same phrase seems to be used in a broader sense, 
mostly in such a situation that subtraction-free rational 
functions and piecewise linear functions appear 
more or less exchangeably; 
it also depends on the author on which side emphasis is put. 
In this paper, following \cite{K} we use 
the word ``tropical'' {\em tentatively} to refer to objects 
concerning subtraction-free rational functions (see Section 1.3).  
This may {\em not\/} be identical to the traditional usage, 
but we could not find a better alternative.  

\par\bigskip

\newpage
\tableofcontents
\newpage
\setcounter{equation}{0}
\section{Preliminaries}
\label{sec:pathrep}
In this section, we give some preliminary remarks 
on the matrix approach to nonintersecting paths. 
In the last part of this section, we also give a summary 
on a canonical procedure for passing from 
subtraction-free rational 
functions to piecewise linear functions. 
In what follows, we fix the ground field $\BK$, and set 
$\BK^\ast=\BK\setminus\br{0}$. 
For a matrix $X=\pmatrix{x^i_j}_{i,j}$ given, we denote 
by 
\begin{equation}
 X^{i_1,\ldots,i_r}_{j_1,\ldots,j_r}
=\pmatrix{x^{i_a}_{j_b}}_{a,b=1}^r,
\quad 
\det X^{i_1,\ldots,i_r}_{j_1,\ldots,j_r}
=\det\pmatrix{x^{i_a}_{j_b}}_{a,b=1}^r
\end{equation}
the $r\times r$ submatrix and 
the $r$-minor determinant of $X$ with row indices 
$i_1,\ldots,i_r$ and column indices $j_1,\ldots,j_r$,
respectively. 
\par\medskip
\subsection{Path representation of minor determinants}
For an $n$-vector $\bx=(x_1,\ldots,x_n)\in(\BK^\ast)^n$ 
given, we introduce the following two matrices 
$E(\bx)$ and $H(\bx)$: 
\begin{equation}
E(\bx)=\diag{\bx}+\Lambda,\quad
H(\bx)=(\diag{\bx}^{-1}-\Lambda)^{-1},
\end{equation}
where $\Lambda=\pmatrix{\delta_{j,i+1}}_{i,j=1}^n$ stands for the 
shift matrix.  
With the notation of matrix units 
$E_{ij}=\pmatrix{\delta_{a,i}\delta_{b,j}}_{a,b=1}^n$, 
these matrices can be written alternatively as
\begin{equation}
E(\bx)=\sum_{i=1}^{n} x_i E_{ii}+\sum_{i=1}^{n-1}E_{i,i+1},
\quad
H(\bx)=\sum_{1\le i\le j\le n}\,x_ix_{i+1}\cdots x_{j}\,E_{ij}.  
\end{equation} 
For a given sequence of $n$-vectors $\bx^1,\ldots,\bx^m$, 
$\bx^i=(x^i_1,\ldots,x^i_n)\in(\BK^\ast)^n$, 
we define 
\begin{eqnarray}\label{eq:EH}
&&E(\bx^1,\ldots,\bx^m)=E(\bx^1)E(\bx^2)\cdots E(\bx^m),
\nonumber\\
&&H(\bx^1,\ldots,\bx^m)=H(\bx^1)H(\bx^2)\cdots H(\bx^m). 
\end{eqnarray}
Note that $H(\bx)=D E(\overline{\bx})^{-1}D^{-1}$, 
$D=\diag{(-1)^{i-1}}_{i=1}^n$,
where 
$\overline{x}=(\overline{x}_1,\ldots,\overline{x}_n)$, 
$\overline{x}_j=\frac{1}{x_j}$; 
we use the notation $\overline{x}$ for $x^{-1}$ in order 
to avoid the conflict with that of upper indices. 
With this notation, the two matrices in \eqref{eq:EH} are 
related as
\begin{equation}
H(\bx^1,\ldots,\bx^m)=DE(\overline{\bx}^m,\ldots,\overline{\bx}^1)D^{-1}.
\end{equation} 
In the following, we propose graphical expressions for 
the minor determinants of 
$E(\bx^1,\ldots,\bx^m)$ and $H(\bx^1,\ldots,\bx^m)$,
in terms of nonintersecting paths. 
\par\medskip
We first consider the case of $E(\bx^1,\ldots,\bx^m)$. 
We represent the matrix $E(\bx)$ by the diagram 
\begin{equation}
\begin{picture}(180,25)(-20,-5)
\multiput(0,0)(3,0){51}{\line(1,0){1}}
\multiput(0,15)(3,0){51}{\line(1,0){1}}
{\thicklines\multiput(0,0)(15,0){11}{\line(0,1){15}}}
\multiput(0,15)(15,0){10}{\line(1,-1){15}}
{\thicklines
\put(60,3){\vector(0,-1){1}}
\put(70,5){\vector(1,-1){1}}
}
\put(-18,5){$\bx :$}
\put(-3,18){\small $1$}
\put(12,18){\small $2$}
\put(27,18){\small $\ldots$}
\put(147,18){\small $n$}
\put(-3,-8){\small $1$}
\put(12,-8){\small $2$}
\put(27,-8){\small $\ldots$}
\put(147,-8){\small $n$}
\put(57,18){\small $i$}
\put(57,-8){\small $i$}
\put(57,-8){\small $i$}
\put(68,-8){\small $i+1$}
\put(102,-8){\small $j$}
\end{picture}
\end{equation}
with weight $x_j$ attached to the $j$-th 
{\em vertical edge} for each $j=1,\ldots,n$,
and weight $1$ to each slanted edge. 
The $(i,j)$-component of $E(\bx)$ can then be 
read off by the weight of paths from $i$ at the top to $j$ 
at the bottom. 
Piling up the diagrams for $E(\bx^1)$,\ldots $E(\bx^m)$ 
all together, we obtain the following diagram 
for $E(\bx^1,\ldots,\bx^m)$. 
\begin{equation}\label{eq:Ex}
\begin{picture}(150,95)(-20,-10)
{\multiput(0,0)(15,0){8}{\line(0,1){75}}}
\multiput(0,75)(3,0){35}{\line(1,0){1}}
\multiput(0,60)(3,0){35}{\line(1,0){1}}
\multiput(0,45)(3,0){35}{\line(1,0){1}}
\multiput(0,30)(3,0){35}{\line(1,0){1}}
\multiput(0,15)(3,0){35}{\line(1,0){1}}
\multiput(0,0)(3,0){35}{\line(1,0){1}}
\multiput(0,75)(15,0){7}{\line(1,-1){15}}
\multiput(0,60)(15,0){7}{\line(1,-1){15}}
\multiput(0,45)(15,0){7}{\line(1,-1){15}}
\multiput(0,30)(15,0){7}{\line(1,-1){15}}
\multiput(0,15)(15,0){7}{\line(1,-1){15}} 
\put(-8,72){\small$0$} 
\put(-8,57){\small$1$} 
\put(-8,42){\small$2$} 
\put(-5,25){$\vdots$}
\put(-10,-2){\small$m$} 
\put(108,64){$\bx^1$} 
\put(108,49){$\bx^2$}
\put(110,25){$\vdots$}
\put(108,4){$\bx^m$}
\put(-1,78){\small$1$}
\put(14,78){\small$2$}
\put(27,78){\small$\ldots$}
\put(45,78){\small$i$}
\put(102,78){\small$n$}
\put(-3,-8){\small$1$}
\put(12,-8){\small$2$}
\put(27,-8){\small$\ldots$}
\put(75,-8){\small$j$}
\put(102,-8){\small$n$}
{\thicklines
\put(45,75){\line(0,-1){30}}
\put(45,45){\line(1,-1){15}}
\put(60,30){\line(0,-1){15}}
\put(60,15){\line(1,-1){15}}
}
{\thicklines
\put(128,50){\vector(0,-1){20}}
\put(128,50){\vector(1,-1){18}}
}
\end{picture}
\end{equation}
Here we make all the edges oriented downward, to the south 
or to the southeast. 
The $(i,j)$-component of $E(\bx^1,\ldots,\bx^m)$ 
is then given as the sum of weights over all paths 
from $i$ at the top to $j$ at the bottom. 
It can also be expressed as 
\begin{equation}\label{eq:Exform}
E(\bx^1,\ldots,\bx^m)^i_j
=\sum_{1\le k_i<k_{i+1}<\cdots<k_{j-1}\le m}\,
\prod_{b=i}^j \prod_{a=k_{b-1}+1}^{k_b-1} x^a_b,
\end{equation} 
where $k_{i-1}=0$ and $k_{j}=m+1$. 
Furthermore, we have

\begin{proposition}\label{prop:EE}
For any choice of row indices $i_1<\cdots<i_r$ and 
column indices $j_1<\ldots<j_r$, the minor determinant 
$\det E(\bx^1,\ldots,\bx^m)^{i_1,\ldots,i_r}_{j_1,\ldots,j_r}$
is expressed as the sum of weights over all 
$r$-tuples  $(\gamma_1,\ldots,\gamma_r)$ 
of nonintersecting paths
$\gamma_k$ from $i_k$ at the top to $j_k$ at the bottom
$(k=1,\ldots,r)$. 
\begin{equation}\label{eq:Exm}
\begin{picture}(290,95)(-160,-10)
\put(-175,30){
$\det E(\bx^1,\ldots,\bx^m)^{i_1,\ldots,i_r}_{j_1,\ldots,j_r}
=\dsum{(\gamma_1,\ldots,\gamma_r)}{}$
}
\multiput(0,0)(15,0){8}{\line(0,1){75}}
\multiput(0,75)(3,0){35}{\line(1,0){1}}
\multiput(0,60)(3,0){35}{\line(1,0){1}}
\multiput(0,45)(3,0){35}{\line(1,0){1}}
\multiput(0,30)(3,0){35}{\line(1,0){1}}
\multiput(0,15)(3,0){35}{\line(1,0){1}}
\multiput(0,0)(3,0){35}{\line(1,0){1}}
\multiput(0,75)(15,0){7}{\line(1,-1){15}}
\multiput(0,60)(15,0){7}{\line(1,-1){15}}
\multiput(0,45)(15,0){7}{\line(1,-1){15}}
\multiput(0,30)(15,0){7}{\line(1,-1){15}}
\multiput(0,15)(15,0){7}{\line(1,-1){15}}
\put(-15,64){$\bx^1$}
\put(-15,49){$\bx^2$}
\put(-12,25){$\vdots$}
\put(-15,4){$\bx^m$}
\put(12,78){\small$i_1$}
\put(27,78){\small$i_2$}
\put(40,78){\small$\ldots$}
\put(57,78){\small$i_r$}
\put(102,78){\small$n$}
\put(27,-8){\small$j_1$}
\put(57,-8){\small$j_2$}
\put(72,-8){\small$\ldots$}
\put(87,-8){\small$j_r$}
\put(102,-8){\small$n$}
{\thicklines
\put(125,50){\vector(0,-1){20}}
\put(125,50){\vector(1,-1){18}}
}
{\thicklines
\put(15,75){\line(0,-1){30}}
\put(15,45){\line(1,-1){15}}
\put(30,30){\line(0,-1){30}}
\put(30,75){\line(0,-1){15}}
\put(30,60){\line(1,-1){15}}
\put(45,45){\line(0,-1){15}}
\put(45,30){\line(1,-1){15}}
\put(60,15){\line(0,-1){15}}
\put(60,75){\line(1,-1){15}}
\put(75,60){\line(0,-1){30}}
\put(75,30){\line(1,-1){15}}
\put(90,15){\line(0,-1){15}}
}
\put(20,25){\small$\gamma_1$}
\put(35,40){\small$\gamma_2$}
\put(78,45){\small$\gamma_r$}
\end{picture}
\end{equation}
\end{proposition}
This proposition is an immediate consequence of the theorem 
of Gessel-Viennot \cite{GV}.  
In our context, however, it is also meaningful to 
understand this passage to nonintersecting paths 
through the multiplicative properties of minor determinants. 
Proposition \ref{prop:EE} is essentially reduced to the 
multiplicative formula
\begin{equation}\label{eq:mult}
\det(X Y)^{i_1,\ldots,i_r}_{j_1,\ldots,j_r}=
\sum_{k_1<\cdots<k_r}\,
\det X^{i_1,\ldots,i_r}_{k_1,\ldots,k_r}
\det Y^{k_1,\ldots,k_r}_{j_1,\ldots,j_r}
\end{equation}
for minor determinants of the product of matrices.
A key step is the following simple lemma. 
Note that the matrix $E(\bx)$ is decomposed in the form
\begin{equation}
E(\bx)=\diag{\bx}(1+\overline{x}_{n-1}E_{n-1,n})\cdots
(1+\overline{x}_{2}E_{2,3})
(1+\overline{x}_{1}E_{1,2}). 
\end{equation}
Also, the minor determinant 
\begin{equation}
\det(1+a E_{k,k+1})^{i_1,\ldots,i_r}_{j_1,\ldots,j_r}
\quad(i_1<\ldots<i_r,\ j_1<\ldots<j_r)
\end{equation}
vanishes unless either 
the two index sets $I=\br{i_1,\ldots,i_r}$ and $J=\br{j_1,\ldots,j_r}$ 
are identical, or 
$J$ is obtained from $I$ by replacing 
$k\in I$ by $k+1$. 
{}From this remark, we have 
\begin{lemma}\label{lem:EE}
For row indices $i_1<\cdots<i_r$ and column indices 
$j_1<\cdots<j_r$ given, the minor determinant 
$\det E(\bx)^{i_1,\ldots,i_r}_{j_1,\ldots,j_r}$ 
vanishes unless 
\begin{equation}
j_{a}=i_{a}\quad\mbox{or}\quad i_{a+1}
\quad\mbox{for all}\quad a=1\ldots,r.
\end{equation}
If this is the case, 
$\det E(\bx)^{i_1,\ldots,i_r}_{j_1,\ldots,j_r}$ 
is the product of $x_{j_a}$ over all $a$ 
such that $j_a=i_a$. 
\end{lemma}
Proposition \ref{prop:EE} is then obtained from 
Lemma \ref{lem:EE} by applying the multiplicative 
formula \eqref{eq:mult} to the decomposition 
$E(\bx^1,\ldots,\bx^m)=E(\bx^1)\cdots E(\bx^m)$. 
Path representations as in \eqref{eq:Exm} can also be 
translated into the language of tableaux; see 
for instance \cite{NNSY}. 

\par\medskip
We now turn to the graphical representation 
of $H(\bx)$ and $H(\bx^1,\ldots,\bx^m)$. 
We represent the matrix $H(\bx)$ by the diagram 
\begin{equation}
\begin{picture}(180,25)(-20,-5)
\put(0,7){\line(1,0){150}}
\multiput(0,0)(15,0){11}{\line(0,1){15}}
\multiput(-2.5,5)(15,0){11}{\small$\bullet$}
\put(-18,5){$\bx :$}
\put(-3,18){\small $1$}
\put(12,18){\small $2$}
\put(27,18){\small $\ldots$}
\put(147,18){\small $n$}
\put(-3,-8){\small $1$}
\put(12,-8){\small $2$}
\put(27,-8){\small $\ldots$}
\put(147,-8){\small $n$}
\put(58,18){\small $i$}
\put(102,-10){\small $j$}
{\thicklines
\put(60,17){\vector(0,-1){10}}
\put(60,7){\line(1,0){45}}
\put(86,7){\vector(1,0){1}}
\put(105,8){\line(0,-1){8}}
\put(105,-2){\vector(0,-1){1}}
}
\end{picture}
\end{equation}
with weight $x_j$ attached to the $j$-th {\em vertex}
($j=1,\ldots,m$). 
Then piling up the diagrams for $H(\bx_1), \ldots, H(\bx_m)$, 
we obtain the $m\times n$ rectangle. 
\begin{equation}\label{eq:HxRec}
\begin{picture}(150,75)(-20,-8)
\multiput(0,0)(15,0){8}{\line(0,1){60}}
\multiput(0,0)(0,15){5}{\line(1,0){105}}
\put(-18,58){$\bx^1$}
\put(-18,43){$\bx^2$}
\put(-15,20){$\vdots$}
\put(-18,-2){$\bx^m$}
\put(-3,63){\small$1$}
\put(12,63){\small$2$}
\put(27,63){\small$\ldots$}
\put(45,63){\small$i$}
\put(102,63){\small$n$}
\put(-3,-8){\small$1$}
\put(12,-8){\small$2$}
\put(27,-8){\small$\ldots$}
\put(90,-8){\small$j$}
\put(102,-8){\small$n$}
{\thicklines
\put(125,45){\vector(0,-1){20}}
\put(125,45){\vector(1,0){20}}
}
\put(42.5,58){\small$\bullet$}
\put(42.5,43){\small$\bullet$}
\put(42.5,28){\small$\bullet$}
\put(57.5,28){\small$\bullet$}
\put(57.5,28){\small$\bullet$}
\put(57.5,13){\small$\bullet$}
\put(72.5,13){\small$\bullet$}
\put(87.5,13){\small$\bullet$}
\put(87.5,-2){\small$\bullet$}
{\thicklines
\put(45,60){\line(0,-1){15}}
\put(45,45){\line(0,-1){13}}
\put(46.5,30){\line(1,0){11.5}}
\put(60,28.5){\line(0,-1){11.5}}
\put(61.5,15){\line(1,0){11.5}}
\put(76.5,15){\line(1,0){11.5}}
\put(90,13.5){\line(0,-1){13.5}}
\put(50,35){$\gamma$}
}
\end{picture}
\end{equation}
In this diagram for $H(\bx_1,\ldots,\bx_m)$, 
for each $a=1,\ldots,m$ and $b=1,\ldots,n$, 
we attach the weight $x^a_b$ 
to the vertex with coordinates $(a,b)$.
This time, the weight of a path $\gamma$
is defined to be the product of all $x^a_b$'s  
attached to the vertices on $\gamma$. 
\begin{proposition}\label{prop:HH}
For any choice of row indices $i_1<\cdots<i_r$ and 
column indices $j_1<\ldots<j_r$, the minor determinant 
$\det H(\bx^1,\ldots,\bx^m)^{i_1,\ldots,i_r}_{j_1,\ldots,j_r}$
is expressed as the sum of weights over all 
$r$-tuples  $(\gamma_1,\ldots,\gamma_r)$ 
of nonintersecting paths
$\gamma_k : (1,i_k) \to (m, j_k)$ \ $(k=1,\ldots,r)$. 
\begin{equation}
\begin{picture}(290,95)(-175,-10)
\put(-175,30){
$\det H(\bx^1,\ldots,\bx^m)^{i_1,\ldots,i_r}_{j_1,\ldots,j_r}
=\dsum{(\gamma_1,\ldots,\gamma_r)}{}$
}
\multiput(0,7)(15,0){8}{\line(0,1){60}}
\multiput(0,67)(0,-15){5}{\line(1,0){105}}
\put(-15,64){$\bx^1$}
\put(-15,49){$\bx^2$}
\put(-12,25){$\vdots$}
\put(-15,4){$\bx^m$}
\put(12,74){\small$i_1$}
\put(27,74){\small$i_2$}
\put(40,74){\small$\ldots$}
\put(57,74){\small$i_r$}
\put(102,74){\small$n$}
\put(27,-4){\small$j_1$}
\put(57,-4){\small$j_2$}
\put(72,-4){\small$\ldots$}
\put(87,-4){\small$j_r$}
\put(102,-4){\small$n$}
{\thicklines
\put(120,45){\vector(0,-1){20}}
\put(120,45){\vector(1,0){20}}
}
{\thicklines
\put(15,70){\line(0,-1){33}}
\put(15,37){\line(1,0){15}}
\put(30,37){\line(0,-1){33}}
\put(30,70){\line(0,-1){18}}
\put(30,52){\line(1,0){15}}
\put(45,52){\line(0,-1){30}}
\put(45,22){\line(1,0){15}}
\put(60,22){\line(0,-1){18}}
\put(60,70){\line(0,-1){3}}
\put(60,67){\line(1,0){15}}
\put(75,67){\line(0,-1){30}}
\put(75,37){\line(1,0){15}}
\put(90,37){\line(0,-1){33}}
}
\put(20,27){\small$\gamma_1$}
\put(35,42){\small$\gamma_2$}
\put(78,42){\small$\gamma_r$}
\end{picture}
\end{equation}
\end{proposition}
The following 
corresponds to Lemma \ref{lem:EE} for $E(\bx)$. 
\begin{lemma}\label{lem:HH}
For row indices $i_1<\cdots<i_r$ and column indices 
$j_1<\cdots<j_r$ given, the minor determinant 
$\det H(\bx)^{i_1,\ldots,i_r}_{j_1,\ldots,j_r}$ 
vanishes unless 
\begin{equation}
i_1\le j_1<i_2\le j_2<\cdots<i_r\le j_r. 
\end{equation}
If this is the case, one has
\begin{equation}
\det H(\bx)^{i_1,\ldots,i_r}_{j_1,\ldots,j_r}
=x_{i_1}\cdots x_{j_1} x_{i_2}\cdots x_{j_2}\cdots
x_{i_r}\cdots x_{j_r}. 
\end{equation}
\end{lemma}
Note also
\begin{equation}
H(\bx)=(1+x_1E_{1,2})(1+x_2 E_{2,3})\cdots(1+x_{n-1} E_{n-1,n})
\,\diag{\bx}. 
\end{equation}
\par\medskip 
In the following, 
we apply the same idea 
to nonintersecting paths in triangles and trapezoids. 
For this purpose,  we define 
\begin{equation}
\Lambda_{\ge k}=\sum_{i=k}^{n-1}E_{i,i+1}
\qquad(k=1,\ldots,n), 
\end{equation}
so that $\Lambda_{\ge1}=\Lambda$ 
and $\Lambda_{\ge n}=0$. 
With these {\em truncated shift matrices}, 
we introduce the following variations of $E(\bx)$ 
and $H(\bx)$:
\begin{equation}
E_k(\bx)=\diag{\bx}+\Lambda_{\ge k},\quad
H_k(\bx)=(\diag{\overline{\bx}}-\Lambda_{\ge k})^{-1}
\end{equation}
for $k=1,\ldots,n$.  
When we use these notations, we will tacitly assume that 
$\bx=(1,\ldots,1,x_k,\ldots,x_n)$, i.e., $x_j=1$ $(j<k)$,
unless otherwise mentioned. 
Under this convention, 
$E_k(\bx)$ and $H_k(\bx)$ are expressed as 
\begin{equation}
E_k(\bx)=\left[\matrix{
1 & 0\cr
0 & E(\bx')
}\right],\quad 
H_k(\bx)=\left[\matrix{
1 & 0\cr
0 & H(\bx')
}\right],
\end{equation}
respectively, where $\bx'=(x_k,x_{k+1},\ldots,x_n)$; 
we will often identify the $(n-k-1)$-vector 
$(x_k,\ldots,x_n)$ with the $n$-vector 
$(1,\ldots,1,x_k,\ldots,x_n)$,
by putting $1$'s in front. 
Assuming that $m\le n$, 
let us consider a sequence of $n$-vectors 
$\bu^1,\ldots,\bu^m$, $\bu^i=(u^i_i,\ldots,u^i_n)$, 
and arrange $u^i_j $ $(i\le j)$ in 
the form 
\begin{equation}\label{eq:U}
U=\pmatrix{u^i_j}_{i\le j}=\left[\matrix{
\smallskip
u^1_1 & u^1_2 & \ldots & u^1_m & \ldots & u^1_n \cr
\smallskip
 & u^2_2 & \ldots & u^2_m &\ldots&u^2_n\cr
\smallskip
&&\ddots &\vdots&&\vdots\cr
&&& u^m_m & \ldots & u^m_n
}\right].
\end{equation}
For such a table $U$ given, we define an $n\times n$ matrix 
$E_U$ by 
\begin{equation} 
E_U=E_1(\bu^1)E_2(\bu^2)\cdots,E_m(\bu^m). 
\end{equation}
The entries of this matrix can be represented by the diagram
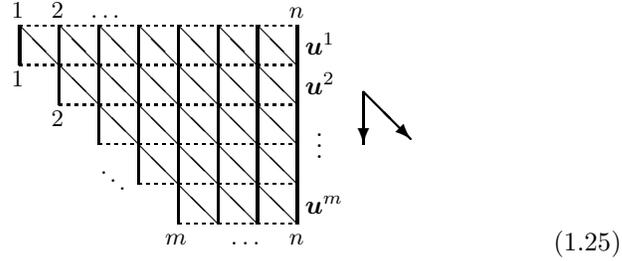
\begin{equation}\label{eq:Eu}
\begin{picture}(150,95)(-20,-10)
{\thicklines
\put(0,75){\line(0,-1){15}}
\put(15,75){\line(0,-1){30}}
\put(30,75){\line(0,-1){45}}
\put(45,75){\line(0,-1){60}}
\put(60,75){\line(0,-1){75}}
\put(75,75){\line(0,-1){75}}
\put(90,75){\line(0,-1){75}}
\put(105,75){\line(0,-1){75}}
}
\multiput(0,75)(3,0){35}{\line(1,0){1}}
\multiput(0,60)(3,0){35}{\line(1,0){1}}
\multiput(15,45)(3,0){30}{\line(1,0){1}}
\multiput(30,30)(3,0){25}{\line(1,0){1}}
\multiput(45,15)(3,0){20}{\line(1,0){1}}
\multiput(60,0)(3,0){15}{\line(1,0){1}}
\multiput(0,75)(15,0){7}{\line(1,-1){15}}
\multiput(15,60)(15,0){6}{\line(1,-1){15}}
\multiput(30,45)(15,0){5}{\line(1,-1){15}}
\multiput(45,30)(15,0){4}{\line(1,-1){15}}
\multiput(60,15)(15,0){3}{\line(1,-1){15}}
\put(108,64){$\bu^1$}
\put(108,49){$\bu^2$}
\put(112,25){$\vdots$}
\put(108,4){$\bu^m$}
\put(-3,78){\small$1$}
\put(12,78){\small$2$}
\put(27,78){\small$\ldots$}
\put(102,78){\small$n$}
\put(-3,52){\small$1$}
\put(12,37){\small$2$}
\put(30,12){\small$\ddots$}
\put(55,-8){\small$m$}
\put(80,-8){\small$\ldots$}
\put(102,-8){\small$n$}
{\thicklines
\put(130,50){\vector(0,-1){20}}
\put(130,50){\vector(1,-1){18}}
}
\end{picture}
\end{equation}
with the weights $u^i_j$ attached to the vertical 
edges;
for each $(i,j)$ with $1\le i\le j\le n$, 
$(E_U)^i_j$ is the sum of weights over all 
paths $\gamma$ from $i$ at the top to $j$ 
along the lower rim. 
The minor determinants of $E_U$ are also 
represented by nonintersecting paths in diagram \eqref{eq:Eu}. 
We also introduce
\begin{equation}
H_{U}=H_m(\bu^m)\cdots H_2(\bu^2) H_1(\bu^1), 
\end{equation}
so that $H_U=DE_{\overline{U}}^{-1}D^{-1}$. 
The diagram for $H_U$ is given by 
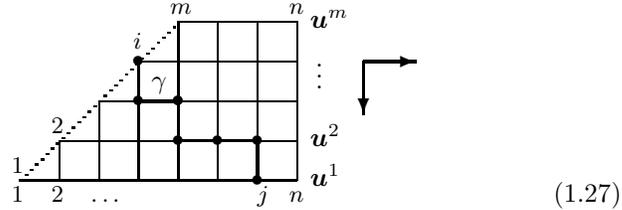
\begin{equation}\label{eq:HuRecDn}
\begin{picture}(150,75)(-20,-8)
\multiput(0,0)(2,2){31}{\line(1,0){1}}
\put(15,0){\line(0,1){15}}
\put(30,0){\line(0,1){30}}
\put(45,0){\line(0,1){45}}
\put(60,0){\line(0,1){60}}
\put(75,0){\line(0,1){60}}
\put(90,0){\line(0,1){60}}
\put(105,0){\line(0,1){60}}
\put(0,0){\line(1,0){105}}
\put(15,15){\line(1,0){90}}
\put(30,30){\line(1,0){75}}
\put(45,45){\line(1,0){60}}
\put(60,60){\line(1,0){45}}
\put(110,58){$\bu^m$}
\put(112,35){$\vdots$}
\put(110,13){$\bu^2$}
\put(110,-2){$\bu^1$}
\put(-3,3){\small$1$}
\put(12,18){\small$2$}
\put(43,50){\small$i$}
\put(57,63){\small$m$}
\put(102,63){\small$n$}
\put(-3,-8){\small$1$}
\put(12,-8){\small$2$}
\put(27,-8){\small$\ldots$}
\put(90,-8){\small$j$}
\put(102,-8){\small$n$}
{\thicklines
\put(130,45){\vector(0,-1){20}}
\put(130,45){\vector(1,0){20}}
}
\put(42.5,43){\small$\bullet$}
\put(42.5,28){\small$\bullet$}
\put(57.5,28){\small$\bullet$}
\put(57.5,28){\small$\bullet$}
\put(57.5,13){\small$\bullet$}
\put(72.5,13){\small$\bullet$}
\put(87.5,13){\small$\bullet$}
\put(87.5,-2){\small$\bullet$}
{\thicklines
\put(45,45){\line(0,-1){13}}
\put(46.5,30){\line(1,0){11.5}}
\put(60,28.5){\line(0,-1){11.5}}
\put(61.5,15){\line(1,0){11.5}}
\put(76.5,15){\line(1,0){11.5}}
\put(90,13.5){\line(0,-1){13.5}}
\put(50,35){$\gamma$}
}
\end{picture}
\end{equation}
or alternatively by
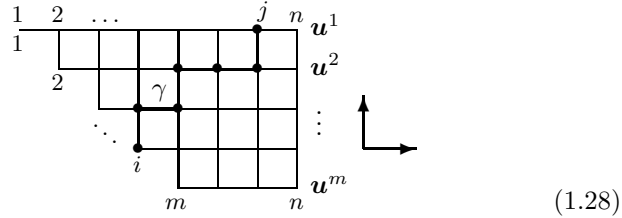
\begin{equation}\label{eq:HuRecUp}
\begin{picture}(150,75)(-20,-8)
\put(15,60){\line(0,-1){15}}
\put(30,60){\line(0,-1){30}}
\put(45,60){\line(0,-1){45}}
\put(60,60){\line(0,-1){60}}
\put(75,60){\line(0,-1){60}}
\put(90,60){\line(0,-1){60}}
\put(105,60){\line(0,-1){60}}
\put(0,60){\line(1,0){105}}
\put(15,45){\line(1,0){90}}
\put(30,30){\line(1,0){75}}
\put(45,15){\line(1,0){60}}
\put(60,0){\line(1,0){45}}
\put(110,58){$\bu^1$}
\put(110,43){$\bu^2$}
\put(112,20){$\vdots$}
\put(110,-2){$\bu^m$}
\put(-3,63){\small$1$}
\put(12,63){\small$2$}
\put(27,63){\small$\ldots$}
\put(90,65){\small$j$}
\put(102,63){\small$n$}
\put(-3,52){\small$1$}
\put(12,37){\small$2$}
\put(27,16){\small$\ddots$}
\put(55,-8){\small$m$}
\put(43,6){\small$i$}
\put(102,-8){\small$n$}
{\thicklines
\put(130,15){\vector(0,1){20}}
\put(130,15){\vector(1,0){20}}
}
\put(42.5,13){\small$\bullet$}
\put(42.5,28){\small$\bullet$}
\put(57.5,28){\small$\bullet$}
\put(57.5,43){\small$\bullet$}
\put(72.5,43){\small$\bullet$}
\put(87.5,43){\small$\bullet$}
\put(87.5,58){\small$\bullet$}
{\thicklines
\put(45,15){\line(0,1){15}}
\put(45,30){\line(1,0){15}}
\put(60,30){\line(0,1){15}}
\put(60,45){\line(1,0){30}}
\put(90,45){\line(0,1){15}}
\put(50,35){$\gamma$}
}
\end{picture}
\end{equation}
with edges oriented upward or rightward.
The minor determinants of $H_U$ are expressed 
in terms of nonintersecting paths in diagram 
\eqref{eq:HuRecDn} or \eqref{eq:HuRecUp}.

\subsection{Minor determinants of triangular matrices}
The matrices $E_U$ discussed above can be thought 
of as {\em canonical forms} of generic upper triangular 
matrices $M$ of the form
\begin{equation}
M=\left[
\begin{picture}(70,40)(8,40)
\put(10,80){\small$1$}
\put(38,80){\small$m$}
\put(70,80){\small$n$}
\put(47,47){\small$a^i_j$}
\put(10,70){$\ast$}
\put(40,70){$\ast$}
\put(50,70){\small$1$}
\put(70,50){\small$1$}
\put(70,40){$\ast$}
\put(70,10){$\ast$}
\put(67,67){\large$0$}
\put(20,20){\Large$0$}
\multiput(16,67)(3,-3){18}{$.$}
\multiput(46,67)(3,-3){8}{$.$}
\multiput(57,66)(3,-3){4}{$.$}
\multiput(16,72)(4,0){6}{$.$}
\multiput(71,37)(0,-4){6}{$.$}
\end{picture}
\right]. 
\end{equation}
Let $M=\pmatrix{a^i_j}_{i,j=1}^n$ be an $n\times n$ 
upper triangular matrix satisfying the condition 
\begin{equation}\label{eq:htm}
a^i_j=0\quad (j<i\ \ \mbox{or}\ \  j>i+m),\quad a^i_j=1\quad(j=i+m);
\end{equation}
when $m=n$, this simply means that $M$ is upper triangular. 
For each $(i,j)$ with $1\le i\le j\le n$, 
introduce the notation of minor determinants
\begin{equation}
Q_{i,j}=Q_{i,j}(M)=\det M^{1,\ldots,j-i+1}_{i,i+1,\ldots,j};
\end{equation} 
when $j-i+1=0$, we set $Q_{i,j}=1$. 
We remark that 
the condition \eqref{eq:htm} for an upper triangular matrix 
$M$ is equivalent to the condition 
\begin{equation}
Q_{m+1,j}=1,
\qquad Q_{i,j}=0\quad(m+1<i\le n),
\end{equation} 
for minor determinants.  
The following proposition is due to 
A.~Berenstein, S.~Fomin and A.~Zelevinsky
\cite{BFZ}. 
\begin{proposition}\label{prop:MEE}
Let $M=\pmatrix{a^i_j}_{i,j=1}^n$ be an $n\times n$ 
upper triangular matrix satisfying the condition \eqref{eq:htm} 
for some $m$ $(1\le m\le n)$.
Suppose that $Q_{i,j}\ne 0$ 
for any $(i,j)$ with $i\le j$ and $i\le m$.  
Then $M$ can be decomposed uniquely in the form
\begin{equation}\label{eq:MEE}
M=E_1(\bv^1)E_2(\bv^2)\cdots E_m(\bv^m), 
\end{equation}
where 
$\bv^i=(1,\ldots,1,v^i_i,\ldots,v^i_n)$, $v^i_j\ne 0$, 
for  $i=1,\ldots,m$. 
Furthermore, $v^i_j$ are determined by
\begin{equation}\label{eq:vinQ}
v^i_i=Q_{i,i},\quad
v^i_j=\frac{Q_{i,j}\,Q_{i+1,j-1}}{Q_{i+1,j}\,Q_{i,j-1}}
\qquad(i<j,\ i\le m ). 
\end{equation}
\end{proposition}

\proof Assume first that $M$ is decomposed as in \eqref{eq:MEE}. 
Then the minor determinants of $M$ are expressed in terms 
of nonintersecting paths in diagram \eqref{eq:Eu} for 
$V=\pmatrix{v^i_j}_{i\le j}$. 
In particular we have
\begin{equation}\label{eq:Qinv}
Q_{i,j}=\det M^{1,\ldots,j-i+1}_{i,\ldots,j} =
\prod_{(a,b) :\, a \ge i, b\le j}\, v^i_j,
\end{equation}
since there is only one $(j-i+1)$-tuple 
of nonintersecting paths relevant to the path representation 
of this case.  Expression \eqref{eq:vinQ} follows 
immediately from \eqref{eq:Qinv}, which also implies  
the uniqueness of decomposition \eqref{eq:MEE}. 
It remains to show that $M$ has a decomposition of the form 
\eqref{eq:MEE} under the condition on $Q_{i,j}$.  
We express $M$ in the form 
\begin{equation}
M=
\left[\matrix{A & B \cr C & D}\right],
\end{equation}
so that $B$ becomes a square matrix of size $n-m+1$:
$B=M^{1,\ldots,n-m+1}_{m,m+1,\ldots,n}$.
We can apply the Gauss decomposition to the matrix 
\begin{equation}
B=\quad\left[
\begin{picture}(60,30)(8,35)
\put(-4,60){\small$1$}
\put(-6,30){\small$m$}
\put(10,60){$\ast$}
\put(60,10){$\ast$}
\put(20,60){\small$1$}
\put(60,20){\small$1$}
\put(10,30){$\ast$}
\put(30,10){$\ast$}
\put(47,47){\Large$0$}
\put(13,13){\large$0$}
\multiput(16,57)(3,-3){15}{$.$}
\multiput(26,57)(3,-3){11}{$.$}
\multiput(16,28)(3,-3){5}{$.$}
\multiput(11,57)(0,-4){6}{$.$}
\multiput(37,12)(4,0){6}{$.$}
\end{picture}
\right]
\end{equation}
since 
\begin{equation}
\det B^{1,\ldots,r}_{1,\ldots,r}=\det M^{1,\ldots,r}_{m,\ldots,m+r-1}
=Q_{m,m+r-1}(M)\ne 0
\end{equation}
for $r=1,\ldots, n-m+1$. 
It also turns out 
that the lower and the upper triangular 
components of the Gauss decomposition $B=B_{<0} B_{\ge0}$ 
are in the form 
\begin{equation}
B_{<0}=\quad\left[
\begin{picture}(60,30)(8,35)
\put(-4,60){\small$1$}
\put(-6,30){\small$m$}
\put(10,60){\small$1$}
\put(60,10){\small$1$}
\put(10,50){$\ast$}
\put(50,10){$\ast$}
\put(10,30){$\ast$}
\put(30,10){$\ast$}
\put(45,45){\Large$0$}
\put(13,13){\large$0$}
\multiput(16,57)(3,-3){15}{$.$}
\multiput(16,47)(3,-3){11}{$.$}
\multiput(16,28)(3,-3){5}{$.$}
\multiput(11,48)(0,-4){4}{$.$}
\multiput(35,12)(4,0){4}{$.$}
\end{picture}
\right],
\quad 
B_{\ge0}=\left[
\begin{picture}(60,30)(8,35)
\put(10,60){$\ast$}
\put(60,10){$\ast$}
\put(20,60){\small$1$}
\put(60,20){\small$1$}
\put(50,50){\Large$0$}
\put(20,20){\Large$0$}
\multiput(16,57)(3,-3){15}{$.$}
\multiput(26,57)(3,-3){11}{$.$}
\end{picture}
\right]. 
\end{equation}
Denoting the diagonal entries of $B_{\ge0}$ 
by $v^m_m,\ldots,v^m_n$, 
we introduce the vector 
$\bv^m=(1,\ldots,1,v^m_m,\ldots,v^m_n)$. 
Then we have the following decomposition of $M$:
\begin{equation}
M=\left[\matrix{A & B \cr C& D}\right]
=\left[\matrix{A & B_{<0} \cr C & D'}\right]
\left[\matrix{1 & 0 \cr 0 & B_{\ge 0}}\right]
= M' E_m(\bv^m),
\end{equation}
where $D'=D B_{\ge0}^{-1}$.  
The matrix $M'$ thus obtained satisfies the condition 
\eqref{eq:htm} with $m$ replaced by $m-1$, 
as can be seen from the decomposition above. 
Also, if $i\le m-1$ and $i\le j$, 
from $M=M' E_m(\bv)$ it follows that 
\begin{eqnarray}
&&
Q_{i,j}(M)=\det M^{1,\ldots,j-i+1}_{i,\ldots,j}
\nonumber\\
&&\phantom{Q_{i,j}(M)}
=\det(M')^{1,\ldots,j-i+1}_{i,\dots,j}
\det E_m(\bv^m)^{i,\ldots,j}_{i,\ldots,j}
\nonumber\\
&&\phantom{Q_{i,j}(M)}
=Q_{i,j}(M') \,v^m_m\cdots v^m_j\ne 0. 
\end{eqnarray}
Hence we can apply the descending induction on $m$ 
to obtain the decomposition  \eqref{eq:MEE}.
\qed

We now translate Proposition \ref{prop:MEE} into a statement 
concerning the decomposition of type $H_U$. 
Let $H$ be an $n\times n$ invertible upper triangular matrix. 
For the decomposition of type $H=H_U$, it is convenient 
to use the following notation
\begin{equation}
\tau^i_j=\tau^i_j(H)=\det H^{1,\ldots,i}_{j-i+1,\ldots,j}
\qquad(1\le i\le j\le n). 
\end{equation}
We also define $\tau^0_j=1$ for any $j$. 
Setting $M=D H^{-1} D^{-1}$, $D=\diag{(-1)^{i-1}}_{i=1}^n$, 
we compare the minor determinants $\tau^i_j=\tau^i_j(H)$ 
and $Q_{i,j}=Q_{i,j}(M)$.  
We remark that 
\begin{equation}
\det H^{1,\ldots,i}_{j-i+1,\ldots,j}
=\frac{\det M^{1,\ldots,j-i,j+1,\ldots,n}_{i+1,\ldots,n}
}{\det M }
=\frac{\det M^{1,\ldots,j-i}_{i+1,\dots,j}}
{\det M^{1,\ldots,j}_{1,\ldots,j}};
\end{equation}
for these particular minor determinants, we have no minus sign.
This implies
\begin{equation}
\tau^i_j=\frac{Q_{i+1,j}}{Q_{1,j}}, \qquad 
Q_{i,j}=\frac{\tau^{i-1}_j}{\tau^j_j}\qquad(i\le j). 
\end{equation} 
Note also that $\tau^j_j=Q_{1,j}^{-1}$. 
Hence we see that the condition \eqref{eq:htm} is equivalent to
\begin{equation}
\tau^i_j=\delta_{i,j} \tau^m_j\qquad(m<i\le j\le n). 
\end{equation}
\begin{proposition}\label{prop:HHH}
Let $H$ be an $n\times n$ 
upper triangular matrix, and suppose that 
the minor determinants $\tau^i_j=\tau^i_j(H)$ $(1\le i\le j\le n)$
satisfy the condition
\begin{equation}\label{eq:condH}
\tau^i_j\ne0 \quad (1\le i\le m),\quad
\tau^i_j=\delta_{i,j} \tau^m_j\quad(m<i\le n), 
\end{equation}
for some $m$ $(1\le m\le n)$. 
Then the matrix $H$ can be decomposed uniquely in the form 
\begin{equation}\label{eq:HH}
H=H_m(\bu^m)\cdots H_2(\bu^2) H_1(\bu^1),
\end{equation}
where $\bu^i=(1,\ldots,1,u^i_i,\ldots,u^i_n)$, 
$u^i_j\ne 0$, for $i=1,\ldots,m$. 
Furthermore $u^i_j$ are determined by
\begin{equation}\label{eq:uintau}
u^i_i=\frac{\tau^i_i}{\tau^{i-1}_i},\qquad
u^i_j=\frac{\tau^i_j\ \tau^{i-1}_{j-1}}{\tau^{i-1}_j\tau^i_{j-1}}
\quad(i<j,\,i\le m). 
\end{equation}
\end{proposition}
Under the condition \eqref{eq:condH}, from Proposition \ref{prop:MEE} 
we have 
\begin{equation}\label{eq:HEE}
DH^{-1}D^{-1}=M=E_1(\overline{\bu}^1) E_2(\overline{\bu}^2)\cdots
E_m(\overline{\bu}^m),
\end{equation}
where we have set $\bu^i=\overline{\bv}^i$ $(i=1,\ldots,m)$. 
Once we have the decomposition \eqref{eq:HH}, 
the minor determinants 
of $H$ is expressed in terms of nonintersecting paths 
in diagram \eqref{eq:HuRecUp} for $U=\pmatrix{u^i_j}_{i,j}$. 
In particular, each $\tau^i_j$ is expressed as 
\begin{equation}\label{eq:tauinu}
\tau^i_j=\det H^{1,\ldots,i}_{j-i+1,\ldots,j}=
\prod_{(a,b); \,a\le i,\,b\le j}\,u^a_b,
\end{equation}
since there is only one $i$-tuple of nonintersecting paths 
relevant to this minor determinant. 
Expression \eqref{eq:uintau} for $u^i_j$ follows immediately 
from \eqref{eq:tauinu}.  

\par\medskip
Proposition \ref{prop:HHH} 
implies the following theorem 
concerning the path representation of minor determinants
of a triangular matrix.
\begin{theorem}\label{thm:H}
Let $H$ be an $n\times n$ upper triangular matrix, and suppose that 
the minor determinants $\tau^i_j=\tau^i_j(H)$ $(i\le j)$
satisfy 
the condition 
\begin{equation}
\tau^i_j\ne 0 \quad (1\le i\le m),\qquad
\tau^i_j =\delta_{i,j}\tau^m_j\quad (m<i\le n)
\end{equation}
for some $m$ $(1\le m\le n)$.  
For each $(i,j)$ with $1\le i\le j\le n$, $i\le m$, define 
\begin{equation}
u^i_i=\frac{\tau^i_i}{\tau^{i-1}_i},\qquad
u^i_j=\frac{\tau^i_j\ \tau^{i-1}_{j-1}}{\tau^{i-1}_j\tau^i_{j-1}}
\quad(i<j,\,i\le m). 
\end{equation}
Then, for any choice of row indices $i_1<\ldots<i_r$ 
and column indices $j_1<\ldots<j_r$, 
the minor determinant $\det H^{i_1,\ldots,i_r}_{j_1,\ldots,j_r}$ 
is expressed as a sum 
\begin{equation}
\begin{picture}(170,60)(-80,-5)
\unitlength=1.2pt
\put(-85,15){$
\det H^{i_1,\ldots,i_r}_{j_1,\ldots,j_r}
=\dsum{(\gamma_1,\ldots,\gamma_r)}{}
$}
\put(0,40){\line(1,0){70}}
\put(0,40){\line(1,-1){40}}
\put(40,0){\line(1,0){30}}
\put(70,40){\line(0,-1){40}}
{\thicklines
\put(15,25){\line(1,0){10}}
\put(25,25){\line(0,1){5}}
\put(25,30){\line(1,0){10}}
\put(35,30){\line(0,1){10}}
\put(25,15){\line(0,1){5}}
\put(25,20){\line(1,0){15}}
\put(40,20){\line(0,1){10}}
\put(40,30){\line(1,0){5}}
\put(45,30){\line(0,1){10}}
\put(45,0){\line(0,1){10}}
\put(45,10){\line(1,0){10}}
\put(55,10){\line(0,1){15}}
\put(55,25){\line(1,0){5}}
\put(60,25){\line(0,1){15}}
\put(8,20){\small $i_1$}
\put(18,10){\small $i_2$}
\put(42,-8){\small $i_r$}
\put(30,44){\small $j_1$}
\put(40,44){\small $j_2$}
\put(60,44){\small $j_r$}
\put(58,5){$U$}
{\thicklines
\put(85,15){\vector(0,1){15}}
\put(85,15){\vector(1,0){15}}
}
}
\end{picture}
\end{equation} 
of weights associated with $U=\pmatrix{u^i_j}_{i,j}$, 
over all $r$-tuples of nonintersecting paths 
$\gamma_k : (\min\br{i_k,m},i_k)\to(1,j_k)$ 
from $i_k$ along the lower rim to $j_k$ 
at the top $(k=1,\ldots,r)$,  
in diagram \eqref{eq:HuRecUp}.
\end{theorem}
\begin{remark}\label{rem:HHH}\rm\ 
Proposition \ref{prop:HHH} for $m=n$ can be reformulated as follows. 
Let us denote 
\begin{equation}
\CB=\br{ B=\pmatrix{b^i_j}_{i,j=1}^n \in GL_n(\BK)\mid 
b^i_j=0\quad(i>j)} 
\end{equation}
the group of all $n\times n$ invertible upper triangular 
matrices.  
For each $U=\pmatrix{u^i_j}_{i,j=1}^n\in\CB$, 
we define $H=\pmatrix{h^i_j}_{i,j=1}^n\in\CB$ 
by setting
\begin{equation}
h^i_j=\sum_{\gamma:(i,i) \to (1,j)}  u_\gamma\quad(i\le j),\quad
h^i_j=0\quad(i>j). 
\end{equation}
We now define two open subsets of $\CB$ as follows:
\begin{eqnarray}
&&\CB_0=\br{U=\pmatrix{u^i_j}_{i,j=1}^n\in\CB\mid u^i_j\ne 0
\quad(i\le j)},
\nonumber\\
&&\CB_\tau=\br{H=\pmatrix{h^i_j}\in\CB\mid \tau^i_j(H)\ne 0
\quad(i\le j)}.
\end{eqnarray}
Then the correspondence $U\mapsto H$ induces the isomorphism 
of affine varieties $h :  \CB_0\, \iso\,\CB_\tau$.
The inverse mapping $H\mapsto U$ is given by
\begin{equation}
u^i_i=\frac{\tau^i_i}{\tau^{i-1}_i},\qquad
u^i_j=\frac{\tau^i_i\ \tau^{i-1}_{j-1}}{\tau^{i-1}_i\tau^i_{j-1}}
\quad(i<j),
\end{equation}
where $\tau^i_j=\tau^i_j(H)$ for $i\le j$.  
Under this correspondence $U\leftrightarrow H$, 
for any choice of row indices $i_1<\ldots<i_r$ and 
column indices $j_1<\ldots<j_r$, 
the minor determinant $\det H^{i_1,\ldots,i_r}_{j_1,\ldots,j_r}$ 
of $H$ 
is expressed as the sum of weights, associated with $U$, 
over all $r$-tuples of nonintersecting paths 
$\gamma_k: (i_k,i_k)\to (1,j_k)$ \ ($k=1,\ldots,r$). 
\end{remark}


\subsection{Passage from tropical  to combinatorial variables}

We now assume that $\BK$ is a field of characteristic 0.  
Consider the ring of polynomials 
$\BK[x]=\BK[\,x_i\ (i\in I)]$ in a set of variables $x=(x_i)_{i\in I}$. 
Denoting by 
\begin{equation}
\BN^{(I)}=\br{\alpha=
(\alpha_i)_{i\in I}
\mid \alpha_i=0 \ \ 
\mbox{except for a finite number of $i$'s}
}
\end{equation}
the set of multi-indices, we use the notation 
of multi-indices $x^\alpha=\prod_{i\in I} x_i^{\alpha_i}$
for the monomials in the $x$-variables. 
Note that any polynomial $a(x)\in\BK[x]$ 
is expressed uniquely in the form
\begin{equation}
a(x)=\sum_{\alpha\in A}\, a_\alpha\,x^\alpha
\qquad(a_\alpha\in \BK^\ast),
\end{equation}
as a sum of monomials over a finite subset 
$A\subset\BN^{(I)}$ of multi-indices, 
with nonzero coefficients. 
In this way, a polynomial $a(x)$ is identified with 
a pair $(A,a)$ of a finite subset $A\subset\BN^{(I)}$ and 
a mapping $a: A\to\BK$. 
Note that $0\in\BK[x]$ and $c\in\BK[x]$  ($c\in \BK^\ast$)  correspond to 
$(\phi,\phi)$ and $(\br{0},c)$, respectively.  

In the following, we fix a multiplicative 
subgroup $\BK_{>0}$ of $\BK^\ast$
such that $c,c'\in \BK_{>0}\ \ \Rightarrow \ \ c+c'\in \BK_{>0}$.  
We say that 
a nonzero rational function $f(x)\in\BK(x)=\BK(x_i\ (i\in I))$
in the $x$-variables is {\em subtraction free} 
(or {\em tropical\/}) with respect to the cone $\BK_{>0}$
if it is expressed as a ratio
\begin{equation} \label{eq:fab}
f(x)=\frac{a(x)}{b(x)}\qquad (\,a(x),  b(x)\in \BK_{>0}[x]\,)
\end{equation}
of two polynomials with coefficients in $\BK_{>0}$. 
We denote by $\BK(x)_{>0}$ the set of all subtraction-free 
rational functions with respect to $\BK_{>0}$. 
It is clear that $\BK(x)_{>0}$ forms again a multiplicative 
subgroup of $\BK(x)^\ast$ closed under the 
addition. 
It is worthwhile to note that all the coefficients 
of a {\em polynomial} $f(x)\in\BK(x)_{>0}$ may {\em not} 
necessarily belong to $\BK_{>0}$: Observe the example
\begin{equation}
f(x,y)=\frac{x^3+y^3}{x+y}=x^2-xy+y^2.
\end{equation}

For a subtraction-free rational function $f=f(x)\in\BK(x)_{>0}$ 
given, 
choose an expression as \eqref{eq:fab}. 
Expressing $a(x)$ and $b(x)$ as
\begin{equation}
a(x)=\sum_{\alpha\in A}\ a_\alpha\, x^\alpha,\quad
b(x)=\sum_{\beta\in B}\ b_\beta\,x^\beta
\end{equation} 
with coefficients in $\BK_{>0}$, 
we define two piecewise linear functions 
$M(f)$ and $m(f)$ on 
$\BR^I$ by 
\begin{equation}
\begin{array}{c}\smallskip
M(f)=
\max\br{\pr{\alpha,x}\mid \alpha\in A}-
\max\br{\pr{\beta,x}\mid \beta\in B},\cr
m(f)=
\min\br{\pr{\alpha,x}\mid \alpha\in A}-
\min\br{\pr{\beta,x}\mid \beta\in B}, 
\end{array}
\end{equation}
where $\pr{\alpha,x}=\sum_{i\in I}\alpha_i x_i$. In this definition, 
we have identified $x=(x_i)_{i\in I}$ with the canonical coordinates 
of $\BR^I$.  It is easily shown that the definition of 
$M(f)$ and $m(f)$ does not depend on the choice of 
expression \eqref{eq:fab}.
Note also that $M(c)=m(c)=0$ for any $c\in \BK_{>0}$. 
\begin{proposition}\label{prop:Mm}
$(1)$ For any subtraction-free rational functions $f, g\in \BK(x)_{>0}$, 
one has
\begin{eqnarray}
&M(fg)=M(f)+M(g),\quad M(\dfrac{f}{g})=M(f)-M(g),\nonumber\\
&M(f+g)=\max\br{M(f), M(g)},
\end{eqnarray}
and 
\begin{eqnarray}
&m(fg)=m(f)+m(g),\quad m(\dfrac{f}{g})=m(f)-m(g),\nonumber\\
&m(f+g)=\min\br{m(f), m(g)}. 
\end{eqnarray} 
$(2)$  Let $\iota: \BK(x)\to \BK(x)$ be the isomorphism 
defined by $\iota(x_i)=x^{-1}_i$ $(i\in I)$.  Then one has 
\begin{equation}
M(f)=m(\iota(f)^{-1}),\quad m(f)=M(\iota(f)^{-1})
\end{equation}
for any $f\in \BK(x)_{>0}$. 
\end{proposition}
This proposition means that the correspondence $f\mapsto M(f)$ 
is nothing but the simple procedure of replacing the operations 
\begin{equation}
a\,b \to a+b,\quad
\frac{a}{b} \to a-b,\quad
a+b\to \max\br{a,b}. 
\end{equation}
Similarly, the correspondence $f\mapsto m(f)$ is the procedure 
\begin{equation}
a\,b \to a+b,\quad
\frac{a}{b} \to a-b,\quad
a+b\to \min\br{a,b}. 
\end{equation} 
The second part of the proposition implies that one can 
interchange ``$\max$'' and ``$\min$'' freely with each other, 
by using the operation $f(x)\to \iota(f)^{-1}=f(x^{-1})^{-1}$. 

Proposition \ref{prop:Mm} guarantees that these procedures 
can be applied consistently to arbitrary subtraction-free rational 
functions to obtain piecewise linear functions.  This passage 
from the subtraction-free rational functions 
to piecewise linear functions, either by $\max$ or $\min$, is called 
in several ways in the literature; it is  
called the {\em ultra-discretization} in the 
context of discrete integrable systems, and also 
the {\em tropicalization} in the context of totally positive 
matrices.  In this paper, we will use the adjective 
``tropical'' for objects and notions concerning 
subtraction-free rational functions, and 
``combinatorial'' for those concerning piecewise linear 
functions.  
It should be noted that there is no {\em canonical} procedure 
in the opposite direction;  
when a combinatorial expression is given, it becomes 
an interesting problem in many occasions to find 
a {\em good\/} counterpart in the tropical setting. 

\par\medskip
The passage from the tropical side to the combinatorial side 
is {\em functorial} in the following sense. 
Consider two fields of rational functions 
$\BK(x)$ in the variables $x=(x_i)_{i\in I}$ and 
$\BK(y)$ in the variables $y=(y_j)_{j\in J}$.  
We say that an isomorphism $\varphi$ form $\BK(y)$ into $\BK(x)$ 
is subtraction free 
if $\varphi(y_j)\in \BK(x)_{>0}$ for all $j\in J$.  
The set of subtraction-free rational functions 
$f_j(x)=\varphi(y_j)$ ($j\in J$) then 
defines a subtraction-free rational mapping  
\begin{equation}
F: y_j=f_j(x)\quad(j\in J) 
\end{equation}
from the affine space $\BK^I$ with coordinates $x=(x_i)_{i\in I}$ 
to $\BK^J$ with coordinates $y=(y_j)_{j\in J}$.  
For such a rational mapping $F$ given, we define 
two piecewise linear mappings $M(F), m(F) : \BR^I\to\BR^J$ 
by setting 
\begin{equation}
M(F):   y_j=M(f_j(x))\quad(j\in J),\qquad
m(F):  y_j=m(f_j(x))\quad(j\in J),
\end{equation}
respectively. 
Then Proposition \ref{prop:Mm} implies
\begin{proposition}
Consider the two subtraction-free rational mappings 
\begin{equation}
F: y_j=f_j(x)\quad(j\in J),\qquad
G: z_k=g_k(y)\quad(k\in K). 
\end{equation}
Then the piecewise linear mappings 
corresponding to the composition $G\circ F$ are given by
\begin{equation}
M(G\circ F)=M(G)\circ M(F),\quad 
m(G\circ F)=m(G)\circ m(F). 
\end{equation}
\end{proposition} 
\setcounter{equation}{0}
\section{Tropical row insertion and tropical tableaux}
In this section we introduce a tropical analogue 
of row insertion by clarifying the internal structure of bumping. 
Combining this with 
the matrix approach of Section 1, 
we give explicit tropical and combinatorial formulas 
for describing the tableau obtained from a word by 
row insertion, 
and for the Sch\"utzenberger involution on the 
set of column strict tableaux. 

\subsection{Row insertion}
Taking the set of letters $\br{1,\ldots,n}$, we consider a column strict 
tableau $T$ of shape $\lambda$, where 
$\lambda=(\lambda_1,\lambda_2,\ldots,\lambda_m)$ 
is a partition with $l(\lambda)\le m$.
For each $i=1,\ldots,m$, 
we define a weakly increasing word
\begin{equation}
w_i=i^{x^i_i}(i+1)^{x^i_{i+1}}\ldots n^{x^i_{n}}\qquad
(x^i_i+\cdots+x^i_n=\lambda_i)
\end{equation}
by reading the $i$-th row of $T$ from left to right,
where $x^i_j$ stands for the number of $j$'s appearing 
in the $i$-th row of $T$ for $i\le j$. 
For a weakly increasing word $v=1^{a_1}2^{a_2}\ldots n^{a_n}$ given, 
consider the tableau $T'=T\leftarrow v$ obtained by the row insertion of 
$v$ into $T$; we denote by 
$w'_i=i^{y^i_i}(i+1)^{y^i_{i+1}}\ldots n^{y^i_{n}}$ 
the weakly increasing word representing 
the $i$-th row of $T'$ for $i=1,\ldots,m'$.  
Our question is : 
{\em How can one describe $y^i_j$ explicitly 
in terms of $x^i_j$'s and $a_j$'s ? }

The bumping procedure $T\leftarrow v$ can be decomposed as follows.
\begin{equation}
\begin{array}{ccc}
& v & \cr
T &\cross & T' \cr
& \phi &
\end{array}
\quad:\quad
\begin{array}{ccc}
  & v=v_1 & \cr
w_1 &\cross & w'_1 \cr
  & v_2 & \cr
w_2 &\cross & w'_2 \cr
& v_3 & \cr
& \vdots & 
\end{array}
\end{equation}
Here $v_1=v$, and for $i=2,3,\ldots$, 
$v_i=i^{a^i_i}(i+1)^{a^i_{i+1}}\ldots n^{a^i_n}$ stands 
for the weakly increasing word consisting of the letters 
that have bumped out from 
$w_{i-1}$ by the row insertion of $v_{i-1}$. 
In what follows, we use the diagram
\begin{equation}\label{eq:rowins}
\begin{array}{ccc}
& v\,&\cr
w &\cross & w' \cr
& v' &
\end{array}
\qquad
\left(\begin{array}{ll}
w=1^{x_1}2^{x_2}\ldots n^{x_n},\quad &v=1^{a_1} 2^{a_2}\ldots n^{a_n}\cr
w'=1^{y_1}2^{y_2}\ldots n^{y_n},\quad &v'=1^{b_1}2^{b_2}\ldots n^{b_n}
\end{array}\right)
\end{equation}
consisting of four weakly increasing words $w$, $v$, $w'$, $v'$, 
to indicate a procedure of inserting a word $v$ into $w$; 
$w'=w\leftarrow v$ denotes the resulting word, and 
$v'$ is the word of the letters bumped out from $w$. 
(We always have $b_1=0$ in this setting.)
Our question is thus reduced to the problem of describing $y_j$ and $b_j$ 
in terms of $x_j$ and $a_j$ in this diagram. 
We also use the diagram of row insertion 
for the corresponding vectors of integers:  
\begin{equation}\label{eq:vects}
\begin{array}{ccc}
& \ba & \cr
\bx & \cross & \by \cr
& \bb 
\end{array}
\qquad
\left(\begin{array}{cc}
\bx=\pmatrix{x_1,\ldots,x_n}, \quad &
\ba=\pmatrix{a_1,\ldots,a_n}\cr 
\by=\pmatrix{y_1,\ldots,y_n},\quad &
\bb=\pmatrix{b_1,\ldots,b_n}
\end{array}\right).
\end{equation}

We now consider the procedure of row insertion as in \eqref{eq:rowins}. 
It is convenient to use the variables
\begin{equation}
\xi_j=x_1+x_2+\cdots+x_j,\quad \eta_j=y_1+y_2+\cdots+y_j\quad(j=1,\ldots,n). 
\end{equation}
Assume first that $v=k^a$ ($k=1,\ldots,n$); in this case, it is easy to see
\begin{eqnarray}
&&\eta_j=\xi_j\quad(j<k),\qquad \eta_k=\xi_k+a,\quad \nonumber\\
&&\eta_j=\max\br{\xi_k+a,\xi_j}=\max\br{\eta_k,\xi_j}\quad(j>k). 
\end{eqnarray}
Applying this result repeatedly for $k=1,\ldots,n$, we obtain 
following recurrence relations for the general case 
$v=1^{a_1}2^{a_2}\ldots n^{a_n}$:
\begin{eqnarray}
&&\eta_1=\xi_1+a_1,\quad 
\eta_2=\max\br{\eta_1,\xi_2}+a_2,
\nonumber\\
&&\eta_3=\max\br{\eta_1,\eta_2,\xi_3}+a_3, \quad \ldots.
\end{eqnarray}
Since $\eta_1\le \eta_2\le\cdots\le\eta_n$, it is equivalent to
\begin{equation}\label{eq:etarec}
\begin{array}{ll}\smallskip
\eta_1=\xi_1+a_1,\quad \\
\eta_j=
\max\br{\eta_{j-1},\xi_j}+a_j\\
\phantom{\eta_j}
=\max\br{\eta_{j-1}+a_j,\xi_j+a_j}
\quad(j=2,\ldots,n). 
\end{array}
\end{equation}
Hence we have
\begin{eqnarray}\label{eq:etaans}
&&\eta_j = 
\max\,\br{\xi_1+a_1+\cdots+a_j, \xi_2+a_2+\cdots+a_j,\ldots, \xi_j+a_j}
\nonumber
\\
&&\phantom{\eta_j}
={\displaystyle\max_{1\le k\le j}}\,
\br{x_1+\cdots+x_k+a_k+\cdots+a_j}
\end{eqnarray}
for $j=1,\ldots,n$. 
Note that 
\begin{equation}\label{eq:yeta}
y_1=\eta_1,\quad y_j=\eta_j-\eta_{j-1}\quad(j=2,\ldots,n),
\end{equation}
and that $b_j$ are determined as $b_1=0$ and 
\begin{equation}\label{eq:baxy}
b_j=a_j+x_j-y_j=a_j+\xi_j-\xi_{j-1}-\eta_j+\eta_{j-1}
\quad(j=2,\ldots,n),
\end{equation}
since the number of $j$'s is conserved during the process. 

\begin{example}\rm Let us consider an example 
of row insertion
\begin{equation}
\begin{array}{ccc}
& v=1245 & \cr
w=\underline{2}2\underline{3}4
\underline{5}\,\underline{\ \,}
&\cross & w'=122445 \cr
& v'=235
\end{array}.
\end{equation}
In terms of the vectors of integers, this procedure is 
expressed as
\begin{equation}
\begin{array}{ccc}
& \ba=(1,1,0,1,1) & \cr
\bx=(0,2,1,1,1)&\cross & \by=(1,2,0,2,1) \cr
& \bb=(0,1,1,0,1)
\end{array}. 
\end{equation}
The numbers 
\begin{equation}
\eta_j=\max_{1\le k\le j} (x_1+\cdots+x_k+a_k+\cdots+a_j)
\end{equation}
can be read off from the table 
\begin{equation}
\begin{picture}(100,20)(0,5)
\put(23,-2){\line(1,0){75}}
\put(23,10){\line(1,0){75}}
\put(23,22){\line(1,0){75}}
\multiput(23,-2)(15,0){6}{\line(0,1){24}}
\put(0,7){$
\begin{array}{c@:ccccc}
\bx \ &\quad 0&2&1&1&1\cr
\ba \ &\quad 1&1&0&1&1
\end{array}
$}
\put(28,24){\small$1$}
\put(43,24){\small$2$}
\put(58,24){\small$3$}
\put(73,24){\small$4$}
\put(88,24){\small$5$}
\end{picture}
\end{equation}
as 
\begin{equation}
\eta_1=1,\quad\eta_2=3,\quad\eta_3=3,\quad\eta_4=5,\quad\eta_5=6. 
\end{equation}
By taking the first difference of this sequence, 
we have $\by=(1,2,0,2,1)$. 
\end{example}

The general procedure 
$T'=T\leftarrow v$ of inserting 
a weakly increasing word  $v$ into a column strict tableau $T$
should be described as a superposition of 
row insertions of type \eqref{eq:rowins}. 
We will make use of the tropical 
analogue of combinatorial (piecewise linear) formulas above
in order to systematize the superposition of row insertions.

\subsection{Tropical row insertion}
We introduce a tropical analogue of combinatorial formulas
for the row insertion \eqref{eq:rowins}. 
We use the same symbols $x_j$, $a_j$, $y_j$, $b_j$ 
as in \eqref{eq:vects} for the {\em tropical variables\/} (indeterminates).  
Introducing the auxiliary variables 
\begin{equation}\label{eq:tropxieta}
\xi_j=x_1\cdots x_j,\quad \eta_j=y_1\cdots y_j\quad(j=1,\ldots,n),
\end{equation}
we define the transformation $(\bx,\ba)\mapsto(\by,\bb)$ by 
\begin{equation}\label{eq:troprowins}
\begin{array}{lll}\smallskip
\eta_1=\xi_1 a_1,\quad&\eta_j=(\eta_{j-1}+\xi_j)a_j\quad&(j=2,\ldots,n),\cr
 y_1=\eta_1,\quad &y_j=\dfrac{\eta_j}{\eta_{j-1}}\quad&(j=2,\ldots,n),\cr
b_1=1,\quad&b_j=a_j\dfrac{x_j}{y_j}
=a_j\dfrac{\xi_j\,\eta_{j-1}}{\xi_{j-1}\eta_j}
\quad&(j=2,\ldots,n).
\end{array}
\end{equation}
We have made these formulas from the combinatorial formulas 
\eqref{eq:etarec}, \eqref{eq:yeta}, \eqref{eq:baxy} by the simple rule of 
replacement
\begin{equation}
\max\br{a,b} \to a+b,\quad a+b \to a\,b\quad a-b \to \frac{a}{b},
\end{equation}
which is sometimes called the {\em tropical variable change}.  
{}From the recurrence relations for $\eta_j$ above, we easily obtain
\begin{eqnarray}\label{eq:tropeta}
&&\eta_j=\xi_1 a_1\cdots a_j+\xi_2 a_2\cdots a_j+\cdots+\xi_j a_j
\nonumber\\
&&\phantom{\eta_j}
= x_1a_1a_2\cdots a_j+x_1x_2a_2\cdots a_j+\cdots+x_1\cdots x_j a_j. 
\end{eqnarray}
{}From this formula, we can recover the combinatorial formula
\eqref{eq:etaans} by the standard procedure as we discussed in 
Section 1.3.
\par\medskip
The tropical transformation $(\bx,\ba)\mapsto (\by,\bb)$ 
we have discussed above 
arises also from the 
system of algebraic equations {\em of discrete Toda type}
\begin{equation}\label{eq:dTodaB}
\begin{array}{lll}\smallskip
a_1x_1=y_1, & a_jx_j=y_jb_j & (j=2,\ldots,n),\cr
\dfrac{1}{a_1}+\dfrac{1}{x_2}=\dfrac{1}{b_2},\quad
& \dfrac{1}{a_j}+\dfrac{1}{x_{j+1}}
=\dfrac{1}{y_j}+\dfrac{1}{b_{j+1}} &(j=2,\ldots,n), 
\end{array}
\end{equation}
for $(y_1,\ldots,y_n)$ and $(b_2,\ldots,b_n)$, 
where we regard $x_j$, $a_j$ as given variables, and $y_j$, $b_j$ 
as unknown functions. 
(For the relationship between \eqref{eq:dTodaB} 
and the discrete Toda equation,  see 
Remark \ref{rem:dToda} below)

\begin{lemma}
The system of algebraic equations \eqref{eq:dTodaB} is equivalent to the 
recurrence formulas \eqref{eq:troprowins} together with \eqref{eq:tropxieta}.
\end{lemma}
\proof
In fact, by eliminating $b_j$ ($j=2,\ldots,n$) from \eqref{eq:dTodaB}, 
and by rewriting the equations in terms of $\xi_j$ and $\eta_j$, 
we obtain
\begin{equation}
\dfrac{\eta_n-\eta_{n-1}a_n}{\xi_na_n}=
\dfrac{\eta_{n-1}-\eta_{n-2}a_{n-1}}{\xi_{n-1}a_{n-1}}=\cdots
=\dfrac{\eta_2-\eta_1 a_2}{\xi_2a_2}=\dfrac{\eta_1}{\xi_1a_1}=1,
\end{equation}
which is equivalent to \eqref{eq:troprowins}.\qed
\noindent
This fact is a key to our matrix approach to tropical combinatorics. 
Note that \eqref{eq:dTodaB} is written as a matrix equation: 
\begin{eqnarray}
&&\left[\matrix{
\overline{a}_1 & 1 &&&\cr
 & \overline{a}_2 & 1 & \cr
&&\ddots&\ddots &\cr
&&&\overline{a}_{n-1}&1\cr
&&&& \overline{a}_n
}\right]
\left[\matrix{
\overline{x}_1 & 1 &&&\cr
 & \overline{x}_2 & 1 & \cr
&&\ddots&\ddots &\cr
&&&\overline{x}_{n-1}&1\cr
&&&& \overline{x}_n
}\right]\nonumber\\
&&=
\left[\matrix{
\overline{y}_1 & 1 &&&\cr
 & \overline{y}_2 & 1 & \cr
&&\ddots&\ddots &\cr
&&&\overline{y}_{n-1}&1\cr
&&&& \overline{y}_n
}\right]
\left[\matrix{
1 & 0 &&&\cr
 & \overline{b}_2 & 1 & \cr
&&\ddots&\ddots &\cr
&&&\overline{b}_{n-1}&1\cr
&&&& \overline{b}_n
}\right],
\end{eqnarray}
where we have used the notation 
$\overline{x}=\frac{1}{x}$.  
By using the notation of Section \ref{sec:pathrep}, 
this equation can be expressed as
\begin{equation}\label{eq:EqEE}
E(\overline{\ba})\,E(\overline{\bx})=
E_1(\overline{\by})\, E_2(\overline{\bb}). 
\end{equation} 
It is also equivalent to
\begin{equation}\label{eq:EqHH}
H(\bx)\,H(\ba)=
H_2(\bb) H_1(\by). 
\end{equation}
Each of these two equations \eqref{eq:EqEE} and 
\eqref{eq:EqHH} can be thought of as a tropical 
expression of the row insertion
\begin{equation}
\begin{array}{ccc}
& \ba & \cr
\bx & \cross & \by \cr
& \bb 
\end{array}
\qquad
\left(\begin{array}{cc}
\bx=\pmatrix{x_1,\ldots,x_n}, \quad &
\ba=\pmatrix{a_1,\ldots,a_n}\cr 
\by=\pmatrix{y_1,\ldots,y_n},\quad &
\bb=\pmatrix{b_1,\ldots,b_n}
\end{array}\right),
\end{equation}
where $b_1=1$.  

We show how the matrix equation \eqref{eq:EqEE} 
is solved by using the path representations of Section 
\ref{sec:pathrep}.  
Setting $H=H(\bx)H(\ba)$, we look at the minor determinants 
$\tau^i_j(H)$ $(i\le j)$. 
By the argument of Section \ref{sec:pathrep}, 
the minor determinants of $H$ are read off from the 
following diagram.
\begin{equation}
\begin{picture}(90,20)(-10,-5) 
\put(0,0){\line(1,0){80}}
\put(0,10){\line(1,0){80}}
\multiput(0,0)(10,0){9}{\line(0,1){10}}
\put(-10,-2){$\ba$}
\put(-10,8){$\bx$}
{\thicklines\put(95,10){\vector(0,-1){10}}}
\end{picture}
\end{equation}
The result for $\tau^i_j=\tau^i_j(H)$ is: 
\begin{equation}
\begin{array}{ll}\smallskip
\tau^1_j=\dsum{k=1}{j}\ x_1\cdots x_k\,a_k\cdots a_j
\quad&(j\ge 1),\\ \smallskip
\tau^2_j=\tau^j_j=x_1\cdots x_j a_1\cdots a_j
\quad&(j\ge 2),\\
\tau^i_j=0
\quad&(3\le i<j\le n).
\end{array}
\end{equation}
By Proposition \ref{prop:HHH}, 
we already know that equation \eqref{eq:EqHH} has 
a unique solution such that
\begin{equation}
\tau^1_j=y_1\ldots y_j \quad(j=1,\ldots,n),\quad
\tau^2_j=y_1\ldots y_j b_1\ldots b_j\quad(j=2,\ldots,n). 
\end{equation}
Namely, $y_j$ and $b_j$ are determined as
\begin{equation}
y_j=\frac{\tau^1_j}{\tau^1_{j-1}}\quad(j=1,\ldots,n),
\quad
b_j=\frac{x_ja_j}{y_j}\quad(j=2,\ldots,n),
\qquad
\end{equation}
consistently with what we have seen before. 
\begin{remark}\label{rem:dToda}\rm 
The system of algebraic equations \eqref{eq:dTodaB} is 
closely related to the {\em discrete Toda equation} 
(\cite{HT}, \cite{SSYY}): 
\begin{equation}
I_i^{t+1} V_i^{t+1} = I_{i+1}^t V_i^t,\quad
I_i^{t+1}+V_{i-1}^{t+1}=I_i^{t}+V_i^{t}
\end{equation}
where $i\in\BZ$ and $t\in\BZ$ stand for 
the discrete coordinates of space and time, respectively, 
and $I^t_i$, $V^t_i$ are the dependent variables. 
If we set 
\begin{equation}
a_i=(I_{i+1}^t)^{-1},\quad x_i=(V^t_i)^{-1},\quad 
y_i=(V^{t+1}_i)^{-1},\quad b_i=(I_i^{t+1})^{-1},
\end{equation}
we have
\begin{equation}
a_i x_i =y_i b_i,\quad 
\frac{1}{a_i}+\frac{1}{x_{i+1}}=\frac{1}{y_i}+\frac{1}{b_{i+1}}
\quad(i\in\BZ). 
\end{equation}
Note also that the discrete Toda equation can be 
expressed as the matrix equation
\begin{equation}
L(t+1)R(t+1)=R(t)L(t)
\end{equation}
for the $\BZ\times\BZ$ matrices
\begin{equation}
L(t)=\sum_{i\in \BZ}{E_{ii}}+\sum_{j\in\BZ} V^t_i E_{i+1,i},
\quad
R(t)=\sum_{i\in\BZ} I^t_i E_{ii}+\sum_{i\in\BZ} E_{i,i+1}. 
\end{equation}
\end{remark}

\subsection{Tropical tableaux}
In the following we discuss the following question 
both in the tropical and the combinatorial setting. 
\par\smallskip\noindent
{\bf Question:}\ 
For a sequence of weakly increasing words $w_1, \ldots, w_m$ 
given, 
find an explicit formula for the column strict tableau 
\begin{equation}
P=P(w)=(\cdots(w_1\leftarrow w_2)\cdots\leftarrow w_m)
\end{equation}
obtained from the word $w=w_1\ldots w_m$ 
by the row insertion. 
\comment{
\par\smallskip\noindent
{\bf Question 2:}\ \ Find an explicit transformation formula 
for the procedure $V=U\leftarrow w$ of inserting a word $w$ 
into a general column strict tableau $U$. 
}
\par\medskip
We can employ \eqref{eq:EqHH} as building blocks 
for the tropical analogue of various combinatorial 
algorithms. 
Let us consider the procedure of successive row insertion 
\begin{equation}
P=(\cdots(w_1\leftarrow w_2)\leftarrow \cdots \leftarrow w_m)
\end{equation}
of weakly increasing word $w_i=1^{x^i_1}\cdots n^{x^i_n}$ 
($i=1,\ldots,n$) to obtain a column strict tableau $P$.  
This procedure can be described by the following diagram. 
\begin{equation}
\begin{array}{cccccccc}
&\bx^1=\bx^{1,1} && \bx^2=\bx^{2,1} & & \bx^3=\bx^{3,1}\cr
\phi&\cross&\by^{1,1} &\cross & \by^{2,1}&\cross &\cdots\cr
&\phi & &\bx^{2,2}&&\bx^{3,2}\cr
&&\phi&\cross&\by^{2,2} &\cross&\cdots\cr
&&&\phi&&\bx^{3,3}\cr
&&&&\phi &\cross &\cdots
\end{array}
\end{equation}
Passing the tropical variables 
$\bx^i=(x^i_1,\ldots,x^i_n)$ \ ($i=1,\ldots,m$), 
we can compute the row insertion above as
\begin{equation}
\arraycolsep=1.5pt
\renewcommand{\arraystretch}{1.4}
\begin{array}{rlllll}
H(\bx^1)&=H_1(\by^{1,1})\cr
H(\bx^1)H(\bx^2)
&=H_1(\by^{1,1}) H(\bx^{2,1})\cr
&=H_2(\by^{2,2})H_1(\by^{2,1})\cr
H(\bx^1)H(\bx^2)H(\bx^3)
&=H_2(\by^{2,2})H_1(\by^{2,1}) H(\bx^3)\cr
&=H_2(\by^{2,2})H_2(\bx^{3,2}) H_1(\by^{3,1})\cr
&=H_3(\by^{3,3})H_2(\by^{3,2}) H_1(\by^{3,1})\cr
&\cdots,
\end{array}
\end{equation}
where $\by^{k,i}=(1,\ldots,1,y^{k,i}_i,\ldots,y^{k,i}_n)$. 
When $m\le n$, 
by setting $\by^{m,i}=\bp^i$ , 
$\bp^i=(1,\ldots,1,p^i_i,\ldots,p^i_n)$ 
($i=1,\ldots,m$), 
we finally obtain
\begin{equation}
H(\bx^1)H(\bx^2)\cdots H(\bx^m)
=H_m(\bp^m)\cdots H_2(\bp^2) H_1(\bp^1),
\end{equation}
In this formula, 
each $p^i_j$ ($i\le j$) 
denotes the tropical variable 
corresponding to the number of $j$'s in the $i$-th row 
of the tableau $P$.  
Namely, we can regard the expression
\begin{equation}
H_P=H_m(\bp^m)\cdots H_2(\bp^2) H_1(\bp^1)
\end{equation}
as representing the {\em tropical tableau} $P=\pmatrix{p^i_j}_{i\le j}$; 
it provides 
the tropical analogue of a general column strict tableau
whose shape is a partition of length $m$. 
\par\medskip
The argument above shows that our
question can be answered by considering 
the matrix equation 
\begin{eqnarray}\label{eq:XH}
&&H(\bx^1)H(\bx^2)\cdots H(\bx^m)
=H_m(\bp^m)\cdots H_2(\bp^2) H_1(\bp^1)\quad(m\le n),
\nonumber\\
&&H(\bx^1)H(\bx^2)\cdots H(\bx^m)
=H_n(\bp^n)\cdots H_2(\bp^2) H_1(\bp^1)\quad\ \,(m\ge n)
\end{eqnarray}
for the unknowns 
$\bp^i=(1,\ldots,1,p^i_i,\ldots,p^i_n)$ 
$(i=1,\ldots,\min\br{m,n})$. 
In the following we regard $x^i_j$ ($i=1,\ldots$, $j=1,\ldots,n$) 
as indeterminates, and look for solutions $p^i_j$ 
of \eqref{eq:XH} in the filed of rational functions in the 
$x$-variables. 

Denoting the left-hand side by $H$, 
consider the minor determinants 
$\tau^i_j(H)=\det H^{1,\ldots,i}_{j-i+1,\ldots,j}$
for $1\le i\le j\le n$. 
By Proposition \ref{prop:HH}, $\tau^i_j(H)$ is expressed 
in terms of the nonintersecting paths in the $m\times n$ 
rectangle associated with 
the matrix $X=\pmatrix{x^i_j}_{i,j}$: 
\begin{equation}\label{eq:tauH}
\begin{picture}(145,60)(-60,-3)
\unitlength=0.8pt
\put(-75,30){$\tau^i_j(H)=\sum$}
\put(80,55){$X$}
\put(0,0){\line(1,0){100}}
\put(0,70){\line(1,0){100}}
\put(0,0){\line(0,1){70}}
\put(100,0){\line(0,1){70}}
\multiput(-1.4,39.5)(2,2){16}{$.$}
\multiput(48.6,-0.5)(2,2){16}{$.$}
\put(-2,73){\small$1$}
\put(8,73){\small$\ldots$}
\put(28,73){\small$i$}
\put(18,-10){\small$j-i+1$}
\put(63,-10){\small$\ldots$}
\put(78,-10){\small$j$}
{\thicklines
\put(0,70){\line(0,-1){30}}
\put(10,70){\line(0,-1){20}}
\put(20,70){\line(0,-1){10}}
\put(30,70){\line(0,-1){0}}
\put(60,0){\line(0,1){10}}
\put(70,0){\line(0,1){20}}
\put(80,0){\line(0,1){30}}
\put(0,40){\line(0,-1){10}}
\put(0,30){\line(1,0){10}}
\put(10,30){\line(0,-1){20}}
\put(10,10){\line(1,0){20}}
\put(30,10){\line(0,-1){10}}
\put(30,0){\line(1,0){20}}
\put(10,50){\line(0,-1){10}}
\put(10,40){\line(1,0){10}}
\put(20,40){\line(0,-1){10}}
\put(20,30){\line(1,0){10}}
\put(30,30){\line(0,-1){10}}
\put(30,20){\line(1,0){10}}
\put(40,20){\line(0,-1){10}}
\put(40,10){\line(1,0){20}}
\put(30,70){\line(1,0){10}}
\put(40,70){\line(0,-1){10}}
\put(40,60){\line(1,0){20}}
\put(60,60){\line(0,-1){20}}
\put(60,40){\line(1,0){20}}
\put(80,40){\line(0,-1){10}}
\put(20,60){\line(1,0){10}}
\put(30,60){\line(0,-1){20}}
\put(30,40){\line(1,0){10}}
\put(40,40){\line(0,-1){10}}
\put(40,30){\line(1,0){20}}
\put(60,30){\line(0,-1){10}}
\put(60,20){\line(1,0){10}}
}
\end{picture}
\end{equation}
When $m<i\le j\le n$, we have 
$\tau^i_j(H)=\delta_{i,j}\tau^m_j(H)$,
$\tau^m_j(H)=\prod_{(a,b);\,b\le j} x^a_b$. 
Hence, by Theorem \ref{thm:H}, we see that 
the matrix equation \eqref{eq:XH} has a unique 
rational solution 
in the $x$-variables; the solution is 
expressed by $\tau^i_j(H)$ above. 
To summarize, we have
\begin{theorem}\label{thm:HxHu}
For $m\times n$ indeterminates $x^i_j$ $(1\le i\le m,\,1\le j\le n)$
given, consider the following matrix equation 
for unknown variables 
$p^i_j$ $(1\le i\le l, 1\le j\le n, i\le j)$, $l=\min\br{m,n}$\,$:$  
\begin{equation}
H(\bx^1) H(\bx^2)
\cdots H(\bx^m)=H_l(\bp^l)\cdots H_2(\bp^2)H_1(\bp^1), 
\end{equation}
where $\bx^i=(x^i_1,\ldots,x^i_n)$ and 
$\bp^i=(1,\ldots,1,p^i_i,\ldots,p^i_n)$. 
This equation has a unique rational solution in the $x$-variables $;$ 
it is given explicitly as
\begin{equation}
p^i_i=\frac{\tau^i_i}{\tau^{i-1}_i},\qquad
p^i_j=\frac{\tau^i_j\ \tau^{i-1}_{j-1}}{\tau^{i-1}_j\tau^i_{j-1}}
\quad(i<j)
\end{equation}
for $1\le i\le l$, $1\le j\le n$, 
with $\tau^i_j$ $(i\le j)$ defined as the sum 
\begin{equation}
\tau^i_j=\sum_{(\gamma_1,\ldots,\gamma_i)}\, 
x_{\gamma_1}\cdots x_{\gamma_i}
\end{equation}
of monomials over all $i$-tuples of nonintersecting paths
$\gamma_k: (1,k)\to(m,j-i+k)$ $(k=1,\ldots,i)$
in the $m\times n$ rectangle. 
Here, the weight $x_\gamma$ of  a path $\gamma$ is the product 
\begin{equation}
x_\gamma=\prod_{(a,b)\in \gamma}\, x^a_b
\end{equation}
of all $x^a_b$'s corresponding to the vertices on $\gamma$.  
\end{theorem}
\noindent
The explicit formula for $p^i_j$ above 
is formulated by A.N.~Kirillov \cite{K}, Theorem 4.23. 

\par\medskip
By the standard passage from subtraction-free rational functions 
to piecewise-linear functions, 
we obtain the following combinatorial formula 
(\cite{K}, Theorem 3.5)
for the column strict tableaux $P$ obtained from a word 
$w=w_1w_2\ldots w_m$ by the row insertion. 
\begin{theorem}
Taking the set of letters $\br{1,\ldots,n}$, 
let $w_1,\ldots,w_m$ be a sequence of weakly increasing words
$w^i=1^{x^i_1}\cdots n^{x^i_n}$ $(i=1,\ldots,m)$. 
Consider the column strict tableau
\begin{equation}
P=P(w)=(\cdots (w_1\leftarrow w_2)\leftarrow\cdots\leftarrow w_m)
\end{equation}
obtained from the word $w=w_1w_2\cdots w_m$ by 
row insertion, 
and denote by $i^{p^i_i}\cdots n^{p^i_n}$ 
the weakly increasing word representing the $i$-th row of 
$P$, for $i=1,\ldots,l$, $l=\min\br{m,n}$. 
Then, for each $(i,j)$, 
the number $p^i_j$ of the letter $j$ in the $i$-th row of $P$
is determined explicitly as 
\begin{equation}\label{eq:uintauComb}
p^i_i=\tau^i_i-\tau^{i-1}_i,
\quad 
p^i_j=\tau^i_j-\tau^{i-1}_j-\tau^i_{j-1}+\tau^{i-1}_{j-1}
\quad(i<j)
\end{equation}
with $\tau^i_j$ $(i\le j)$ defined as the maximum 
\begin{equation}\label{eq:taumax}
\tau^i_j=\max_{(\gamma_1,\ldots,\gamma_i)}\, 
(x_{\gamma_1}+\cdots +x_{\gamma_i})
\end{equation}
of weights over all $i$-tuples of nonintersecting paths
$\gamma_k: (1,k)\to(m,j-i+k)$ $(k=1,\ldots,i)$
in the $m\times n$ rectangle\,$;$ we set $\tau^0_j=0$ for all $j$.
Here, the weight $x_\gamma$ of  a path $\gamma$ is the sum  
\begin{equation}
x_\gamma=\sum_{(a,b)\in \gamma}\, x^a_b
\end{equation}
of all $x^a_b$'s corresponding to the vertices on $\gamma$.  
\end{theorem}
Note also that each $\tau^i_j$ represents the sum
of $p^a_b$'s in the region $a\le i$, $b\le j$:
\begin{equation}
\tau^i_j=\sum_{(a,b):\,a\le i,\,b\le j} p^a_b\qquad (i\le j).
\end{equation}

\begin{example}\rm 
We give an example with $m=3$, $n=4$ by taking the word
\begin{equation}
w=2234134411224=2234|1344|11224,
\end{equation}
which corresponds to the column strict tableau
\begin{equation}
P=P(w)=\tbl{1&1&1&2&2&4&4\cr2&2&3&3&4\cr4}.
\end{equation} 
{}From $w$, we first construct the $3\times 4$ matrix 
$X=\pmatrix{x^i_j}_{i,j}$ 
by counting the number of $j$'s in the $i$-th block 
of $w$:
\begin{equation}
X=\left[\matrix{
0 & 2 & 1 & 1\cr
1 & 0 & 1 & 2\cr
2 & 2 & 0 & 1
}\right]. 
\end{equation} 
The numbers $\tau^i_j$ of \eqref{eq:taumax} are determined 
from $X$ by statistics of nonintersecting paths: 
\begin{equation}
\btau=\pmatrix{\tau^i_j}_{i,j}=\left[\matrix{
3 & 5 & 5 & 7\cr
&7& 9& 12\cr
&&9&13
}\right]. 
\end{equation}
For instance, $\tau^2_3$ is computed as the maximum of 
weights over three pairs of nonintersecting paths
$(\gamma_1,\gamma_2)$ such that 
$\gamma_1:(1,1)\to(3,2)$ and $\gamma_2:(1,2)\to(3,3)$\,: 
\begin{equation}
X=
\begin{picture}(70,25)
\put(0,0){$\left[\matrix{
0 & 2 & 1 & 1\cr
1 & 0 & 1 & 2\cr
2 & 2 & 0 & 1
}\right].$}
\put(11,30){\vector(0,-1){8}}
\put(26,30){\vector(0,-1){8}}
\put(26,-16){\vector(0,-1){8}}
\put(41,-16){\vector(0,-1){8}}
\multiput(11,22)(0,-2){16}{\line(0,-1){1}}
\multiput(11,-9)(2,0){8}{\line(1,0){1}}
\multiput(26,-9)(0,-2){4}{\line(0,-1){1}}
\multiput(26,22)(0,-2){10}{\line(0,-1){1}}
\multiput(26,3)(2,0){8}{\line(1,0){1}}
\multiput(41,3)(0,-2){10}{\line(0,-1){1}}
\end{picture}
\vspace{14pt}
\end{equation} 
{}From 
\begin{equation}
\begin{array}{l}
(0+1+2+2)+(2+0+1+0)=8 \cr
(0+1+2+2)+(2+1+1+0)=9 \cr
(0+1+0+2)+(2+1+1+0)=7
\end{array}
\end{equation}
we get
\begin{equation}
\tau^2_3=\max\br{8,9,7}=9. 
\end{equation}
According to \eqref{eq:uintauComb}, 
we compute $p^i_j$ by taking the {\em discrete Laplacian}
of $\tau^i_j$: 
\begin{equation}
\bp=\pmatrix{p^i_j}_{i,j}=\left[\matrix{
3&2&0&2\cr
&2&2&1\cr
&&0&1
}\right].
\end{equation}
Then $p^i_j$ gives the number of $j$'s in the $i$-th row 
of the tableau $P$ above. 
\end{example}

\subsection{Tropical Sch\"utzenberger involution}
We now recall the Sch\"utzenberger involution on 
the set of column strict tableaux.  
Taking the set $\br{1,2\ldots,n}$ of letters 
as before, we define an involution $k\mapsto k^\ast$
on $\br{1,2\ldots,n}$
by $k^\ast=n-k+1$ for $k=1,\ldots,n$.
For a word $w=k_1\,k_2\ldots k_l$ consisting of letters
in $\br{1,\ldots,n}$ given, we define the word $w^\ast$ by
\begin{equation}
w^\ast=k_l^\ast\ldots k_2^\ast \,k_1^\ast,
\end{equation}
by applying $k\mapsto k^\ast$ to each letter, and then 
by reversing the order.  
Let us denote by $P=P(w)$ the column strict tableau 
obtained from a word $w$.  
Since the involution $w\mapsto w^\ast$ on the set 
of words preserve the Knuth equivalence, 
it induces an involution $P\mapsto \SI{P}$ on the 
set of column strict tableaux such that 
\begin{equation}
P(w^\ast)=\SI{P(w)}
\end{equation}
for any word $w$; 
we call this involution $P\mapsto \SI{P}$ the 
{\em Sch\"utzenberger involution}. 
It is well known that $P$ and $\SI{P}$ has the 
same shape, and that the column strict tableau 
$\SI{P}$ is obtained from $P$ as the evacuation 
tableau by a successive application of 
{\em jeu de taquin}. 
We remark that, when the word $w=k_1\ldots k_n$ 
represents a permutation in $\BS_n$, 
$w^\ast$ is the conjugation of $w$ by the longest 
element of $\BS_n$. 
(Sch\"utzenberger's algorithm for column 
strict tableaux can be obtained essentially 
from \cite{S}, Theorem 3.9.4, for instance. 
In \cite{S}, 
Sch\"utzenberger's algorithm is formulated for  
permutations, but it is not difficult to extend it 
to that for words.)

\begin{example}\label{ex:SchP}\rm
Consider the word $w=42213132$ with $n=4$. 
In this case, we have $w^\ast=32424331$. 
\begin{equation}
P=P(w)=\tbl{1&1&2&3\cr2&2&3\cr4},
\quad
\SI{P}=P(w^\ast)=\tbl{1&2&3&3\cr2&4&4\cr3}. 
\end{equation}
\end{example}
\par\medskip
Let us decompose a given word $w$ into a chain of  
weakly increasing words:
\begin{equation}
w=w_1w_2\ldots w_m,\qquad
w_i=1^{x^i_1}2^{x^i_2}\ldots n^{x^i_n}\quad(i=1,\ldots,m). 
\end{equation}
Then we have 
\begin{equation}
w^\ast=w_n^\ast \ldots w_2^\ast\,w_1^\ast,\qquad
w_i^\ast=1^{x^i_n}2^{x^i_{n-1}}\ldots n^{x^i_1}\quad(i=1,\ldots,m). 
\end{equation} 
Setting $P=P(w)$, $\SI{P}=P(w^\ast)$, 
we denote by $p^i_j$ and by $\widetilde{p}^i_j$ the number of $j$'s 
in the $i$-th row of $P$ and $\SI{P}$, respectively.  
Passing to the tropical variables, we have the 
matrix equation
\begin{eqnarray}\label{eq:SchInv}
&&H(\bx^1)H(\bx^2)\cdots H(\bx^m)
=H_P=H_l(\bp^n)\cdots H_2(\bp^2)H_1(\bp^1),
\nonumber\\
&&H(\bx^m_\ast)\cdots H(\bx^2_\ast)H(\bx^1_\ast)
=H_{\SI{P}}=H_l(\widetilde{\bp}^n)\cdots 
H_2(\widetilde{\bp}^2)H_1(\widetilde{\bp}^1),
\end{eqnarray}
where $l=\min\br{m,n}$ and, 
for a vector $\bx=(x_1,x_2,\ldots,x_n)$ given, 
$\bx_\ast=(x_n,\ldots,x_2,x_1)$ 
denotes the vector obtained by reversing the order. 
In the following, we denote by
\begin{equation}
J_n=\pmatrix{\delta_{i+j,n+1}}_{i,j=1}^n
\end{equation}
the permutation matrix representing the 
longest element of $\BS_n$. 
Since $J_n\Lambda J_n =\Lambda^{\trp}$, from 
\begin{equation}
H(\bx)=(\diag{\overline{\bx}}-\Lambda)^{-1},
\quad
\bx=(x_1,x_2,\ldots,x_n)
\end{equation}
we have 
\begin{equation}
J_nH(\bx)^{\trp}J_n=(\diag{\overline{\bx}_\ast}-\Lambda)^{-1},
\quad
\bx_\ast=(x_n,\ldots,x_2,x_1),
\end{equation}
namely, $J_n H(\bx)^{\trp}\,J_n=H(\bx_\ast)$.  
Hence, \eqref{eq:SchInv} implies  
$J_n H_P^\trp J_n =H_{\SI{P}}$, namely, 
\begin{equation}\label{eq:UtoV}
J_n H_1(\bp^1)^\trp\,H_2(\bp^2)^\trp\cdots H_n(\bp^n)^\trp J_n
=H_n(\widetilde{\bp}^n)\cdots 
H_2(\widetilde{\bp}^2)H_1(\widetilde{\bp}^1). 
\end{equation}

Supposing that $m\le n$, we take two general 
tropical tableaux
\begin{equation}
H_U=H_m(\bu^m)\cdots H_1(\bu^1),\quad
H_V=H_m(\bv^m)\cdots H_1(\bv^1),
\end{equation}
where 
$\bu^i=(u^i_i,\ldots,u^i_n)$ and 
$\bv^i=(v^i_i,\ldots,v^i_n)$ for $i=1,\ldots,m$. 
In view of \eqref{eq:UtoV}, 
we define the {\em tropical Sch\"utzenberger involution} 
to be the birational transformation $U\mapsto V$ 
defined through the matrix equation
$J_n H_U^\trp J_n=H_V$ :
\begin{equation}\label{eq:SchUV}
J_n H_1(\bu^1)^\trp \cdots H_m(\bu^m)^\trp J_n
=H_m(\bv^m)\cdots H_1(\bv^1). 
\end{equation}
We now propose to solve this matrix equation for 
$v^i_j$ ($i\le j$), regarding 
$u^i_j$ ($i\le j$) as indeterminates. 
Note that
\begin{equation}
J_n \Lambda_{\ge k}^\trp J_n =\Lambda_{\le n-k+1},
\quad
\Lambda_{\le k}=\sum_{i=1}^{k-1} E_{i,i+1},
\end{equation}
for $k=1,\ldots,n$. 
Hence, \eqref{eq:SchUV} can be written as
\begin{equation}
\begin{array}{l}\smallskip
(\diag{\overline{\bu}^1_\ast}-\Lambda_{\le n})^{-1}
(\diag{\overline{\bu}^2_\ast}-\Lambda_{\le n-1})^{-1}
\cdots
(\diag{\overline{\bu}^m_\ast}-\Lambda_{\le n-m+1})^{-1}
\cr
=
(\diag{\overline{\bv}^m}-\Lambda_{\ge m})^{-1}
\cdots
(\diag{\overline{\bv}^2}-\Lambda_{\ge 2})^{-1}
(\diag{\overline{\bv}^1}-\Lambda_{\ge 1})^{-1}. 
\end{array}
\end{equation}
This equality is expressed graphically as follows. 
For each $1\le i\le j\le n$,
\begin{equation}
\begin{picture}(95,70)(-35,-5)
\put(-35,25){$\dsum{}{} $}
\put(0,40){\line(1,0){20}}
\put(0,30){\line(1,0){30}}
\put(0,20){\line(1,0){40}}
\put(0,10){\line(1,0){50}}
\put(0,0){\line(1,0){60}}
\put(50,10){\line(0,-1){10}}
\put(40,20){\line(0,-1){20}}
\put(30,30){\line(0,-1){30}}
\put(20,40){\line(0,-1){40}}
\put(10,40){\line(0,-1){40}}
\put(0,40){\line(0,-1){40}}
\multiput(18.5,39.5)(2,-2){21}{$.$}
\put(-13,39){$\bu^{m}_\ast$}
\put(-10,25){$\vdots$}
\put(-13,9){$\bu^2_\ast$}
\put(-13,-3){$\bu^1_\ast$}
\put(-2,-12){\small $1$}
\put(8,-12){\small $i$}
\put(58,-12){\small $n$}
\put(38,27){\small $j$}
\put(58,4){\small $n$}
{\thicklines
\put(35,40){\vector(1,0){15}}
\put(35,40){\vector(0,1){15}}
\put(10,-3){\line(0,1){13}}
\put(10,10){\line(1,0){20}}
\put(30,10){\line(0,1){10}}
\put(30,20){\line(1,0){10}}
\put(40,20){\line(0,1){3}}
}
\end{picture}
\qquad
\begin{picture}(100,70)(-20,5)
\put(-30,35){$=\dsum{}{}$}
\put(0,60){\line(1,0){60}}
\put(10,50){\line(1,0){50}}
\put(20,40){\line(1,0){40}}
\put(30,30){\line(1,0){30}}
\put(40,20){\line(1,0){20}}
\put(10,60){\line(0,-1){10}}
\put(20,60){\line(0,-1){20}}
\put(30,60){\line(0,-1){30}}
\put(40,60){\line(0,-1){40}}
\put(50,60){\line(0,-1){40}}
\put(60,60){\line(0,-1){40}}
\multiput(-1.5,59.5)(2,-2){21}{$.$}
\put(65,58){$\bv^{1}$}
\put(65,48){$\bv^{2}$}
\put(67,28){$\vdots$}
\put(65,18){$\bv^m$}
\put(-5,52){\small$1$}
\put(5,40){\small$i$}
\put(-5,63){\small$1$}
\put(38,66){\small$j$}
\put(58,63){\small$n$}
{\thicklines
\put(10,10){\vector(1,0){15}}
\put(10,10){\vector(0,1){15}}
\put(10,47){\line(0,1){3}}
\put(10,50){\line(1,0){20}}
\put(30,50){\line(0,1){10}}
\put(30,60){\line(1,0){10}}
\put(40,60){\line(0,1){3}}
}
\end{picture}
\end{equation}
or equivalently,
\begin{equation}
\begin{picture}(100,60)(-20,10)
\put(-25,30){$\dsum{}{}$}
\put(0,60){\line(1,0){60}}
\put(10,50){\line(1,0){50}}
\put(20,40){\line(1,0){40}}
\put(30,30){\line(1,0){30}}
\put(40,20){\line(1,0){20}}
\put(10,60){\line(0,-1){10}}
\put(20,60){\line(0,-1){20}}
\put(30,60){\line(0,-1){30}}
\put(40,60){\line(0,-1){40}}
\put(50,60){\line(0,-1){40}}
\put(60,60){\line(0,-1){40}}
\multiput(-1.5,59.5)(2,-2){21}{$.$}
\put(65,58){$\bu^{1}$}
\put(65,48){$\bu^{2}$}
\put(67,25){$\vdots$}
\put(65,18){$\bu^m$}
\put(-5,52){\small$1$}
\put(15,28){\small$j^\ast$}
\put(50,7){$U$}
\put(-5,63){\small$1$}
\put(48,66){\small$i^\ast$}
\put(58,63){\small$n$}
{\thicklines
\put(20,20){\vector(-1,0){15}}
\put(20,20){\vector(0,-1){15}}
\put(20,37){\line(0,1){3}}
\put(20,40){\line(1,0){10}}
\put(30,40){\line(0,1){10}}
\put(30,50){\line(1,0){20}}
\put(50,50){\line(0,1){13}}
}
\end{picture}
\qquad
\begin{picture}(100,60)(-20,10)
\put(-35,30){$=\dsum{}{}$}
\put(0,60){\line(1,0){60}}
\put(10,50){\line(1,0){50}}
\put(20,40){\line(1,0){40}}
\put(30,30){\line(1,0){30}}
\put(40,20){\line(1,0){20}}
\put(10,60){\line(0,-1){10}}
\put(20,60){\line(0,-1){20}}
\put(30,60){\line(0,-1){30}}
\put(40,60){\line(0,-1){40}}
\put(50,60){\line(0,-1){40}}
\put(60,60){\line(0,-1){40}}
\multiput(-1.5,59.5)(2,-2){21}{$.$}
\put(65,58){$\bv^{1}$}
\put(65,48){$\bv^{2}$}
\put(67,25){$\vdots$}
\put(65,18){$\bv^m$}
\put(-5,52){\small$1$}
\put(5,40){\small$i$}
\put(50,7){$V$}
\put(-5,63){\small$1$}
\put(38,66){\small$j$}
\put(58,63){\small$n$}
{\thicklines
\put(10,10){\vector(1,0){15}}
\put(10,10){\vector(0,1){15}}
\put(10,47){\line(0,1){3}}
\put(10,50){\line(1,0){20}}
\put(30,50){\line(0,1){10}}
\put(30,60){\line(1,0){10}}
\put(40,60){\line(0,1){3}}
}
\end{picture}
\end{equation}
where $i^\ast=n-i+1$ and $j^\ast=n-j+1$. 
Hence we have
\begin{theorem}\label{thm:tropSch}
For the indeterminates $u^i_j$ $(1\le i\le j \le n)$ given, 
consider the matrix equation
\begin{equation}
J_n H_1(\bu^1)^\trp\,H_2(\bu^2)^\trp\cdots H_m(\bu^m)^\trp J_n
=H_m(\bv^m)\cdots 
H_2(\bv^2)H_1(\bv^1)
\end{equation}
for $v^i_j$ $(1\le i\le j\le n)$, 
where $m\le n$, and  
$\bu^i=(u^i_i,\ldots,u^i_n)$,
$\bv^i=(v^i_i,\ldots,v^i_n)$ for $i=1,\ldots,m$.
This equation has a unique rational solution\,$;$ it is given by
\begin{equation}
v^i_i=\frac{\sigma^i_i}{\sigma^{i-1}_i},
\quad 
v^i_j=\frac{\sigma^i_j\ \sigma^{i-1}_{j-1}}
{\sigma^{i-1}_j\sigma^i_{j-1}}
\quad(i<j),
\end{equation}
with $\sigma^i_j$ $(i\le j)$ defined as the sum 
\begin{equation}
\sigma^i_j=\sum_{(\gamma_1,\ldots,\gamma_i)}\ 
u_{\gamma_1}\cdots u_{\gamma_i}
\end{equation}
of weights over all $i$-tuples of nonintersecting paths 
\begin{equation}
\gamma_k: (1,n-i+k)\to(\min\br{m,n-j+k},n-j+k)
\end{equation} 
$(k=1,\ldots,i)$,
where the weight $u_\gamma$ of a path $\gamma$ is the 
product of all $u^a_b$'s corresponding to the vertices of $\gamma$.  
\end{theorem} 
Graphically, 
$\sigma^i_j$ ($i\le j$) in the explicit formula above 
is expressed as follows. 
\begin{equation}
\begin{picture}(150,70)(-65,10)
\put(-70,40){$\sigma^i_j=\dsum{(\gamma_1,\ldots,\gamma_i)}{}$}
\put(-10,70){\line(1,0){80}}
\put(70,70){\line(0,-1){60}}
\put(-10,70){\line(1,-1){60}}
\put(50,10){\line(1,0){20}}
\put(80,20){$U$}
\put(-15,62){\small$1$}
\put(2,40){\small$j^\ast$}
\put(12,30){\small$\ddots$}
\put(40,5){\small$m$}
\put(-15,73){\small$1$}
\put(38,75){\small$i^\ast$}
\put(48,75){\small$\ldots$}
\put(68,75){\small$n$}
\multiput(38.5,69.5)(2,-2){15}{$.$}
{\thicklines
\put(40,73){\line(0,-1){3}}
\put(50,73){\line(0,-1){13}}
\put(60,73){\line(0,-1){23}}
\put(70,73){\line(0,-1){33}}
\put(40,70){\line(-1,0){10}}
\put(30,70){\line(0,-1){10}}
\put(30,60){\line(-1,0){20}}
\put(10,60){\line(0,-1){10}}
\put(50,60){\line(-1,0){10}}
\put(40,60){\line(0,-1){10}}
\put(40,50){\line(-1,0){10}}
\put(30,50){\line(0,-1){10}}
\put(30,40){\line(-1,0){10}}
\put(60,50){\line(-1,0){10}}
\put(50,50){\line(0,-1){20}}
\put(50,30){\line(-1,0){20}}
\put(70,40){\line(0,-1){10}}
\put(70,30){\line(-1,0){10}}
\put(60,30){\line(0,-1){10}}
\put(60,20){\line(-1,0){20}}
\put(10,50){\line(0,-1){3}}
\put(20,40){\line(0,-1){3}}
\put(30,30){\line(0,-1){3}}
\put(40,20){\line(0,-1){3}}
}
\end{picture}
\end{equation}
The explicit formula above for the tropical Sch\"utzenberger involution 
is proposed by A.N.~Kirillov \cite{K}, Theorem 4.18. 
\par\medskip

By returning to the combinatorial variables, we obtain the 
following explicit formula for the Sch\"utzenberger involution
(with $m=n$), due to H.~Knight and A.~Zelevinsky \cite{KZ}
(see also \cite{BFZ}, \cite{Z}).
 
\begin{theorem} 
Let $P$ be a column strict tableau and denote by 
$p^i_j$ the number of $j$'s in the $i$-th row of $P$
for $1\le i\le j\le n$.  
Let $\SI{P}$ be the column strict tableau obtained from $P$ by 
applying  
the Sch\"utzenberger involution, and 
denote by 
$\widetilde{p}^i_j$ the number of $j$'s in the $i$-th row of $\SI{P}$
for $1\le i\le j\le n$.  
Then $\widetilde{p}^i_j$ are determined from $p^i_j$ by 
the following explicit formula\,$:$ 
\begin{equation}
\widetilde{p}^i_i=\sigma^i_i -\sigma^{i-1}_i,
\quad 
\widetilde{p}^i_j=\sigma^i_j-\sigma^{i-1}_j 
- \sigma^i_{j-1}+\sigma^{i-1}_{j-1}
\quad(i<j),
\end{equation}
with $\sigma^i_j$ $(1\le i\le j\le n)$ defined as the maximum 
\begin{equation}
\sigma^i_j=\dmax{(\gamma_1,\ldots,\gamma_i)}\ 
(p_{\gamma_1}+\cdots+p_{\gamma_i})
\end{equation}
of weights over all $i$-tuples of nonintersecting paths 
$\gamma_k: (1,n-i+k)\to(n-j+k,n-j+k)$ \ $(k=1,\ldots,i)$,
where the weight $p_\gamma$ of a path $\gamma$ is the 
sum of all $p^a_b$'s corresponding to the vertices of $\gamma$.  
\end{theorem} 
\begin{example}\rm
In the case of $P$ and $\SI{P}$ of Example \ref{ex:SchP},
$\bp=\pmatrix{p^i_j}_{i\le j}$ and 
$\widetilde{\bp}=\pmatrix{\widetilde{p}^i_j}_{i\le j}$ are 
given by
\begin{equation}
\bp=\left[\matrix{
2&1&1&0 \cr
 & 2 &1&0 \cr
&&0&1\cr
&&&0
}\right],\quad
\widetilde{\bp}=\left[\matrix{
1&1&2&0 \cr
 & 1 &0&2 \cr
&&1&0\cr
&&&0
}\right].
\end{equation}
The table $\widetilde{\bp}$ can be determined through 
$\bsigma=\pmatrix{\sigma^i_j}_{i\le j}$: 
\begin{equation}
\bsigma=\left[\matrix{
1& 2& 4& 4\cr
& 3 & 5& 7 \cr
&& 6& 8\cr
&&&8 
}\right].
\end{equation}
\end{example}

\begin{remark} \rm\ 
With the notation of Remark \ref{rem:HHH}, 
the {tropical Sch\"utzenberger involution} can be 
formulated as follows. 
We define the involution $\theta : \CB \to\CB$ by setting 
\begin{equation}
\theta(H)=J_n H^\trp J_n \qquad(H\in\CB). 
\end{equation}
Then the isomorphism $h : \CB_0 \iso \CB_\tau$ induces 
a birational involution $U\mapsto \SI{U}$ on $\CB_0$ such that 
\begin{equation}
h(\SI{U})=\theta(h(U))=J_n h(U)^\trp\,J_n
\end{equation}
for generic $U\in\CB_0$.  
We already gave the explicit formula for $V=\SI{U}$
in Theorem \ref{thm:tropSch}.  
Note also that the inverse correspondence $V\mapsto U$ 
is given by the same formula. 
\end{remark}
\setcounter{equation}{0}
\section{Tropical RSK correspondence}

\subsection{Variations of RSK correspondence}
\label{ssec:varRSK}
Let $A=\pmatrix{a^i_j}_{i,j}\in\Mat_{m,n}(\BN)$ be an
$m\times n$ matrix of nonnegative integers.
\begin{equation}
A=\left[\matrix{
a^1_1 & a^1_2 & \ldots & a^1_n\cr
a^2_1 & a^2_2 & \ldots & a^2_n\cr
\vdots &\vdots &  &\vdots\cr
a^m_1 & a^m_2 & \ldots & a^m_n
}\right]
\end{equation}
Setting 
$w_i=1^{a^i_1}\ldots n^{a^i_n}$ for 
$i=1,\ldots,m$, 
we denote by $P=P(w)$ the column strict tableau 
obtained from the word $w=w_1\ldots w_m$ by 
row insertion.  
Similarly, setting 
$w'_j=1^{a^1_j}\ldots m^{a^m_j}$ for 
$j=1,\ldots,n$, 
we denote by $Q=P(w')$ the column strict tableau 
obtained from the word $w'=w'_1\ldots w'_n$ by 
row insertion.  
The two tableaux $P$ and $Q$ have the same shape,
and 
the correspondence $A\mapsto (P,Q)$ induces 
a bijection between the set of all $m\times n$ 
matrices of nonnegative integers and the 
set of pairs $(P,Q)$ of column strict tableaux 
of a same shape, 
$P$ with contents in $\br{1,\ldots,n}$ and $Q$ 
with contents in $\br{1,\ldots,m}$.
This bijection $A\mapsto (P,Q)$ is called the 
{\em Robinson-Schensted-Knuth correspondence}
({\em RSK correspondence}, for short). 
In this context, the matrix $A$ is sometimes called 
the {\em transportation matrix}. 
By combining this {\em standard} RSK correspondence with 
the Sch\"utzenberger involution, 
we have the following four variations of 
RSK correspondences: 
\begin{equation}
\begin{array}{ll}\smallskip
A \ \mapsto\ (P,Q),\qquad
& A \ \mapsto\ (P,\SI{Q}),\cr
A \ \mapsto\ (\SI{P},Q),\qquad
& A\ \mapsto\ (\SI{P},\SI{Q}).
\end{array}
\end{equation}

Let us denote by $p^i_j$, $q^i_j$, $\widetilde{p}^i_j$ and 
$\widetilde{q}^i_j$  
the number of 
$j$'s in the $i$-th row of the tableaux $P$,  $Q$, $\SI{P}$ and 
$\SI{Q}$, respectively.  
The common shape $\lambda=(\lambda_1,\lambda_2,\ldots,\lambda_l)$,
$l=\min\br{m,n}$, of these four tableaux is given by 
\begin{equation}
\lambda_i=p^i_i+\cdots+p^i_n=q^i_i+\cdots+q^i_m
=\widetilde{p}^i_i+\cdots+\widetilde{p}^i_n
=\widetilde{q}^i_i+\cdots+\widetilde{q}^i_m
\end{equation}
for $i=1,\ldots,l$. 
\comment{
The Gelfand-Tsetlin pattern for $P$ 
will be denoted by 
\begin{equation}
\bmu=\left[\matrix{
\mu^{(n)}_1\quad\mu^{(n)}_2\ \qquad\ \ldots \qquad\ \mu^{(n)}_n\cr
\mu^{(n-1)}_1\,\mu^{(n-1)}_2\ \ldots\,\mu^{(n-1)}_{n-1}\cr
\cdots\cr
\mu^{(1)}_1
}\right],
\end{equation}
where $\mu^{(j)}_i=p^i_i+\cdots+p^i_j$ for $i\le j$. 
Similarly, we denote the Gelfand-Tsetlin patterns for 
$Q$, $\SI{P}$ and $\SI{Q}$ by $\bnu$, $\widetilde{\bmu}$
and $\widetilde{\bnu}$, respectively. 
}
\begin{example}\rm
Consider the transportation matrix 
\begin{equation}
A=\left[\matrix{
0&2&1&1\cr
1&0&1&2\cr
2&2&0&1
}\right]. 
\end{equation}
The tableaux $P$, $Q$ for $A$, and their counterparts under the 
Sch\"utzenberger involution are determined as follows: 
\begin{eqnarray}
&&
\,P=P(2234|1344|11224)=\tbl{1&1&1&2&2&4&4\cr2&2&3&3&4\cr4}, 
\quad
\bp=\left[\matrix{
3&2&0&2\cr
&2&2&1\cr
&&0&1
}\right]
\nonumber\\
&&
\,Q=P(233|1133|12|1223)=\tbl{1&1&1&1&2&2&3\cr2&2&3&3&3\cr3},
\quad
\bq=\left[\matrix{
4&2&1\cr
&2&3\cr
&&1
}\right]
\nonumber\\
&&
\SI{P}=P(13344|1124|1233)=\tbl{1&1&1&1&2&3&3\cr2&3&4&4&4\cr3},
\quad
\widetilde{\bp}=\left[\matrix{
4&1&2&0\cr
&1&1&3\cr
&&1&0
}\right]
\nonumber\\
&&
\SI{Q}=P(1223|23|1133|112)=\tbl{1&1&1&1&1&2&3\cr2&2&2&3&3\cr3},
\quad
\widetilde{\bq}=\left[\matrix{
5&1&1\cr
&3&2\cr
&&1
}\right]
\end{eqnarray}
\end{example}

In the arguments of this section, the correspondence 
$A\mapsto (P,\SI{Q})$ 
will 
play the essential role, rather than the ordinary RSK correspondence. 
For this reason, we use the following convention.  
Let $X=\pmatrix{x^i_j}_{i,j}$ be an $m\times n$ matrix of 
nonnegative integers. 
We denote 
the $i$-th row of $X$ by $\bx^i$, 
and the $j$-th column of $X$ by $\bx_j$\,: 
\begin{equation}
X=\left[\matrix{
x^1_1 & x^1_2 & \ldots & x^1_n\cr
x^2_1 & x^2_2 & \ldots & x^2_n\cr
\vdots &\vdots &  &\vdots\cr
x^m_1 & x^m_2 & \ldots & x^m_n
}\right]
=
\left[\matrix{
\bx^1 \cr \bx^2 \cr \vdots \cr \bx^m
}\right]
=
\big[\,\bx_1\ \bx_2\ \ldots\  \bx_n\,\big]. 
\end{equation}
{}From $X$, we construct the column strict tableau 
\begin{equation}
U=P(w_mw_{m-1}\cdots w_1),\quad 
w_i=1^{x^i_1}2^{x^i_2}\cdots n^{x^i_n}
\end{equation}
by reading the rows $\bx^m,\ldots,\bx^1$ 
of $X$ {\em from bottom to top}, and 
\begin{equation}
V=P(w'_nw'_{n-1}\cdots w'_1),\quad 
w'_j=1^{x^1_j}2^{x^2_j}\cdots m^{x^m_j},
\end{equation}
by reading the columns $\bx_n,\ldots,\bx_1$ 
{\em from right to left}. 
If we set $X=J_m A$, these $U$ and $V$ 
correspond to $P$ and $\SI{Q}$, determined 
from $A$, so that
\begin{equation} 
U=P, \quad V=\SI{Q},\quad \SI{U}=\SI{P},\quad \SI{V}=Q. 
\end{equation}
In what follows, we refer to this particular 
correspondence $X\mapsto(U,V)$ as the {\em \RSKs correspondence}.  
\par\medskip
Before the discussion of tropical RSK correspondences, 
we formulate an isomorphism theorem concerning  
the path representation of generic matrices.

\subsection{A fundamental isomorphism}
We first fix a notation of {\em special} minor determinants. 
For an $m\times n$ matrix $\Phi=\pmatrix{\varphi^i_j}_{i,j}$ given, 
we introduce the notation 
\begin{equation}
\tau^i_j(\Phi)=\left\{
\begin{array}{ll}\medskip
\det\Phi^{i-j+1,\ldots,i}_{1,\ldots,j}\quad &(i\ge j),\cr
\det\Phi^{1,\ldots,i}_{j-i+1,\ldots,j}\quad &(i\le j),
\end{array}\right.
\end{equation} 
of the minor determinant of $\Phi$ corresponding to the 
largest square in the rectangle 
$\br{1,\ldots,m}\times\br{1,\ldots,n}$  
whose right-bottom corner is located at $(i,j)$.
\begin{equation}
\begin{picture}(200,90)(-20,0)
\put(-60,45){$\tau^i_j(\Phi) : $}
\put(0,20){\line(0,1){60}}
\put(70,20){\line(0,1){60}}
\put(0,20){\line(1,0){70}}
\put(0,80){\line(1,0){70}}
\put(0,30){\line(1,0){30}}
\put(0,60){\line(1,0){30}}
\put(30,80){\line(0,-1){50}}
\multiput(0,56)(4,0){8}{$\cdot$}
\multiput(2,54)(4,0){7}{$\cdot$}
\multiput(0,52)(4,0){8}{$\cdot$}
\multiput(2,50)(4,0){7}{$\cdot$}
\multiput(0,48)(4,0){8}{$\cdot$}
\multiput(2,46)(4,0){7}{$\cdot$}
\multiput(0,44)(4,0){8}{$\cdot$}
\multiput(2,42)(4,0){7}{$\cdot$}
\multiput(0,40)(4,0){8}{$\cdot$}
\multiput(2,38)(4,0){7}{$\cdot$}
\multiput(0,36)(4,0){8}{$\cdot$}
\multiput(2,34)(4,0){7}{$\cdot$}
\multiput(0,32)(4,0){8}{$\cdot$}
\multiput(2,30)(4,0){7}{$\cdot$}
\multiput(0,28)(4,0){8}{$\cdot$}
\put(-6,76){\small$1$}
\put(-6,28){\small$i$}
\put(-10,18){\small$m$}
\put(-2,83){\small$1$}
\put(28,84){\small$j$}
\put(68,83){\small$n$}
\put(24,4){$i\ge j$}
\put(52,28){$\Phi$}
\put(120,20){\line(0,1){60}}
\put(190,20){\line(0,1){60}}
\put(120,20){\line(1,0){70}}
\put(120,80){\line(1,0){70}}
\put(140,50){\line(0,1){30}}
\put(170,50){\line(0,1){30}}
\put(120,50){\line(1,0){50}}
\multiput(140,76)(4,0){8}{$\cdot$}
\multiput(142,74)(4,0){7}{$\cdot$}
\multiput(140,72)(4,0){8}{$\cdot$}
\multiput(142,70)(4,0){7}{$\cdot$}
\multiput(140,68)(4,0){8}{$\cdot$}
\multiput(142,66)(4,0){7}{$\cdot$}
\multiput(140,64)(4,0){8}{$\cdot$}
\multiput(142,62)(4,0){7}{$\cdot$}
\multiput(140,60)(4,0){8}{$\cdot$}
\multiput(142,58)(4,0){7}{$\cdot$}
\multiput(140,56)(4,0){8}{$\cdot$}
\multiput(142,54)(4,0){7}{$\cdot$}
\multiput(140,52)(4,0){8}{$\cdot$}
\multiput(142,50)(4,0){7}{$\cdot$}
\multiput(140,48)(4,0){8}{$\cdot$}
\put(114,76){\small$1$}
\put(114,48){\small$i$}
\put(110,18){\small$m$}
\put(118,83){\small$1$}
\put(168,84){\small$j$}
\put(188,83){\small$n$}
\put(144,4){$i\le j$}
\put(172,28){$\Phi$}
\end{picture}
\end{equation}
For convenience, we define $\tau^0_k(\Phi)=\tau^k_0(\Phi)=0$ for any $k$. 
\comment{
Setting $\tau(\Phi)=\pmatrix{\tau^i_j(\Phi)}_{i,j}$, we obtain 
a regular mapping
\begin{equation}
\tau\,:\, \Mat_{m,n}(\BK) \ \to\  \Mat_{m,n}(\BK).
\end{equation}
As we will see later, this mapping induces the isomorphism
\begin{equation} 
\tau\,:\, \Mat_{m,n}(\BK)_\tau \ \iso\  \Mat_{m,n}(\BK^\ast),\quad
\BK^\ast=\BK\setminus \br{0}, 
\end{equation}
where 
}
We define the subset $\Mat_{m,n}(\BK)_\tau$ 
of  $\Mat_{m,n}(\BK)$ by
\begin{equation}
\Mat_{m,n}(\BK)_\tau=\br{\Phi\in\Mat_{m,n}(\BK)\mid 
\tau^i_j(\Phi)\ne 0\ \ \mbox{for all}\ \ (i,j)}.
\end{equation}

For an $m\times n$ matrix $X=\pmatrix{x^i_j}_{i,j}$ given, we now 
construct an $m\times n$ matrix $\Phi=\pmatrix{\varphi^i_j}_{i,j}$
by using the paths on the lattice 
$\br{1,\ldots m}\times\br{1,\ldots,n}$.  
When we refer to a path in the rectangular lattice, we mean 
a shortest path joining two vertices, 
without specifying the orientation of edges. 
As before, 
for each path $\gamma : (a,b) \to (c,d)$,
we define the weight $x_\gamma$ of $\gamma$, 
associated with $X$, to be the product
of all $x^i_j$'s corresponding the vertices on $\gamma$.  
\begin{equation}
\begin{picture}(80,80)(0,0)
\put(0,0){\line(1,0){80}}
\put(0,70){\line(1,0){80}}
\put(0,0){\line(0,1){70}}
\put(80,0){\line(0,1){70}}
{\thicklines
\put(15,20){\line(1,0){20}}
\put(35,20){\line(0,1){10}}
\put(35,30){\line(1,0){20}}
\put(55,30){\line(0,1){20}}
}
\put(5,10){\small$(a,b)$}
\put(45,55){\small$(c,d)$}
\put(13,18){\scriptsize$\bullet$}
\put(53,48){\scriptsize$\bullet$}
\put(50,20){$\gamma$}
\put(-6,66){\small$1$}
\put(-10,-2){\small$m$}
\put(-2,74){\small$1$}
\put(77,74){\small$n$}
\put(5,55){$X$}
\end{picture}
\end{equation}
With this definition of weight, we define the 
matrix $\Phi=\pmatrix{\varphi^i_j}_{i,j}$ 
by setting 
\begin{equation}\label{eq:xtophi}
\begin{picture}(250,60)(0,10)
\put(0,35){$\varphi^i_j=\varphi^i_j(X)
=\dsum{\gamma: (i,1)\to(1,j)}{}\, x_\gamma 
=\dsum{\gamma}{}\ $}
\put(160,10){\line(1,0){70}}
\put(160,70){\line(1,0){70}}
\put(160,10){\line(0,1){60}}
\put(230,10){\line(0,1){60}}
{\thicklines
\put(160,30){\line(1,0){20}}
\put(180,30){\line(0,1){15}}
\put(180,45){\line(1,0){15}}
\put(195,45){\line(0,1){25}}
\put(195,70){\line(1,0){10}}
}
\put(153,27){\small$i$}
\put(203,76){\small$j$}
\put(158,28){\scriptsize$\bullet$}
\put(203,68){\scriptsize$\bullet$}
\put(205,25){$X$}
\end{picture}
\end{equation}
for each $(i,j)\in\br{1,\ldots,m}\times\br{1,\ldots,n}$, 
where the summation is taken over all paths from $(i,1)$ to $(1,j)$.  
The mapping 
$X\mapsto\Phi$ thus obtained will be denoted by 
\begin{equation}
\varphi\ :\ \Mat_{m,n}(\BK) \to \Mat_{m,n}(\BK),\quad
\varphi(X)=\pmatrix{\varphi^i_j(X)}_{i,j}. 
\end{equation}
This mapping $\varphi$ provides us with a device 
for generating nonintersecting paths on 
the lattice $\br{1,\ldots,m}\times\br{1,\ldots,n}$. 
In fact, from the theorem of Gessel-Viennot \cite{GV}, it follows that, 
for any choice of column indices $i_1<\cdots<i_r$ and 
row indices $j_1<\cdots<j_r$, 
the corresponding minor determinant of $\Phi$ is expressed as a sum  
\begin{equation}\label{eq:phinipaths}
\det \Phi^{i_1,\ldots,i_r}_{j_1,\ldots,j_r} = 
\sum_{(\gamma_1,\ldots,\gamma_r)} x_{\gamma_1}\!\cdots x_{\gamma_r}
\end{equation}
of the product of weights over all $r$-tuples $(\gamma_1,\ldots,\gamma_r)$ 
of nonintersecting paths $\gamma_k : (i_k,1) \to (1,j_k)$ ($k=1,\ldots,r$). 
Graphically, this summation can be expressed as follows.
\begin{equation}
\begin{picture}(170,80)(-90,5)
\put(-90,38){$\det \Phi^{i_1,\ldots,i_r}_{j_1,\ldots,j_r} = \sum $}
\put(60,20){$X$}
\put(0,10){\line(1,0){80}}
\put(0,80){\line(1,0){80}}
\put(0,10){\line(0,1){70}}
\put(80,10){\line(0,1){70}}
{\thicklines
\put(-3,60){\line(1,0){3}}
\put(0,60){\line(1,0){10}}
\put(10,60){\line(0,1){5}}
\put(10,65){\line(1,0){10}}
\put(20,65){\line(0,1){15}}
\put(20,80){\line(1,0){5}}
\put(25,80){\line(0,1){3}}
\put(-3,45){\line(1,0){3}}
\put(0,45){\line(1,0){5}}
\put(5,45){\line(0,1){5}}
\put(5,50){\line(1,0){25}}
\put(30,50){\line(0,1){10}}
\put(30,60){\line(1,0){10}}
\put(40,60){\line(0,1){20}}
\put(40,80){\line(0,1){3}}
\put(-3,20){\line(1,0){3}}
\put(0,20){\line(1,0){10}}
\put(10,20){\line(0,1){15}}
\put(10,35){\line(1,0){30}}
\put(40,35){\line(0,1){10}}
\put(40,45){\line(1,0){15}}
\put(55,45){\line(0,1){20}}
\put(55,65){\line(1,0){10}}
\put(65,65){\line(0,1){15}}
\put(65,80){\line(0,1){3}}
\put(-12,60){\small$i_1$}
\put(-12,45){\small$i_2$}
\put(-10,28){\small$\vdots$}
\put(-12,18){\small$i_r$}
\put(23,86){\small$j_1$}
\put(38,86){\small$j_2$}
\put(48,86){\small$\ldots$}
\put(65,86){\small$j_r$}
}
\end{picture}
\end{equation}
Let us look at the special minor determinants 
$\tau^i_j(\Phi)$ introduced above. 
Notice that, for each $(i,j)$, there is only one $r$-tuple 
($r=\min\br{i,j}$) of nonintersecting paths relevant to the summation, 
so that 
the minor determinant $\tau^i_j(\Phi)$ reduces to the product 
\begin{equation}\label{eq:xtotau}
\tau^i_j(\Phi)=\prod_{(a,b); \,a\le i,\, b\le j} x^a_b. 
\end{equation}
This implies that, if $x^i_j\ne 0$ for all $i,j$, then one has 
$\tau^i_j(\Phi)\ne0$ for all $i,j$.
{}From formula \eqref{eq:xtotau}, it is also clear 
that the entries $x^i_j$ of the matrix $X$ are recovered 
as the ratios of minor determinants of $\Phi$
\begin{equation}\label{eq:phitox}
x^i_j=\frac{\tau^i_j\ \tau^{i-1}_{j-1}\ }
{\tau^{i-1}_j\,\tau^i_{j-1}},
\qquad \tau^i_j=\tau^i_j(\Phi),
\end{equation}
provided that $x^i_j\ne 0$ for all $i,j$. 
The correspondence $X\mapsto\Phi$ defined by \eqref{eq:xtophi} 
induces a mapping
\begin{equation}
\varphi\ :\ \Mat_{m,n}(\BK^\ast)\ \to\ \Mat_{m,n}(\BK)_\tau. 
\end{equation}
As we have seen above, if $\Phi=\varphi(X)$, then 
the matrix $X$ is recovered by the formula \eqref{eq:phitox}. 
\begin{theorem}\label{thm:A}
The correspondence $X\mapsto\Phi$ defined by \eqref{eq:xtophi} 
induces an isomorphism of 
affine varieties
\begin{equation}
\varphi\ :\ \Mat_{m,n}(\BK^\ast) \ \iso\  \Mat_{m,n}(\BK)_\tau. 
\end{equation}
\end{theorem}
\proof
Since $\varphi$ has a left inverse defined by \eqref{eq:phitox}, 
we have only to show that $\varphi$ is surjective. 
For each $\Phi\in \Mat_{m,n}(\BK)_\tau$, we construct 
an $X\in\Mat_{m,n}(\BK^\ast)$ such that $\varphi(X)=\Phi$, 
by the induction on $m$. 
For this purpose, we first investigate the inductive structure 
of the mapping $\varphi$. 
Assuming that $\varphi(X)=\Phi$, set
\begin{equation}
\psi^i_j=\sum_{\gamma:(i,1)\to(2,j)}\ x_\gamma
\qquad(2\le i\le m, 1\le j\le n). 
\end{equation}
Then $\varphi^i_j$ are determined as
\begin{equation}\label{eq:phipsi}
\varphi^1_j=x^1_1\,x^1_2\cdots x^1_j,\qquad
\varphi^i_j=\sum_{k=1}^j \psi^i_k \,x^1_k\,x^1_{k+1}\cdots x^1_j
\quad(i=2,\ldots,m),
\end{equation}
for all $j$. 
In view of this, we consider the $n\times n$ upper triangular  matrix 
$H(\bx^1)$, $\bx^1=\pmatrix{x^1_1,\ldots,x^1_n}$, 
associated with the first row of $X$.  
Then the condition \eqref{eq:phipsi} is equivalent to the matrix equation
\begin{equation}\label{eq:phipsih}
\Phi=\Psi\,H(\bx^1),\quad \Psi=\left[\matrix{
1 & 0 &\ldots & 0\cr
\psi^2_1& \psi^2_2 & \ldots & \psi^2_n \cr
\vdots &\vdots&&\vdots\cr
\psi^m_1& \psi^m_2 & \ldots & \psi^m_n 
}\right].
\end{equation}
Let us show that any $\Phi\in \Mat_{m,n}(\BK)_\tau$ can be 
decomposed in this form by choosing $x^1_j$ and $\psi^i_j$ appropriately.  
The condition to be satisfied by the first row 
$(x^1_1,\ldots,x^1_n)$ of $X$ is:
\begin{equation}\label{eq:phihi}
\pmatrix{\Phi H(\bx^1)^{-1}}^1_j=\delta_{1,j}\qquad(j=1,\ldots,n).
\end{equation}
Since $H(\bx^1)^{-1}=\diag{\overline{\bx}^1}-\Lambda$, 
it is easily seen that \eqref{eq:phihi} is equivalent to
\begin{equation}
x^1_1=\varphi^1_1,\quad x^1_j
=\frac{\varphi^1_j}{\varphi^1_{j-1}}\ \ (j=2,\ldots,n).
\end{equation}
Since $\varphi^1_j=\tau^1_j(\Phi)\ne 0$, 
we can define $x^1_j$ ($j=1,\ldots,n$)
as above.  Then the matrix $\Psi=\Phi H(\bx^1)^{-1}$ has 
the first row $(1,0,\ldots,0)$;
hence, $H(\bx^1)$ and $\Psi$ satisfy the condition \eqref{eq:phipsih}.  
Define $\Phi'$ to be the $(m-1)\times n$ matrix obtained from $\Psi$ by 
removing the first row. 
We will verify that $\Phi'\in\Mat_{m-1,n}(\BK)_\tau$ so that 
$\Phi'$ can be expressed as $\Phi'=\varphi(X')$ by the induction 
hypothesis.   
Then, setting 
\begin{equation}
X=\left[\matrix{x^1_1& \ldots & x^1_n \cr \cr & X' &}\right],
\quad X'=\left[\matrix{
x^2_1 & \ldots & x^2_n \cr
\vdots &&\vdots\cr
x^m_1 & \ldots & x^m_n
}\right], 
\end{equation}
we must have $\varphi(X)=\Phi$, which will complete 
the proof of Theorem \ref{thm:A}. 
We now examine the minor determinants of $\Phi = \Psi H(\bx^1)$.  
Let $(i,j)\in\br{1,\ldots,m}\times\br{1,\ldots,n}$ and assume $i>j$. 
Then it is clear 
\begin{equation}
\tau^i_j(\Phi)=\det\Phi^{i-j+1,\ldots,i}_{1,\ldots,j}
=\det\Psi^{i-j+1,\ldots,i}_{1,\ldots,j}
x^1_1\ldots x^1_j,
\end{equation}
since $H$ is upper triangular. 
Hence we have 
\begin{equation}
\tau^{i-1}_j(\Phi')=\det\Psi^{i-j+1,\ldots,i}_{1,\ldots,j}
=\frac{1}{x^1_1\ldots,x^1_j}
\tau^i_j(\Phi)\ne 0\quad(i>j).
\end{equation}
Next assume $i\le j$.  In this case, we have
\begin{equation}
\tau^i_j(\Phi)=
\det\Phi^{1,\ldots,i}_{j-i+1,\ldots,j}
=\sum_{k_1<\ldots<k_i}\det \Psi^{1,\ldots,i}_{k_1,\ldots,k_i} 
\det H(\bx^1)^{k_1,\ldots,k_i}_{j-i+1,\ldots,j}. 
\end{equation}
Since the first row of $\Psi$ is $\pmatrix{1,0,\ldots,0}$, we have 
$\det\Psi^{1,\ldots,i}_{k_1,\ldots,k_i} =0$ unless $k_1=1$.  
When $k_1=1$, from Lemma \ref{lem:HH} it follows that 
$\det H(\bx^1)^{1,k_2,\ldots,k_i}_{j-i+1,\ldots,j}=0$ unless 
$(k_2,\ldots,k_i)=(j-i+2,\ldots,j)$. 
Since $\det H(\bx^1)^{1,j-i+2,\ldots,j}_{j-i+1,\ldots,j}
=x^1_1\,x^1_2\cdots x^1_j$,
we finally obtain
\begin{equation}
\tau^i_j(\Phi)=\det\Psi^{1,\ldots,i}_{1,j-i+2,\ldots,j}\,x^1_1\cdots x^1_j.
\end{equation}
Hence we have 
\begin{equation}
\tau^{i-1}_j(\Phi')=\det\Psi^{1,\ldots,i}_{1,j-i+2,\ldots,j}
=\frac{1}{x^1_1\cdots x^1_j} \tau^i_j(\Phi)\ne 0\quad(i\le j).  
\end{equation}
This argument implies $\tau^i_j(\Phi')\ne0$ for all $(i,j)$
with $1\le i\le m-1$ and $1\le j\le n$ 
as desired.  This completes the proof of Theorem \ref{thm:A}. \qed

It is convenient for our purpose to restate Theorem \ref{thm:A} as follows. 
\begin{theorem}\label{thm:B}
Let $\Phi=\pmatrix{\varphi^i_j}_{i,j}$ be an $m\times n$ matrix 
with coefficients 
in $\BK$ such that $\tau^i_j(\Phi)\ne 0$ for all $(i,j)$. 
For such a matrix $\Phi$ given, define the $m\times n$ 
matrix $X=\pmatrix{x^i_j}_{i,j}$ 
by setting 
\begin{equation}\label{eq:xphi}
x^i_j=\frac{\tau^i_j\ \tau^{i-1}_{j-1}\ }
{\tau^{i-1}_j\,\tau^i_{j-1}},
\qquad \tau^i_j=\tau^i_j(\Phi). 
\end{equation}
Then, 
for any choice of column indices $i_1<\cdots<i_r$ and 
row indices $j_1<\cdots<j_r$, 
the corresponding minor determinant of $\Phi$ is expressed as a sum  
\begin{equation}\label{eq:phiminpaths}
\det \Phi^{i_1,\ldots,i_r}_{j_1,\ldots,j_r} = 
\sum_{(\gamma_1,\ldots,\gamma_r)} x_{\gamma_1}\!\cdots x_{\gamma_r}
\end{equation}
of the product of weights over all $r$-tuples $(\gamma_1,\ldots,\gamma_r)$ 
of nonintersecting paths $\gamma_k : (i_k,1) \to (1,j_k)$ $(k=1,\ldots,r)$ 
in the $m\times n$ rectangle.
\end{theorem}

\begin{remark}\rm
An $m\times n$ real matrix $\Phi$ ($\BK=\BR$) is said 
to be {\em totally positive}
if $\det \Phi^{i_1,\ldots,i_r}_{j_1,\ldots,j_r}>0$ 
for any choice of row indices 
$i_1<\cdots<i_r$ and column indices $j_1<\cdots<j_r$.  
Theorem \ref{thm:A} implies 
that, if $\tau^i_j(\Phi)>0$ for all $(i,j)$, then 
$\Phi$ is already totally positive.  
In fact, if this condition is satisfied, all the 
$x^i_j$'s are positive; hence, any minor determinant 
$\det \Phi^{i_1,\ldots,i_r}_{j_1,\ldots,j_r}$ is positive 
since it is expressed 
as a sum of weights associated with $X$ over $r$-tuples 
of nonintersecting paths 
as in \eqref{eq:phiminpaths}. 
Let us denote by $\Mat_{m,n}(\BR)_{\mbox{\scriptsize tot.pos.}}$ 
the open subset
of $\Mat_{m,n}(\BR)$ consisting of all totally positive matrices.  
Then Theorem \ref{thm:A} also implies that $\varphi$ induces the isomorphism 
\begin{equation}
\varphi\ :\ \Mat_{m,n}(\BR_{>0}) \ \iso\ 
\Mat_{m,n}(\BR)_{\mbox{\scriptsize tot.pos}}. 
\end{equation}
In particular, $\Mat_{m,n}(\BR)_{\mbox{\scriptsize tot.pos}}$ is isomorphic  
to $\BR_{>0}^{mn}$ as a real analytic manifold. 
For the theory of totally positive matrices, we refer the 
reader to \cite{BFZ}.
\end{remark}

We apply the fundamental isomorphism 
of Theorem \ref{thm:A} to formulating   
a prototype of subtraction-free birational 
involution 
on the space $\Mat_{m,n}(\BK^\ast)$ of matrices. 

For each $\Phi\in \Mat_{m,n}(\BK)$, we define 
the matrix $\Phi^\vee$ by setting
\begin{equation}
\Phi^\vee =J_m\, \Phi\, J_n, 
\end{equation}
where $J_m=\pmatrix{\delta_{i+j,m+1}}_{i,j=1}^m$ 
and $J_n=\pmatrix{\delta_{i+j,n+1}}_{i,j=1}^n$ 
are the permutation matrices representing 
to the longest element of $\BS_m$ and $\BS_n$,
respectively. 
This correspondence $\Phi\mapsto\Phi^\vee$  
defines an involution on the space $\Mat_{m,n}(\BK)$ 
of $m\times n$ matrices.
Then, via the isomorphism 
\begin{equation}
\varphi : \ \Mat_{m,n}(\BK^\ast)\ \iso \ \Mat_{m,n}(\BK)_\tau,
\end{equation}
we obtain a birational involution $X\mapsto \iota(X)$ 
on $\Mat_{m,n}(\BK^\ast)$ such that
\begin{equation}
\varphi(\iota(X))=\varphi(X)^\vee=J_m\,\varphi(X)\,J_n
\end{equation}
for {\em generic} $X\in\Mat_{m,n}(\BK^\ast)$. 

To be more explicit, 
let us consider two matrices $X, Y\in\Mat_{m,n}(\BK^\ast)$,
and set $\Phi=\varphi(X)$ and $\Psi=\varphi(Y)$. 
If we impose the relation 
$\Psi=\Phi^\vee$ between $\Phi$ and $\Psi$, 
it induces a birational correspondence 
between $X$ and $Y=\iota(X)$.  
As we will see below, this correspondence $X\leftrightarrow Y$ 
provides the essential ingredient of the \RSKs 
correspondence. 
Recall that $X=\pmatrix{x^i_j}_{i,j}$ 
is recovered from $\Phi=\pmatrix{\varphi^i_j}_{i,j}$
by the formula
\begin{equation}
x^i_j=\frac{\tau^i_j\ \tau^{i-1}_{j-1}}{\tau^{i-1}_{j} \tau^i_{j-1}},
\qquad \tau^i_j=\tau^i_j(\Phi), 
\end{equation}
and $Y=\pmatrix{y^i_j}_{i,j}$ from $\Psi=\pmatrix{\psi^i_j}_{i,j}$ by 
\begin{equation}
y^i_j=\frac{\sigma^i_j\ \sigma^{i-1}_{j-1}}
{\sigma^{i-1}_{j} \sigma^i_{j-1}},
\qquad \sigma^i_j=\tau^i_j(\Psi). 
\end{equation}
We now look at the determinant $\sigma^i_j$.  
Since $\Psi=\Phi^\vee$, we have
\begin{equation}
\sigma^i_j=\tau^i_j(\Psi)=\det \Psi^{i-r+1,\ldots,i}_{j-r+1,\ldots,j}
=\det \Phi^{m-i+1,\ldots,m-i+r}_{n-j+1,\ldots,n-j+r},
\end{equation}
where $r=\min\br{i,j}$. 
Hence, each $\sigma^i_j$ is expressed as the sum
\begin{equation}
\sigma^i_j=\sum_{(\gamma_1,\ldots,\gamma_r)}
\ x_{\gamma_1}\cdots x_{\gamma_r}
\end{equation}
of weights associated with $X$, over all $r$-tuples of 
nonintersecting paths 
\begin{equation}
\gamma_k: (m-i+k,1)\to(1,n-j+k) \quad(l=1,\ldots,r). 
\end{equation}
Graphically, $\sigma^i_j$ can be expressed as follows. 
\begin{equation}
\begin{picture}(150,90)(-75,-15)
\unitlength=0.8pt
\put(-98,40){$\sigma^i_j=\sum$}
\put(105,40){$,\ \mbox{or}\ \sum$}
\put(30,-20){$(\,i\ge j\,)$}
\put(70,10){$X$}
\put(0,0){\line(1,0){100}}
\put(0,80){\line(1,0){100}}
\put(0,0){\line(0,1){80}}
\put(100,0){\line(0,1){80}}
{\thicklines
\put(70,84){\line(0,-1){4}}
\put(80,84){\line(0,-1){14}}
\put(90,84){\line(0,-1){24}}
\put(100,84){\line(0,-1){34}}
\multiput(70,77)(2,-2){15}{$\cdot$}
\multiput(80,77)(2,-2){10}{$\cdot$}
\multiput(90,77)(2,-2){5}{$\cdot$}
\put(30,87){\scriptsize$n-j+1$}
\put(80,88){\scriptsize$\ldots$}
\put(100,87){\scriptsize$n$}
\put(-4,50){\line(1,0){4}}
\put(-4,40){\line(1,0){14}}
\put(-4,30){\line(1,0){24}}
\put(-4,20){\line(1,0){34}}
\multiput(0,47)(2,-2){15}{$\cdot$}
\multiput(0,37)(2,-2){10}{$\cdot$}
\multiput(0,27)(2,-2){5}{$\cdot$}
\put(-50,48){\scriptsize$m-i+1$}
\put(-26,30){\scriptsize$\vdots$}
\put(-50,16){\scriptsize$m-i+j$}
\put(0,50){\line(0,1){10}}
\put(0,60){\line(1,0){20}}
\put(20,60){\line(0,1){10}}
\put(20,70){\line(1,0){30}}
\put(50,70){\line(0,1){10}}
\put(50,80){\line(1,0){20}}
\put(10,40){\line(0,1){10}}
\put(10,50){\line(1,0){30}}
\put(40,50){\line(0,1){10}}
\put(40,60){\line(1,0){20}}
\put(60,60){\line(0,1){10}}
\put(60,70){\line(1,0){20}}
\put(20,30){\line(1,0){10}}
\put(30,30){\line(0,1){10}}
\put(30,40){\line(1,0){20}}
\put(50,40){\line(0,1){10}}
\put(50,50){\line(1,0){30}}
\put(80,50){\line(0,1){10}}
\put(80,60){\line(1,0){10}}
\put(30,20){\line(1,0){10}}
\put(40,20){\line(0,1){10}}
\put(40,30){\line(1,0){30}}
\put(70,30){\line(0,1){10}}
\put(70,40){\line(1,0){30}}
\put(100,40){\line(0,1){10}}
}
\end{picture}
\qquad\qquad
\begin{picture}(120,90)(-40,-15)
\unitlength=0.8pt
\put(30,-20){$(\,i\le j\,)$}
\put(80,20){$X$}
\put(0,0){\line(1,0){100}}
\put(0,80){\line(1,0){100}}
\put(0,0){\line(0,1){80}}
\put(100,0){\line(0,1){80}}
{\thicklines
\put(40,84){\line(0,-1){4}}
\put(50,84){\line(0,-1){14}}
\put(60,84){\line(0,-1){24}}
\put(70,84){\line(0,-1){34}}
\multiput(40,77)(2,-2){15}{$\cdot$}
\multiput(50,77)(2,-2){10}{$\cdot$}
\multiput(60,77)(2,-2){5}{$\cdot$}
\put(2,87){\scriptsize$n-j+1$}
\put(50,88){\scriptsize$\ldots$}
\put(68,87){\scriptsize$n-j+i$}
\put(-4,30){\line(1,0){4}}
\put(-4,20){\line(1,0){14}}
\put(-4,10){\line(1,0){24}}
\put(-4,0){\line(1,0){34}}
\multiput(0,27)(2,-2){15}{$\cdot$}
\multiput(0,17)(2,-2){10}{$\cdot$}
\multiput(0,07)(2,-2){5}{$\cdot$}
\put(-50,28){\scriptsize$m-i+1$}
\put(-12,10){\scriptsize$\vdots$}
\put(-15,-3){\scriptsize$m$}
\put(0,30){\line(0,1){10}}
\put(0,40){\line(1,0){20}}
\put(20,40){\line(0,1){10}}
\put(20,50){\line(1,0){10}}
\put(30,50){\line(0,1){30}}
\put(30,80){\line(1,0){10}}
\put(10,20){\line(0,1){10}}
\put(10,30){\line(1,0){20}}
\put(30,30){\line(0,1){10}}
\put(30,40){\line(1,0){10}}
\put(40,40){\line(0,1){20}}
\put(40,60){\line(1,0){10}}
\put(50,60){\line(0,1){10}}
\put(20,10){\line(1,0){10}}
\put(30,10){\line(0,1){10}}
\put(30,20){\line(1,0){20}}
\put(50,20){\line(0,1){20}}
\put(50,40){\line(1,0){10}}
\put(60,40){\line(0,1){20}}
\put(30,0){\line(1,0){10}}
\put(40,0){\line(0,1){10}}
\put(40,10){\line(1,0){20}}
\put(60,10){\line(0,1){10}}
\put(60,20){\line(1,0){10}}
\put(70,20){\line(0,1){30}}
}
\end{picture}
\end{equation}
{}From symmetry of the construction, $x^i_j$ are recovered from $y^i_j$ 
by the same procedure. 

\begin{theorem}\label{thm:C}
Let $X=\pmatrix{x^i_j}_{i,j}$, 
$Y=\pmatrix{y^i_j}_{i,j}$ be two $m\times n$ matrices
such that $x^i_j\ne0$, $y^i_j\ne0$ for all $i,j$. 
Setting $\Phi=\varphi(X)$, $\Psi=\varphi(Y)$, 
suppose that $\Phi$ and $\Psi$ are 
related as $\Psi=\Phi^\vee$ $:$
\begin{equation}
X \ \stackrel{\varphi}{\rightarrow}\  \Phi \ 
\stackrel{\vee}{\leftrightarrow} 
\ \Psi \ \stackrel{\varphi}{\leftarrow} Y.
\end{equation} 
Then, for each $(i,j)$, 
$y^i_j$ is expressed as follows in terms of $X$ $:$
\begin{equation}\label{eq:xtoy}
y^i_j=\frac{\sigma^i_j\ \sigma^{i-1}_{j-1}}{\sigma^{i-1}_j\sigma^i_{j-1}},
\qquad
\sigma^i_j=
\sum_{(\gamma_1,\ldots,\gamma_r)} x_{\gamma_1}\cdots x_{\gamma_r},
\end{equation}
where $r=\min\br{i,j}$, and the summation is taken over all $r$-tuples of 
nonintersecting paths 
$\gamma_k: (m-i+k,1)\to(1,n-j+k)$  $(k=1,\ldots,r)$. 
Conversely, each $x^i_j$ is expressed as follows in terms of $Y$ $:$ 
\begin{equation}\label{eq:ytox}
x^i_j=\frac{\tau^i_j\ \tau^{i-1}_{j-1}}{\tau^{i-1}_j\tau^i_{j-1}},
\qquad
\tau^i_j=
\sum_{(\gamma_1,\ldots,\gamma_r)} y_{\gamma_1}\cdots y_{\gamma_r},
\end{equation}
summed over the same set of $r$-tuples of nonintersecting paths as above. 
\end{theorem}

Note that 
the transformation from $X=\pmatrix{x^i_j}_{i,j}$ to 
$Y=\pmatrix{y^i_j}_{i,j}$ 
in Theorem \ref{thm:C} is realized as a subtraction-free birational mapping 
from $\Mat_{m,n}(\BK^\ast)$ to itself; this birational mapping 
is in fact an involution on $\Mat_{m,n}(\BK^\ast)$. 
Passing to the piecewise linear functions, we obtain
\begin{theorem}\label{thm:D} 
For each $m\times n$ matrix 
$X=\pmatrix{x^i_j}_{i,j}\in\Mat_{m,n}(\BR)$, 
define an $m\times n$ matrix  
$Y=\pmatrix{y^i_j}_{i,j}$ by
\begin{equation}
y^i_j=\sigma^i_j-\sigma^{i-1}_j-\sigma^i_{j-1}+
\sigma^{i-1}_{j-1},
\qquad
\sigma^i_j=
\max_{(\gamma_1,\ldots,\gamma_r)} 
(x_{\gamma_1}+\cdots +x_{\gamma_r}),
\end{equation}
where $r=\min\br{i,j}$, and the maximum is taken over all $r$-tuples of 
nonintersecting paths 
$\gamma_k: (m-i+k,1)\to(1,n-j+k)$  $(k=1,\ldots,r)$\,$;$ 
the weight of a path $\gamma$ is the sum of all $x^a_b$'s 
corresponding to the vertices of $\gamma$.
Then the piecewise linear mapping $X\mapsto Y$ is an involution 
on $\Mat_{m,n}(\BR)$. 
\end{theorem}

\subsection{Tropical \RSKs correspondence}

Theorem \ref{thm:C} is an essential ingredient of the  
tropical RSK correspondences.   
Regarding $x^i_j$ as indeterminates, 
we now work within the field of rational functions 
$\BK(x)$ in $mn$ variables $x^i_j$ ($1\le i\le m, 1\le j\le n$). 
In what follows, we assume that $m\le n$ to fix the idea. 
\par\medskip 
Consider the $m\times n$ matrix 
$X=\pmatrix{x^i_j}_{i,j}$ regarding $x^i_j$ as 
indeterminates. 
We denote 
the $i$-th row of $X$ by $\bx^i$, 
and the $j$-th column of $X$ by $\bx_j$ : 
\begin{equation}
X=\left[\matrix{
x^1_1 & x^1_2 & \ldots & x^1_n\cr
x^2_1 & x^2_2 & \ldots & x^2_n\cr
\vdots &\vdots &  &\vdots\cr
x^m_1 & x^m_2 & \ldots & x^m_n
}\right]
=
\left[\matrix{
\bx^1 \cr \bx^2 \cr \vdots \cr \bx^m
}\right]
=
\big[\,\bx_1\ \bx_2\ \ldots\  \bx_n\,\big]. 
\end{equation}
{}From the matrix $X=\pmatrix{x^i_j}_{i,j}$, we construct 
four tropical tableaux 
\begin{equation}
U=\pmatrix{u^i_j}_{i\le j},\quad
V=\pmatrix{v^i_j}_{i\le j},\quad
\SI{U}=\pmatrix{\widetilde{u}^i_j}_{i\le j},\quad
\SI{V}=\pmatrix{\widetilde{v}^i_j}_{i\le j} 
\end{equation}
as follows: 
\begin{eqnarray}\label{eq:HxHuv}
& 
H(\bx^m)\cdots H(\bx^2)H(\bx^1)=
H_m(\bu^m) \cdots H_2(\bu^2)H_1(\bu^1)=H_U,\nonumber\\
&
H(\bx_n) \cdots H(\bx_2) H(\bx_1)=
H_m(\bv^m) \cdots H_2(\bv^2)
H_1(\bv^1)=H_{V}\nonumber\\
&
H(\bx^1_\ast) H(\bx^2_\ast)\cdots H(\bx^m_\ast)=
H_m(\widetilde{\bu}^m) \cdots H_2(\widetilde{\bu}^2)
H_1(\widetilde{\bu}^1)=H_{\SI{U}},\nonumber\\
&
H(\bx_1^\ast) H(\bx_2^\ast)\cdots H(\bx_n^\ast)=
H_m(\widetilde{\bv}^m) \cdots 
H_2(\widetilde{\bv}^2)H_1(\widetilde{\bv}^1)=H_{\SI{V}},
\end{eqnarray}
where $\bx^i_\ast=(x^i_n,\ldots,x^i_2,x^i_1)$ and 
$\bx_j^\ast=(x^m_j,\ldots,x^2_j,x^1_j)$;
$H_U$, $H_{\SI{U}}$ are $n\times n$ matrices, 
and $H_V$, $H_{\SI{V}}$ are $m\times m$ matrices.
Note that $U$ and $\SI{U}$ 
(resp. $V$ and $\SI{V}$) 
are transformed into each other by the tropical 
Sch\"utzenberger involution. 
We also introduce the {\em tropical Gelfand-Tsetlin pattern} $\bmu$ 
associated with the tropical tableau $U$ as 
\begin{equation}
\bmu=\left[\matrix{
\mu^{(n)}_1\quad\mu^{(n)}_2\ \qquad\ \ldots \qquad\ \mu^{(n)}_n\cr
\mu^{(n-1)}_1\,\mu^{(n-1)}_2\ \ldots\,\mu^{(n-1)}_{n-1}\cr
\cdots\cr
\mu^{(1)}_1
}\right],
\end{equation}
where, for $i\le j$, we define 
$\mu^{(j)}_i=u^i_i \cdots u^i_j$ ($i\le m$) 
and $\mu^{(j)}_i=1$ ($i>m$).  
\par\medskip
Applying Theorem \ref{thm:HxHu} to $A=J_m X$, 
we already know that the variables $u^i_j$ ($i\le j$) are 
determined by
\begin{equation}
u^i_i=\frac{\tau^i_i(H_U)}{\tau^{i-1}_i(H_U)},
\quad
u^i_j=\frac{\tau^i_j(H_U)\ \tau^{i-1}_{j-1}(H_U)}
{\tau^{i-1}_j(H_U)\tau^i_{j-1}(H_U)}
\quad (i<j),
\end{equation}
with
\begin{equation}
\begin{picture}(175,60)(-85,-3)
\unitlength=0.8pt
\put(-80,30){$\tau^i_j(H_U)=\sum$}
\put(80,55){$A$}
\put(0,0){\line(1,0){100}}
\put(0,70){\line(1,0){100}}
\put(0,0){\line(0,1){70}}
\put(100,0){\line(0,1){70}}
\multiput(-1.4,39.5)(2,2){16}{$.$}
\multiput(48.6,-0.5)(2,2){16}{$.$}
\put(-2,73){\small$1$}
\put(8,73){\small$\ldots$}
\put(28,73){\small$i$}
\put(18,-10){\small$j-i+1$}
\put(63,-10){\small$\ldots$}
\put(78,-10){\small$j$}
{\thicklines
\put(0,70){\line(0,-1){30}}
\put(10,70){\line(0,-1){20}}
\put(20,70){\line(0,-1){10}}
\put(30,70){\line(0,-1){0}}
\put(60,0){\line(0,1){10}}
\put(70,0){\line(0,1){20}}
\put(80,0){\line(0,1){30}}
\put(0,40){\line(0,-1){10}}
\put(0,30){\line(1,0){10}}
\put(10,30){\line(0,-1){20}}
\put(10,10){\line(1,0){20}}
\put(30,10){\line(0,-1){10}}
\put(30,0){\line(1,0){20}}
\put(10,50){\line(0,-1){10}}
\put(10,40){\line(1,0){10}}
\put(20,40){\line(0,-1){10}}
\put(20,30){\line(1,0){10}}
\put(30,30){\line(0,-1){10}}
\put(30,20){\line(1,0){10}}
\put(40,20){\line(0,-1){10}}
\put(40,10){\line(1,0){20}}
\put(30,70){\line(1,0){10}}
\put(40,70){\line(0,-1){10}}
\put(40,60){\line(1,0){20}}
\put(60,60){\line(0,-1){20}}
\put(60,40){\line(1,0){20}}
\put(80,40){\line(0,-1){10}}
\put(20,60){\line(1,0){10}}
\put(30,60){\line(0,-1){20}}
\put(30,40){\line(1,0){10}}
\put(40,40){\line(0,-1){10}}
\put(40,30){\line(1,0){20}}
\put(60,30){\line(0,-1){10}}
\put(60,20){\line(1,0){10}}
}
\end{picture}
\begin{picture}(145,60)(-30,-3)
\unitlength=0.8pt
\put(-40,30){$=\sum$}
\put(105,30){$.$}
\put(80,15){$X$}
\put(0,0){\line(1,0){100}}
\put(0,70){\line(1,0){100}}
\put(0,0){\line(0,1){70}}
\put(100,0){\line(0,1){70}}
\multiput(-1.4,29.5)(2,-2){16}{$.$}
\multiput(48.6,69.5)(2,-2){16}{$.$}
\put(-2,-10){\small$1$}
\put(8,-10){\small$\ldots$}
\put(28,-10){\small$i$}
\put(18,74){\small$j-i+1$}
\put(63,74){\small$\ldots$}
\put(78,74){\small$j$}
{\thicklines
\put(0,0){\line(0,1){30}}
\put(10,0){\line(0,1){20}}
\put(20,0){\line(0,1){10}}
\put(30,0){\line(0,1){0}}
\put(60,70){\line(0,-1){10}}
\put(70,70){\line(0,-1){20}}
\put(80,70){\line(0,-1){30}}
\put(0,30){\line(0,1){10}}
\put(0,40){\line(1,0){10}}
\put(10,40){\line(0,1){20}}
\put(10,60){\line(1,0){20}}
\put(30,60){\line(0,1){10}}
\put(30,70){\line(1,0){20}}
\put(10,20){\line(0,1){10}}
\put(10,30){\line(1,0){10}}
\put(20,30){\line(0,1){10}}
\put(20,40){\line(1,0){10}}
\put(30,40){\line(0,1){10}}
\put(30,50){\line(1,0){10}}
\put(40,50){\line(0,1){10}}
\put(40,60){\line(1,0){20}}
\put(30,0){\line(1,0){10}}
\put(40,0){\line(0,1){10}}
\put(40,10){\line(1,0){20}}
\put(60,10){\line(0,1){20}}
\put(60,30){\line(1,0){20}}
\put(80,30){\line(0,1){10}}
\put(20,10){\line(1,0){10}}
\put(30,10){\line(0,1){20}}
\put(30,30){\line(1,0){10}}
\put(40,30){\line(0,1){10}}
\put(40,40){\line(1,0){20}}
\put(60,40){\line(0,1){10}}
\put(60,50){\line(1,0){10}}
}
\end{picture}
\end{equation}
Notice that $\tau^i_j(H_U)$ for $i\le j$ coincides with
\begin{equation}
\begin{picture}(220,60)(-135,-3)
\unitlength=0.8pt
\put(-175,30){
$\det\Phi^{m-i+1,\ldots,m}_{j-i+1,\ldots,j}=\sum$}
\put(105,30){$.$}
\put(80,15){$X$}
\put(0,0){\line(1,0){100}}
\put(0,70){\line(1,0){100}}
\put(0,0){\line(0,1){70}}
\put(100,0){\line(0,1){70}}
\multiput(-1.4,29.5)(2,-2){16}{$.$}
\multiput(48.6,69.5)(2,-2){16}{$.$}
\put(-10,-4){\small$m$}
\put(-6,10){\small$\vdots$}
\put(-50,25){\small$m-i+1$}
\put(18,74){\small$j-i+1$}
\put(63,74){\small$\ldots$}
\put(78,74){\small$j$}
{\thicklines
\put(10,20){\line(-1,0){10}}
\put(20,10){\line(-1,0){20}}
\put(30,0){\line(-1,0){30}}
\put(60,70){\line(0,-1){10}}
\put(70,70){\line(0,-1){20}}
\put(80,70){\line(0,-1){30}}
\put(0,30){\line(0,1){10}}
\put(0,40){\line(1,0){10}}
\put(10,40){\line(0,1){20}}
\put(10,60){\line(1,0){20}}
\put(30,60){\line(0,1){10}}
\put(30,70){\line(1,0){20}}
\put(10,20){\line(0,1){10}}
\put(10,30){\line(1,0){10}}
\put(20,30){\line(0,1){10}}
\put(20,40){\line(1,0){10}}
\put(30,40){\line(0,1){10}}
\put(30,50){\line(1,0){10}}
\put(40,50){\line(0,1){10}}
\put(40,60){\line(1,0){20}}
\put(30,0){\line(1,0){10}}
\put(40,0){\line(0,1){10}}
\put(40,10){\line(1,0){20}}
\put(60,10){\line(0,1){20}}
\put(60,30){\line(1,0){20}}
\put(80,30){\line(0,1){10}}
\put(20,10){\line(1,0){10}}
\put(30,10){\line(0,1){20}}
\put(30,30){\line(1,0){10}}
\put(40,30){\line(0,1){10}}
\put(40,40){\line(1,0){20}}
\put(60,40){\line(0,1){10}}
\put(60,50){\line(1,0){10}}
}
\end{picture}
\end{equation}
Hence we have
\begin{equation}
\tau^i_j(H_U)
=\det\Phi^{m-i+1,\ldots,m}_{j-i+1,\ldots,j}
=\det\Psi^{1,\ldots,i}_{n-j+1,\ldots,n-j+i}
=\tau^i_{n-j+i}(\Psi). 
\end{equation}
This implies 
\begin{equation}
\sigma^i_j=\tau^i_j(\Psi)
=\tau^i_{n-j+i}(H_U)=\prod_{(a,b);\,a\le i,\,b\le n-j+i}u^a_b
\qquad(i\le j).
\end{equation} 
Similarly, for $i\le j$, we have
\begin{equation}
\tau^i_j(H_{V})=\det\Phi^{j-i+1,\ldots,j}_{n-i+1,\ldots,i}
=\det\Psi^{m-j+1,\ldots,m-j+i}_{1,\ldots,i}
=\tau^{m-j+i}_i(\Psi),
\end{equation}
hence, for $i\ge j$, 
\begin{equation}
\sigma^i_j=\tau^i_j(\Psi)
=\tau^{j}_{m-i+j}(H_{V})
=\prod_{(a,b);\,a\le j,\,b\le m-i+j}v^a_b
\qquad(i\ge j).
\end{equation}
Summarizing the argument above, we have
\begin{equation}\label{eq:sigmauv}
\sigma^i_j=\tau^i_j(\Psi)=\left\{
\begin{array}{ll}\smallskip
{\displaystyle\prod_{(a,b);\,a\le i,\,b\le n-j+i}u^a_b}
\quad& (i\le j),\cr
{\displaystyle\prod_{(a,b);\,a\le j,\,b\le m-i+j}v^a_b}
& (i\ge j),
\end{array}
\right.
\end{equation}
for all $(i,j)$ with $1\le i\le m$ and $1\le j\le n$.  
Conversely, $u^i_j$ and $v^i_j$ are determined as
\begin{equation}\label{eq:ubysigma}
u^i_i=\frac{\sigma^i_n}{\sigma^{i-1}_{n-1}},
\quad
u^i_j=\frac{\sigma^i_{n-j+i}\ \sigma^{i-1}_{n-j+i}}
{\sigma^{i-1}_{n-j+i-1}\sigma^{i}_{n-j+i+1}}\quad(i<j)
\end{equation}
and
\begin{equation}\label{eq:vbysigma}
v^i_i=\frac{\sigma^m_i}{\sigma^{m-1}_{i-1}},
\quad
v^i_j=\frac{\sigma^{m-j+i}_{i}\ \sigma^{m-j+i}_{i-1}}
{\sigma^{m-j+i-1}_{i-1}\sigma^{m-j+i-1}_{i}}\quad(i<j).
\end{equation}

\begin{remark}\rm 
It should be noted 
that the upper (resp. lower)
triangular components of 
the $m\times n$ matrix $S=\pmatrix{\sigma^i_j}_{i,j}$ are 
determined from $U=\pmatrix{u^i_j}_{i\le j}$
(resp. $V=\pmatrix{v^i_j}_{i\le j}$), 
and {\em vice versa}. 
Formula \eqref{eq:sigmauv} is also equivalent to
\begin{equation}
\frac{\sigma^i_j}{\sigma^{i-1}_{j-1}}
=\left\{
\begin{array}{ll}\smallskip
u^i_i\cdots u^i_{n-j+i}=\mu^{(n-j+i)}_i
\quad& (i\le j),\cr
v^j_j\cdots v^j_{m-i+j}=\nu^{(m-i+j)}_j
& (i\ge j),
\end{array}
\right.
\end{equation}
where $\mu^{(j)}_i$ and $\nu^{(j)}_i$ 
are the tropical variables representing the 
Gelfand-Tsetlin pattern of $U$ and $V$, respectively. 
Namely, the $m\times n$ matrix 
\begin{equation}
\begin{picture}(260,65)
\put(0,30){
$\pmatrix{\dfrac{\sigma^i_j}{\sigma^{i-1}_{j-1}}}_{i,j}
=\left[\tbl{\smallskip
\,&\lambda_1 & \mu^{(n-1)}_1 &  &
&\mu^{(n-m)}_1 &\ldots &\mu^{(1)}_1 \cr
\medskip
&\nu^{(m-1)}_1\,&\lambda_2 & & &\cr
\smallskip
& & & &&&&\cr
&\nu^{(1)}_1 & \ldots &\nu^{(m-1)}_{m-1}\,&
\lambda_m \ & \mu^{(n-1)}_{m} & \ldots \ & \mu^{(m)}_m 
}
\right]$,}
\multiput(130,45)(5,-3){11}{$.$}
\multiput(125,32)(5,-3){6}{$.$}
\multiput(100,32)(5,-3){6}{$.$}
\multiput(80,30)(0,-4){4}{$.$}
\multiput(190,45)(0,-4){8}{$.$}
\multiput(240,45)(0,-4){8}{$.$}
\multiput(140,53)(4,0){8}{$.$}
\end{picture}
\end{equation}
defined by the ratios of $\sigma^i_j$, 
is obtained by glueing the two Gelfand-Tsetlin patterns $\bmu$ 
and $\bnu$ at the main diagonal, where 
the diagonal entries
\begin{equation}
\lambda_i = \mu^{(n)}_i =\nu^{(m)}_i
\quad(i=1,\ldots,m)
\end{equation}
are the tropical variables representing the common shape 
of $U$ and $V$. 
\end{remark}

{}From \eqref{eq:sigmauv}, we obtain the following 
expression for $y^i_j$:
\begin{equation}\label{eq:uvtoy}
y^i_j=\dfrac{\sigma^i_j\ \sigma^{i-1}_{j-1}}
{\sigma^{i-1}_{j}\sigma^{i}_{j-1}}
=\left\{\begin{array}{ll}
\dfrac{u^1_{n-j+i}\cdots u^{i-1}_{n-j+i}}
{u^1_{n-j+i+1}\cdots u^{i}_{n-j+i+1}} & (i<j),\cr
\dfrac{\lambda_i}{u^1_n\cdots u^{i-1}_n v^1\cdots v^{i-1}_m}&(i=j),\cr
\dfrac{v^1_{m-i+j}\cdots v^{j-1}_{m-i+j}}
{v^1_{m-i+j+1}\cdots u^{j}_{m-i+j+1}} &(i>j).
\end{array}
\right.
\end{equation}
Hence we have
\begin{theorem} 
Under the assumption of Theorem \ref{thm:C},
let $U=\pmatrix{u^i_j}_{i\le j}$, $V=\pmatrix{v^i_j}_{i\le j}$ be 
the tropical tableaux defined by the tropical row insertions 
\begin{equation}
H_U=H(\bx^m)\cdots H(\bx^2)H(\bx^1),\quad
H_{V}=
H(\bx_n) \cdots H(\bx_2) H(\bx_1),
\end{equation}
respectively.  
Then $u^i_j$ and $v^i_j$ are expressed as 
\eqref{eq:ubysigma} and \eqref{eq:vbysigma},
respectively, in terms 
of $\sigma^i_j$ defined in Theorem \ref{thm:C}.  
Conversely, 
the matrix $X=\pmatrix{x^i_j}_{i,j}$ is recovered from 
the tropical tableaux $U$ and $V$ by the formula
\eqref{eq:ytox} with $y^i_j$ defined by 
\eqref{eq:uvtoy}. 
\end{theorem}

Passing to the combinatorial variables, we obtain 
the explicit inversion formula for the \RSKs correspondence. 

\begin{theorem}\label{thm:CIRSKs}
Let $X=\pmatrix{x^i_j}_{i,j}$ be an $m\times n$ matrix 
of nonnegative integers.  Consider the two column strict tableaux 
$U$ and $V$ obtained by the row insertion
\begin{eqnarray}
&U=(\cdots(w_m\leftarrow w_{m-1})\leftarrow\cdots\leftarrow w_1),
&w_i=1^{x^i_1}2^{x^i_2}\cdots n^{x^i_n}
\nonumber\\
&V=(\cdots(w'_n\leftarrow w'_{n-1})\leftarrow\cdots\leftarrow w'_1),
&w'_j=1^{x^1_j}2^{x^2_j}\cdots m^{x^m_j}. 
\end{eqnarray} 
Denote by $u^i_j$ $(\mbox{resp.}\  v^i_j)$ be the number of $j'$s in the 
$i$-th row of $U$ $(\mbox{resp.}\ V )$.  
Then $u^i_j$ and $v^i_j$ are expressed as 
\begin{eqnarray}
&&u^i_i={\sigma^i_n}-{\sigma^{i-1}_{n-1}},
\nonumber\\
&&
u^i_j=\sigma^i_{n-j+i}-
\sigma^{i-1}_{n-j+i-1}-
\sigma^{i}_{n-j+i+1}+
\sigma^{i-1}_{n-j+i}\ \ (i<j),
\\
&&v^i_i={\sigma^m_i}-{\sigma^{m-1}_{i-1}},
\nonumber\\
&&v^i_j=
\sigma^{m-j+i}_{i}-
\sigma^{m-j+i-1}_{i-1}-
\sigma^{m-j+i-1}_{i}+
\sigma^{m-j+i}_{i-1}
\ \ (i<j)
\end{eqnarray}
in terms of $\sigma^i_j$ defined in Theorem \ref{thm:D}. 
Let $\lambda=(\lambda_1,\ldots,\lambda_l)$,
$l=\min\br{m,n}$, be the common shape of $U$ and $V$,
and set
\begin{equation}
y^i_j=\left\{
\begin{array}{llll}
\medskip
\sum_{k=1}^{i-1} u^k_{n-j+i}-
\sum_{k=1}^{i} u^k_{n-j+i+1}&(i<j),\cr
\medskip
\lambda_i-\sum_{k=1}^{i-1}u^k_n- \sum_{k=1}^{i-1}v^k_m &(i=j),\cr
\sum_{k=1}^{j-1} v^k_{m-i+j}-
\sum_{k=1}^{j} v^k_{m-i+j+1}&(i>j),
\end{array}
\right.
\end{equation}
for each $i=1,\ldots,m$ and $j=1,\ldots,n$. 
Then the matrix $X$ is recovered from $U$ and $V$ by the formulas
\begin{equation}
x^i_j=\tau^i_j-\tau^{i-1}_j-\tau^i_{j-1}+
\tau^{i-1}_{j-1},
\qquad
\tau^i_j=
\max_{(\gamma_1,\ldots,\gamma_r)} 
(y_{\gamma_1}+\cdots +y_{\gamma_r}),
\end{equation}
where $r=\min\br{i,j}$, and the maximum is taken over all $r$-tuples of 
nonintersecting paths 
$\gamma_k: (m-i+k,1)\to(1,n-j+k)$  $(k=1,\ldots,r)$
in the $m\times n$ rectangle\,$;$ 
the weight of a path $\gamma$ is the sum of all $y^a_b$'s 
corresponding to the vertices of $\gamma$.
\end{theorem} 

An explicit inversion formula for the usual RSK correspondence 
is obtained by combining Theorem \ref{thm:CIRSKs} and 
the Sch\"utzenberger involution. 
The inversion formula discussed above has an obvious  
theoretical meaning, but is somewhat indirect. 
We will discuss in the following subsection a different type of 
inversion formulas for the four variations of 
RSK correspondences which recovers the transportation 
matrix directly from the corresponding tableaux. 

\subsection{Inverse tropical RSK via the Gauss decomposition} 
Keeping the notations $X$, $\Phi$, $\Psi$ as before, 
we now consider the ``Gauss decomposition'' of the 
$m\times n$ matrix 
$\Psi$ ($m\le n$):
\begin{equation}
\Psi=\Psi_- \, \Psi_0\, \Psi_+, 
\end{equation}
where $\Psi_+$, $\Psi_0$ and $\Psi_-$ are 
a $m\times n$ upper unitriangular matrix, 
a $m\times m$ diagonal matrix, 
and a $m\times m$ 
lower unitriangular matrix, respectively: 
\begin{equation}
(\Psi_+)^i_j=\delta_{i,j}\quad(i\ge j), 
\quad
(\Psi_0)^i_j=0\quad(i\ne j),
\quad
(\Psi_-)^i_j=\delta_{i,j}\quad(i\le j).
\end{equation}
Nontrivial entries of these matrices are 
given explicitly as follows:
\begin{equation}
\tbl{
\medskip
(\Psi_+)^i_j
&=\dfrac{\det\Psi^{1,\ldots,i}_{1,\ldots,i-1,j}}
{\det\Psi^{1,\ldots,i}_{1,\ldots,i}}
&=\dfrac{\det\Phi^{m-i+1,\ldots,m}_{n-j+1,n-i+2,\ldots,n}}
{\det\Phi^{m-i+1,\ldots,m}_{n-i+1,\ldots,n}}
\quad&(i\le j),\\
\medskip
(\Psi_0)^i_i
&=\dfrac{\det\Psi^{1,\ldots,i}_{1,\ldots,i}}
{\det\Psi^{1,\ldots,i-1}_{1,\ldots,i-1}}
&=\dfrac{\det\Phi^{m-i+1,\ldots,m}_{n-i+1,\ldots,n}}
{\det\Phi^{m-i+2,\ldots,m}_{n-i+2,\ldots,n}}
\quad&(i=j),\\
(\Psi_-)^i_j
&=\dfrac{\det\Psi^{1,\ldots,j-1,i}_{1,\ldots,j}}
{\det\Psi^{1,\ldots,j}_{1,\ldots,j}}
&=\dfrac{\det\Phi^{m-i+1,m-j+2,\ldots,m}_{n-j+1,\ldots,n}}
{\det\Phi^{m-j+1,\ldots,m}_{n-j+1,\ldots,n}}
\quad&(i\ge j). 
}
\end{equation}
Comparing the path representations of $\Phi$ and 
$H_U$, $H_{\SI{U}}$, 
$H_V$, $H_{\SI{V}}$,  we have 
\begin{equation}\label{eq:PhiU}
\det\Phi^{m-r+1,\ldots,m}_{l_1,\ldots,l_r}
=\det(H_U)^{1,\ldots,r}_{l_1,\ldots,l_r}
=\det(H_{\SI{U}})^{l_1^\ast,\ldots,l_r^\ast}_{n-r+1,\ldots,n},
\end{equation}
where $1\le l_1<\ldots<l_r\le n$ and $l_i^\ast=n-l_i+1$, and 
\begin{equation}\label{eq:PhiV}
\det\Phi^{k_1,\ldots,k_r}_{n-r+1,\ldots,n}
=\det(H_V)^{1,\ldots,r}_{k_1,\ldots,k_r}
=\det(H_{\SI{V}})^{k_1^\ast,\ldots,k_r^\ast}_{m-r+1,\ldots,m},
\end{equation}
for $1\le k_1<\ldots<k_r\le m$ with $k_i^\ast=m-k_i+1$.  
{}From these formulas, we see for instance,
\begin{equation}
(\Psi_0)^i_i
=u^i_i\cdots u^i_n
=\widetilde{u}^i_i\cdots \widetilde{u}^i_n
=v^i_i\cdots v^i_m
=\widetilde{v}^i_i\cdots \widetilde{v}^i_m
\end{equation}
for $i=1,\ldots,m$.  
We set $\lambda_i=(\Psi_0)^i_i$, so that the vector 
$\lambda=(\lambda_1,\ldots,\lambda_m)$ 
represents the common shape of $U$, $\SI{U}$, $V$ and $\SI{V}$. 
By using formulas \eqref{eq:PhiU}, \eqref{eq:PhiV}, we can represent 
$\Psi_+$, $\Psi_0$, $\Psi_-$ in terms of the tropical 
tableaux. 
\begin{proposition}\label{prop:PsiUV}
$(0)$ For $i=1,\ldots,m$,
\begin{equation}
(\Psi_0)^i_i=\lambda_i
=u^i_i\cdots u^i_n
=\widetilde{u}^i_i\cdots \widetilde{u}^i_n
=v^i_i\cdots v^i_m
=\widetilde{v}^i_i\cdots \widetilde{v}^i_m. 
\end{equation}
$(1)$ For $i\le j$, $(\Psi_+)^i_j$ is expressed in terms of $U$ 
as the sum 
\begin{equation}\label{eq:PsiU}
\begin{picture}(210,70)(-115,0)
\put(-115,25){$
(\Psi_+)^i_j=\dsum{\gamma:(i,n)\to(1,n-j+1)}{}
$}
\put(0,60){\line(1,0){70}}
\put(10,50){\line(1,0){60}}
\put(20,40){\line(1,0){50}}
\put(30,30){\line(1,0){40}}
\put(40,20){\line(1,0){30}}
\multiput(8.5,59.5)(0,-2){6}{$.$}
\multiput(18.5,59.5)(0,-2){11}{$.$}
\multiput(28.5,59.5)(0,-2){16}{$.$}
\multiput(38.5,59.5)(0,-2){21}{$.$}
\multiput(48.5,59.5)(0,-2){21}{$.$}%
\multiput(58.5,59.5)(0,-2){21}{$.$}%
\multiput(68.5,59.5)(0,-2){21}{$.$}%
\put(10,60){\line(1,-1){40}}%
\put(20,60){\line(1,-1){40}}%
\put(30,60){\line(1,-1){40}}
\put(40,60){\line(1,-1){30}}
\put(50,60){\line(1,-1){20}}
\put(60,60){\line(1,-1){10}}
\put(-8,58){\small$1$}
\put(2,48){\small$2$}
\put(72,56){\small$1$}
\put(75,37){\small$i$}
\put(73,17){\small$m$}
\put(30,17){\small$m$}
\put(53,5){$\overline{U}$}
\put(-3,65){\small$0$}
\put(7,65){\small$1$}
\put(27,66){\small$j^\ast$}
\put(68,65){\small$n$}
{\thicklines
\put(100,40){\vector(-1,0){15}}
\put(100,40){\vector(-1,1){12}}
\put(30,60){\line(0,1){3}}
\put(30,60){\line(1,-1){10}}
\put(40,50){\line(1,0){20}}
\put(60,50){\line(1,-1){10}}
\put(70,40){\line(1,0){3}}
}
\end{picture}
\end{equation}
of weights over all path $\gamma: (i,n)\to(1,n-j+1)$,   
with weight 
$\overline{u}^a_b=\frac{1}{u^a_b}$ assigned to the horizontal edge 
connecting $(a,b-1)$ and $(a,b)$, for each $a\le b$. 
In terms  of $\SI{U}$, $(\Psi_+)^i_j$ is expressed as 
\begin{equation}\label{eq:PsiSU}
\begin{picture}(210,70)(-135,-5)
\put(-140,25){$
(\Psi_+)^i_j=\lambda_i^{-1}\dsum{\gamma:(i,n)\to(\min\br{j,m},j)}{}
$}
\put(0,60){\line(1,0){60}}
\put(10,50){\line(1,0){50}}
\put(20,40){\line(1,0){40}}
\put(30,30){\line(1,0){30}}
\put(40,20){\line(1,0){20}}
\put(10,60){\line(0,-1){10}}
\put(20,60){\line(0,-1){20}}
\put(30,60){\line(0,-1){30}}
\put(40,60){\line(0,-1){40}}
\put(50,60){\line(0,-1){40}}%
\put(60,60){\line(0,-1){40}}%
\multiput(-1.5,59.5)(2,-2){21}{$.$}
\put(-5,52){\small$1$}
\put(64,56){\small$1$}
\put(65,47){\small$i$}
\put(63,17){\small$m$}%
\put(30,17){\small$m$}%
\put(47,2){$\SI{U}$}%
\put(-5,63){\small$1$}
\put(20,25){\small$j$}
\put(58,63){\small$n$}
{\thicklines
\put(95,50){\vector(-1,0){15}}
\put(95,50){\vector(0,-1){15}}
\put(60,50){\line(1,0){3}}
\put(60,50){\line(-1,0){10}}
\put(50,50){\line(0,-1){10}}
\put(50,40){\line(-1,0){20}}
\put(30,40){\line(0,-1){10}}
\put(30,30){\line(-1,0){3}}
}
\end{picture}
\end{equation}
$(2)$ For $i\ge j$, $(\Psi_-)^i_j$ is expressed in terms of $V$ 
as the sum 
\begin{equation}\label{eq:PsiV}
\begin{picture}(210,70)(-115,-5)
\put(-115,25){$
(\Psi_-)^i_j=\dsum{\gamma:(1,m-i+1)\to(j,m)}{}
$}
\put(0,60){\line(1,0){70}}
\put(10,50){\line(1,0){60}}
\put(20,40){\line(1,0){50}}
\put(30,30){\line(1,0){40}}
\put(40,20){\line(1,0){30}}
\put(50,10){\line(1,0){20}}
\put(60,0){\line(1,0){10}}
\multiput(8.5,59.5)(0,-2){6}{$.$}
\multiput(18.5,59.5)(0,-2){11}{$.$}
\multiput(28.5,59.5)(0,-2){16}{$.$}
\multiput(38.5,59.5)(0,-2){21}{$.$}
\multiput(48.5,59.5)(0,-2){26}{$.$}
\multiput(58.5,59.5)(0,-2){31}{$.$}
\multiput(68.5,59.5)(0,-2){31}{$.$}
\put(10,60){\line(1,-1){60}}
\put(20,60){\line(1,-1){50}}
\put(30,60){\line(1,-1){40}}
\put(40,60){\line(1,-1){30}}
\put(50,60){\line(1,-1){20}}
\put(60,60){\line(1,-1){10}}
\put(-8,58){\small$1$}
\put(2,48){\small$2$}
\put(72,56){\small$1$}
\put(75,37){\small$j$}
\put(73,-3){\small$m$}
\put(50,-3){\small$m$}
\put(80,12){$\overline{V}$}
\put(-3,65){\small$0$}
\put(7,65){\small$1$}
\put(27,66){\small$i^\ast$}
\put(68,65){\small$m$}
{\thicklines
\put(90,50){\vector(1,0){15}}
\put(90,50){\vector(1,-1){12}}
\put(30,60){\line(0,1){3}}
\put(30,60){\line(1,-1){10}}
\put(40,50){\line(1,0){20}}
\put(60,50){\line(1,-1){10}}
\put(70,40){\line(1,0){3}}
}
\end{picture}
\end{equation}
of weights over all path $\gamma: (m-i+1,1)\to(j,m)$,   
with weight 
$\overline{v}^a_b=\frac{1}{v^a_b}$ assigned to the horizontal edge 
connecting $(a,b-1)$ and $(a,b)$, for each $a\le b$. 
In terms  of $\SI{V}$, $(\Psi_-)^i_j$ is expressed as 
\begin{equation}\label{eq:PsiSV}
\begin{picture}(210,70)(-115,-5)
\put(-110,25){$
(\Psi_-)^i_j=\lambda_j^{-1}\dsum{\gamma:(i,i)\to(j,m)}{}
$}
\put(0,60){\line(1,0){60}}
\put(10,50){\line(1,0){50}}
\put(20,40){\line(1,0){40}}
\put(30,30){\line(1,0){30}}
\put(40,20){\line(1,0){20}}
\put(50,10){\line(1,0){10}}
\put(10,60){\line(0,-1){10}}
\put(20,60){\line(0,-1){20}}
\put(30,60){\line(0,-1){30}}
\put(40,60){\line(0,-1){40}}
\put(50,60){\line(0,-1){50}}
\put(60,60){\line(0,-1){60}}
\multiput(-1.5,59.5)(2,-2){31}{$.$}
\put(-5,52){\small$1$}
\put(64,56){\small$1$}
\put(65,47){\small$j$}
\put(63,-3){\small$m$}
\put(50,-3){\small$m$}
\put(67,12){$\SI{V}$}
\put(-5,63){\small$1$}
\put(20,25){\small$i$}
\put(58,63){\small$m$}
{\thicklines
\put(80,40){\vector(1,0){15}}
\put(80,40){\vector(0,1){15}}
\put(60,50){\line(1,0){3}}
\put(60,50){\line(-1,0){10}}
\put(50,50){\line(0,-1){10}}
\put(50,40){\line(-1,0){20}}
\put(30,40){\line(0,-1){10}}
\put(30,30){\line(-1,0){3}}
}
\end{picture}
\end{equation}
\end{proposition}

\begin{figure}
%
%
$$
\begin{picture}(270,120)(-48,-15)
\unitlength=1.6pt
\put(-25,20){$\Gamma_{U,V} :$}
{\thicklines
\put(50,0){\line(1,0){30}}
\put(50,10){\line(1,0){40}}
\put(50,20){\line(1,0){50}}
\put(50,30){\line(1,0){60}}
\put(50,40){\line(1,0){70}}
\put(50,50){\line(1,0){80}}
\put(50,0){\line(-1,0){10}}
\put(50,10){\line(-1,0){20}}
\put(50,20){\line(-1,0){30}}
\put(50,30){\line(-1,0){40}}
\put(50,40){\line(-1,0){50}}
\put(50,50){\line(-1,0){60}}
}
\put(-10,50){\line(1,-1){50}}
\put(0,50){\line(1,-1){40}}
\put(10,50){\line(1,-1){30}}
\put(20,50){\line(1,-1){20}}
\put(30,50){\line(1,-1){10}}
\put(60,50){\line(-1,-1){10}}
\put(70,50){\line(-1,-1){20}}
\put(80,50){\line(-1,-1){30}}
\put(90,50){\line(-1,-1){40}}
\put(100,50){\line(-1,-1){50}}
\put(110,50){\line(-1,-1){50}}
\put(120,50){\line(-1,-1){50}}
\put(130,50){\line(-1,-1){50}}
\put(122,52){\small$\overline{u}^1_2$}
\put(112,52){\small$\overline{u}^1_3$}
\put(52,52){\small$\overline{u}^1_n$}
\put(123,34){\small$\overline{u}^{2}_3$}
\put(122,36){\line(-2,1){8}}
\put(93,0){\small$\overline{u}^{m-1}_m$}
\put(92,6){\line(-2,1){8}}
\put(53,-6){\small$\overline{u}^m_n$}
\put(73,-6){\small$\overline{u}^m_{m+1}$}
\put(-8,52){\small$\overline{v}^1_2$}
\put(2,52){\small$\overline{v}^1_3$}
\put(32,52){\small$\overline{v}^1_m$}
\put(-10,30){\small$\overline{v}^{2}_3$}
\put(-3,36){\line(2,1){8}}
\put(16,0){\small$\overline{v}^{m-1}_m$}
\put(27,6){\line(2,1){8}}
%
\put(42,52){\small$\lambda_1$}
\put(42,42){\small$\lambda_2$}
\put(42,2){\small$\lambda_m$}
\multiput(-10,50)(10,0){6}{\line(0,1){9}}
\multiput(50,50)(10,0){9}{\line(0,1){9}}
\put(-12,60){\small$a_m$}
\put(18,60){\small$\cdots$}
\put(28,60){\small$a_2$}
\put(38,60){\small$a_1$}
\put(48,60){\small$b_1$}
\put(58,60){\small$b_2$}
\put(68,60){\small$\cdots$}
\put(128,60){\small$b_n$}
\multiput(-10,53)(10,0){6}{\vector(0,-1){0}}
\multiput(50,56)(10,0){9}{\vector(0,1){0}}
\multiput(-3,50)(10,0){14}{\vector(1,0){0}}
\multiput(7,40)(10,0){12}{\vector(1,0){0}}
\multiput(17,30)(10,0){10}{\vector(1,0){0}}
\multiput(27,20)(10,0){8}{\vector(1,0){0}}
\multiput(37,10)(10,0){6}{\vector(1,0){0}}
\multiput(47,0)(10,0){4}{\vector(1,0){0}}
\multiput(36,44)(-10,0){5}{\vector(1,-1){0}}
\multiput(36,34)(-10,0){4}{\vector(1,-1){0}}
\multiput(36,24)(-10,0){3}{\vector(1,-1){0}}
\multiput(36,14)(-10,0){2}{\vector(1,-1){0}}
\multiput(36,4)(-10,0){1}{\vector(1,-1){0}}
\multiput(56,46)(10,0){8}{\vector(1,1){0}}
\multiput(56,36)(10,0){7}{\vector(1,1){0}}
\multiput(56,26)(10,0){6}{\vector(1,1){0}}
\multiput(56,16)(10,0){5}{\vector(1,1){0}}
\multiput(56,6)(10,0){4}{\vector(1,1){0}}
\end{picture}
$$
%
%
$$
\begin{picture}(270,120)(-38,-15)
\unitlength=1.6pt
\put(-20,25){$\Gamma_{U,\sSI{V}}: $}
{\thicklines
\put(50,0){\line(1,0){30}}
\put(50,10){\line(1,0){40}}
\put(50,20){\line(1,0){50}}
\put(50,30){\line(1,0){60}}
\put(50,40){\line(1,0){70}}
\put(50,50){\line(1,0){80}}
}
\put(50,10){\line(-1,0){10}}
\put(50,20){\line(-1,0){20}}
\put(50,30){\line(-1,0){30}}
\put(50,40){\line(-1,0){40}}
\put(50,50){\line(-1,0){50}}
\put(50,50){\line(0,-1){50}}
\put(40,50){\line(0,-1){40}}
\put(30,50){\line(0,-1){30}}
\put(20,50){\line(0,-1){20}}
\put(10,50){\line(0,-1){10}}
\put(60,50){\line(-1,-1){10}}
\put(70,50){\line(-1,-1){20}}
\put(80,50){\line(-1,-1){30}}
\put(90,50){\line(-1,-1){40}}
\put(100,50){\line(-1,-1){50}}
\put(110,50){\line(-1,-1){50}}
\put(120,50){\line(-1,-1){50}}
\put(130,50){\line(-1,-1){50}}
\put(122.5,52){\small$\overline{u}^1_2$}
\put(112.5,52){\small$\overline{u}^1_3$}
\put(52.5,52){\small$\overline{u}^1_n$}
\put(122,30){\small$\overline{u}^{2}_3$}
\put(124,35){\line(-2,1){8}}
\put(93,0){\small$\overline{u}^{m-1}_m$}
\put(94,5){\line(-2,1){8}}
\put(53,-6){\small$\overline{u}^m_n$}
\put(73,-6){\small$\overline{u}^m_{m+1}$}
\put(-3,53){\small$\tilde{v}^1_1$}
\put(7,53){\small$\tilde{v}^1_2$}
\put(42,52){\small$\tilde{v}^1_m$}
\put(3,41){\small$\tilde{v}^{2}_2$}
\put(41,1){\small$\tilde{v}^{m}_m$}
\multiput(48.5,48.5)(0,-10){6}{\small$\bullet$}
\multiput(38.5,48.5)(0,-10){5}{\small$\bullet$}
\multiput(28.5,48.5)(0,-10){4}{\small$\bullet$}
\multiput(18.5,48.5)(0,-10){3}{\small$\bullet$}
\multiput(8.5,48.5)(0,-10){2}{\small$\bullet$}
\multiput(-1.5,48.5)(0,-10){1}{\small$\bullet$}
%
\multiput(0,50)(10,-10){6}{\line(-1,-1){7}}
\multiput(-2,48)(10,-10){6}{\vector(1,1){0}}
\multiput(50,50)(10,0){9}{\line(0,1){10}}
\multiput(50,57)(10,0){9}{\vector(0,1){0}}
\put(-11,40){\small$a_1$}
\put(-1,30){\small$a_2$}
\multiput(9,20)(2,-2){3}{\small$.$}
\put(39,-10){\small$a_m$}
\put(48,61){\small$b_1$}
\put(58,61){\small$b_2$}
\put(68,61){\small$\cdots$}
\put(128,61){\small$b_n$}
\multiput(7,50)(10,0){13}{\vector(1,0){0}}
\multiput(17,40)(10,0){11}{\vector(1,0){0}}
\multiput(27,30)(10,0){9}{\vector(1,0){0}}
\multiput(37,20)(10,0){7}{\vector(1,0){0}}
\multiput(47,10)(10,0){5}{\vector(1,0){0}}
\multiput(57,0)(10,0){3}{\vector(1,0){0}}
\multiput(10,47)(10,0){5}{\vector(0,1){0}}
\multiput(20,37)(10,0){4}{\vector(0,1){0}}
\multiput(30,27)(10,0){3}{\vector(0,1){0}}
\multiput(40,17)(10,0){2}{\vector(0,1){0}}
\multiput(50,7)(10,0){1}{\vector(0,1){0}}
\multiput(56.5,46.5)(10,0){8}{\vector(1,1){0}}
\multiput(56.5,36.5)(10,0){7}{\vector(1,1){0}}
\multiput(56.5,26.5)(10,0){6}{\vector(1,1){0}}
\multiput(56.5,16.5)(10,0){5}{\vector(1,1){0}}
\multiput(56.5,6.5)(10,0){4}{\vector(1,1){0}}
\end{picture}
$$
%
%
$$
\begin{picture}(270,120)(-38,-15)
\unitlength=1.6pt
\put(-20,25){$\Gamma_{\sSI{U},V}: $}
\put(50,0){\line(1,0){30}}
\put(50,10){\line(1,0){40}}
\put(50,20){\line(1,0){50}}
\put(50,30){\line(1,0){60}}
\put(50,40){\line(1,0){70}}
\put(50,50){\line(1,0){80}}
{\thicklines
\put(50,10){\line(-1,0){10}}
\put(50,20){\line(-1,0){20}}
\put(50,30){\line(-1,0){30}}
\put(50,40){\line(-1,0){40}}
\put(50,50){\line(-1,0){50}}
}
\put(40,50){\line(1,-1){10}}
\put(30,50){\line(1,-1){20}}
\put(20,50){\line(1,-1){30}}
\put(10,50){\line(1,-1){40}}
\put(0,50){\line(1,-1){50}}
\put(50,50){\line(0,-1){50}}
\put(60,50){\line(0,-1){50}}
\put(70,50){\line(0,-1){50}}
\put(80,50){\line(0,-1){50}}
\put(90,50){\line(0,-1){40}}
\put(100,50){\line(0,-1){30}}
\put(110,50){\line(0,-1){20}}
\put(120,50){\line(0,-1){10}}
\put(128,52){\small$\tilde{u}^1_1$}
\put(118.5,52){\small$\tilde{u}^1_2$}
\put(51,52){\small$\tilde{u}^1_n$}
\put(121,40){\small$\tilde{u}^{2}_2$}
\put(81,0){\small$\tilde{u}^{m}_m$}
\put(51,2){\small$\tilde{u}^m_n$}
\put(3,53){\small$\overline{v}^1_2$}
\put(13,53){\small$\overline{v}^1_3$}
\put(42,52){\small$\overline{v}^1_m$}
\put(2,32){\small$\overline{v}^{2}_3$}
\put(8,35.5){\line(2,1){7}}
\put(30,1){\small$\overline{v}^{m-1}_m$}
\put(38,5.5){\line(2,1){7}}
\multiput(48.5,48.5)(0,-10){6}{\small$\bullet$}
\multiput(58.5,48.5)(0,-10){6}{\small$\bullet$}
\multiput(68.5,48.5)(0,-10){6}{\small$\bullet$}
\multiput(78.5,48.5)(0,-10){6}{\small$\bullet$}
\multiput(88.5,48.5)(0,-10){5}{\small$\bullet$}
\multiput(98.5,48.5)(0,-10){4}{\small$\bullet$}
\multiput(108.5,48.5)(0,-10){3}{\small$\bullet$}
\multiput(118.5,48.5)(0,-10){2}{\small$\bullet$}
\multiput(128.5,48.5)(0,-10){1}{\small$\bullet$}
\multiput(0,50)(10,0){6}{\line(0,1){8}}
\multiput(0,53)(10,0){6}{\vector(0,-1){0}}
\multiput(130,50)(-10,-10){6}{\line(1,-1){7}}
\multiput(70,0)(-10,0){3}{\line(1,-1){7}}
\multiput(135,45)(-10,-10){6}{\vector(1,-1){0}}
\multiput(55,-5)(10,0){3}{\vector(1,-1){0}}
\put(48,60){\small$a_1$}
\put(38,60){\small$a_2$}
\multiput(28,62)(2,0){3}{\small$.$}
\put(-2,60){\small$a_m$}
\put(138,40){\small$b_1$}
\put(128,30){\small$b_2$}
\multiput(118,20)(2,2){3}{\small$.$}
\put(88,-10){\small$b_m$}
\put(72,-10){\small$\cdots$}
\put(58,-10){\small$b_n$}
\multiput(7,50)(10,0){13}{\vector(1,0){0}}
\multiput(17,40)(10,0){11}{\vector(1,0){0}}
\multiput(27,30)(10,0){9}{\vector(1,0){0}}
\multiput(37,20)(10,0){7}{\vector(1,0){0}}
\multiput(47,10)(10,0){5}{\vector(1,0){0}}
\multiput(57,0)(10,0){3}{\vector(1,0){0}}
\multiput(6,44)(10,0){5}{\vector(1,-1){0}}
\multiput(16,34)(10,0){4}{\vector(1,-1){0}}
\multiput(26,24)(10,0){3}{\vector(1,-1){0}}
\multiput(36,14)(10,0){2}{\vector(1,-1){0}}
\multiput(46,4)(10,0){1}{\vector(1,-1){0}}
\multiput(50,44)(10,0){8}{\vector(0,-1){0}}
\multiput(50,34)(10,0){7}{\vector(0,-1){0}}
\multiput(50,24)(10,0){6}{\vector(0,-1){0}}
\multiput(50,14)(10,0){5}{\vector(0,-1){0}}
\multiput(50,4)(10,0){4}{\vector(0,-1){0}}
\end{picture}
$$
%
%
$$
\begin{picture}(270,110)(-48,-15)
\unitlength=1.6pt
\put(-30,10){$\Gamma_{\sSI{U},\sSI{V}}: $}
\put(50,0){\line(1,0){30}}
\put(50,10){\line(1,0){40}}
\put(50,20){\line(1,0){50}}
\put(50,30){\line(1,0){60}}
\put(50,40){\line(1,0){70}}
\put(50,50){\line(1,0){80}}
\put(30,10){\line(-1,0){10}}
\put(30,20){\line(-1,0){20}}
\put(30,30){\line(-1,0){30}}
\put(30,40){\line(-1,0){40}}
\put(30,50){\line(-1,0){50}}
\multiput(30,50)(0,-10){6}{\line(1,0){20}}
\put(30,50){\line(0,-1){50}}
\put(20,50){\line(0,-1){40}}
\put(10,50){\line(0,-1){30}}
\put(0,50){\line(0,-1){20}}
\put(-10,50){\line(0,-1){10}}
\put(50,50){\line(0,-1){50}}
\put(60,50){\line(0,-1){50}}
\put(70,50){\line(0,-1){50}}
\put(80,50){\line(0,-1){50}}
\put(90,50){\line(0,-1){40}}
\put(100,50){\line(0,-1){30}}
\put(110,50){\line(0,-1){20}}
\put(120,50){\line(0,-1){10}}
\put(125.5,53){\small$\tilde{u}^1_1$}
\put(115.5,53){\small$\tilde{u}^1_2$}
\put(48,53){\small$\tilde{u}^1_{n}$}
\put(122,40){\small$\tilde{u}^{2}_2$}
\put(51,2){\small$\tilde{u}^m_{n}$}
\put(82,0){\small$\tilde{u}^m_{m}$}
\put(-23,53){\small$\tilde{v}^1_1$}
\put(-13,53){\small$\tilde{v}^1_2$}
\put(27,53){\small$\tilde{v}^1_{m}$}
\put(-17,41){\small$\tilde{v}^{2}_2$}
\put(21,0){\small$\tilde{v}^m_{m}$}
\put(38,52){\small$\overline{\lambda}_1$}
\put(38,42){\small$\overline{\lambda}_2$}
\put(38,-8){\small$\overline{\lambda}_m$}
\multiput(38.5,48.5)(0,-10){6}{\small$\bullet$}
\multiput(28.5,48.5)(0,-10){6}{\small$\bullet$}
\multiput(18.5,48.5)(0,-10){5}{\small$\bullet$}
\multiput(8.5,48.5)(0,-10){4}{\small$\bullet$}
\multiput(-1.5,48.5)(0,-10){3}{\small$\bullet$}
\multiput(-11.5,48.5)(0,-10){2}{\small$\bullet$}
\multiput(-21.5,48.5)(0,-10){1}{\small$\bullet$}
\multiput(48.5,48.5)(0,-10){6}{\small$\bullet$}
\multiput(58.5,48.5)(0,-10){6}{\small$\bullet$}
\multiput(68.5,48.5)(0,-10){6}{\small$\bullet$}
\multiput(78.5,48.5)(0,-10){6}{\small$\bullet$}
\multiput(88.5,48.5)(0,-10){5}{\small$\bullet$}
\multiput(98.5,48.5)(0,-10){4}{\small$\bullet$}
\multiput(108.5,48.5)(0,-10){3}{\small$\bullet$}
\multiput(118.5,48.5)(0,-10){2}{\small$\bullet$}
\multiput(128.5,48.5)(0,-10){1}{\small$\bullet$}
%
\multiput(-20,50)(10,-10){6}{\line(-1,-1){7}}
\multiput(-22,48)(10,-10){6}{\vector(1,1){0}}
\multiput(130,50)(-10,-10){6}{\line(1,-1){7}}
\multiput(70,0)(-10,0){3}{\line(1,-1){7}}
\multiput(135,45)(-10,-10){6}{\vector(1,-1){0}}
\multiput(75,-5)(-10,0){3}{\vector(1,-1){0}}
\put(-31,40){\small$a_1$}
\put(-21,30){\small$a_2$}
\multiput(-11,20)(2,-2){3}{\small$.$}
\put(19,-10){\small$a_m$}
\put(138,39){\small$b_1$}
\put(128,29){\small$b_2$}
\multiput(123,22)(-2,-2){3}{\small$.$}
\put(88,-11){\small$b_m$}
\put(72,-11){\small$\ldots$}
\put(58,-11){\small$b_n$}
\multiput(-13,50)(10,0){15}{\vector(1,0){0}}
\multiput(-3,40)(10,0){13}{\vector(1,0){0}}
\multiput(7,30)(10,0){11}{\vector(1,0){0}}
\multiput(17,20)(10,0){9}{\vector(1,0){0}}
\multiput(27,10)(10,0){7}{\vector(1,0){0}}
\multiput(37,0)(10,0){5}{\vector(1,0){0}}
\multiput(-10,47)(10,0){5}{\vector(0,1){0}}
\multiput(0,37)(10,0){4}{\vector(0,1){0}}
\multiput(10,27)(10,0){3}{\vector(0,1){0}}
\multiput(20,17)(10,0){2}{\vector(0,1){0}}
\multiput(30,7)(10,0){1}{\vector(0,1){0}}
\multiput(50,43)(10,0){8}{\vector(0,-1){0}}
\multiput(50,33)(10,0){7}{\vector(0,-1){0}}
\multiput(50,23)(10,0){6}{\vector(0,-1){0}}
\multiput(50,13)(10,0){5}{\vector(0,-1){0}}
\multiput(50,3)(10,0){4}{\vector(0,-1){0}}
\end{picture}
$$
\caption{\label{fig:PathPsi}Path representations of $\Psi$}
\end{figure}
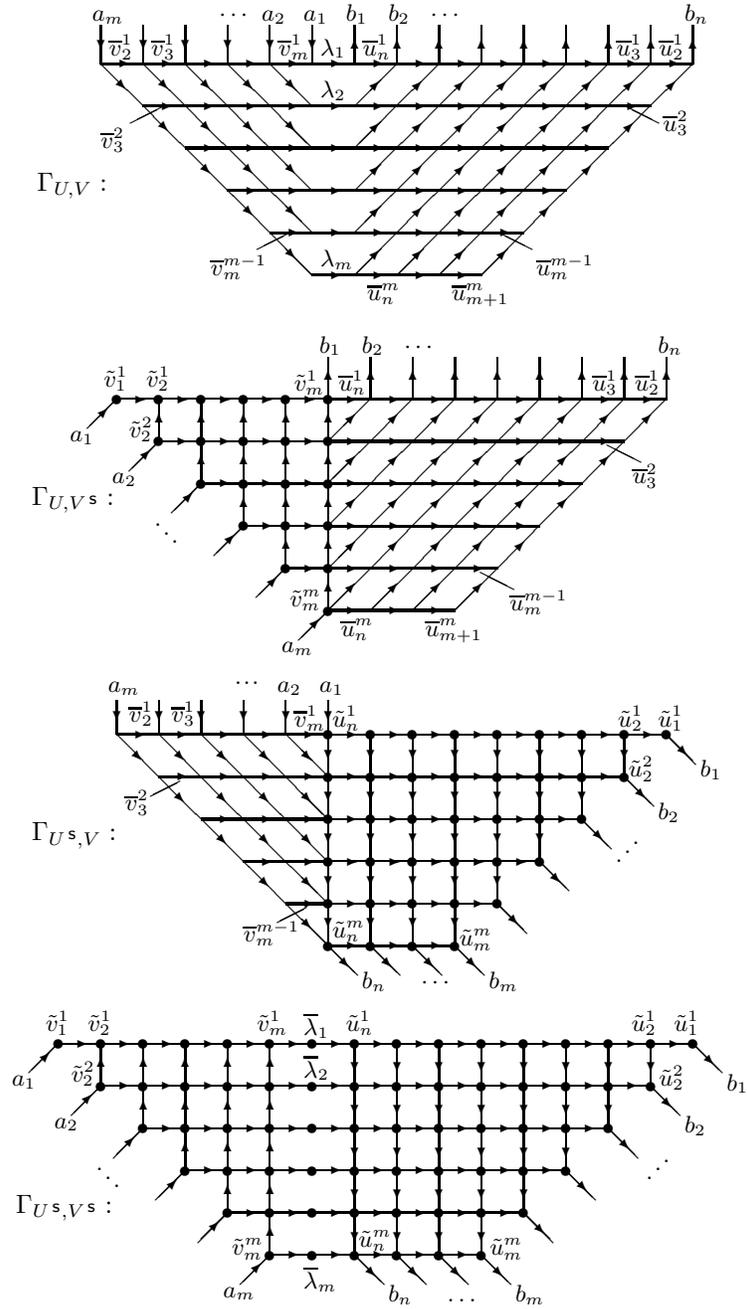

This proposition implies 
that the matrix $\Psi=\Psi_-\Psi_0\Psi_+$, 
as well as $\Phi=\Psi^\vee$, 
is completely recovered from each of the four pairs of 
tropical tableaux 
\begin{equation}\label{eq:UVs}
(U,V),\quad (U,\SI{V}),\quad (\SI{U},V),\quad (\SI{U},\SI{V}). 
\end{equation} 
By combining the graphical representations in Proposition 
\ref{prop:PsiUV}, 
we can construct a path representation of 
$\Psi$, associated with each pair in \eqref{eq:UVs}. 
According to 
\begin{equation}
\Psi^i_j=\sum_{k=1}^m (\Psi_-)^i_k\,(\Psi_0)^k_k\,(\Psi_+)^k_j,
\end{equation}
we glue the diagrams  of \eqref{eq:PsiU} or \eqref{eq:PsiSU}
for $\Psi_+$ and 
\eqref{eq:PsiV} or \eqref{eq:PsiSV} for $\Psi_-$. 
We show in Figure \ref{fig:PathPsi} the diagrams
\begin{equation}
\Gamma=\Gamma_{U,V},\ \
\Gamma_{U,\sSI{V}},\ \
\Gamma_{\sSI{U},V},\ \
\Gamma_{\sSI{U},\sSI{V}} 
\end{equation}
obtained in this way.
In diagram $\Gamma$, the orientation of the edges
are indicated by arrows. 
We assign the weights, associated with the pair of tropical tableaux, 
to the thick edges and the vertices marked by $\bullet$. 
For each path $\gamma$ in $\Gamma$, 
we define the weight $\wt(\gamma)$
to be the product of weights attached to all the edges 
and the vertices.  
In Figure \ref{fig:PathPsi},
$a_1,\ldots,a_m$ and $b_1,\ldots,b_n$ indicate 
the entrances and the exits for the path representation 
of $\Psi$, respectively. 
Namely, for each $(i,j)$, $\psi^i_j$ is expressed as 
the sum 
 \begin{equation}
\psi^i_j=\sum_{\gamma: a_i \to b_j} \wt(\gamma)
\end{equation}
of weights defined as above, 
over all paths $\gamma: a_i\to b_j$ in $\Gamma$. 
Recall that the matrix $X=\pmatrix{x^i_j}$ is determined 
from $\Phi$ through the minor determinants 
$\tau^i_j=\tau^i_j(\Phi)$ of $\Phi=\Psi^\vee$. 
Hence each $x^i_j$ is determined as 
\begin{equation}
x^i_j=\frac{\tau^i_j\ \ \tau^{i-1}_{j-1}}{\tau^{i-1}_j\tau^i_{j-1}},
\qquad
\quad \tau^i_j=\sum_{(\gamma_1,\ldots,\gamma_r)}\ 
\wt(\gamma_1)\cdots\wt(\gamma_r), 
\end{equation} 
where the summation is taken over all $r$-tuples ($r=\min\br{i,j}$) 
of nonintersecting paths 
\begin{equation}
\gamma_k : a_{m-i+k} \to b_{n-j+k}\qquad(k=1,\ldots,r)
\end{equation}
in $\Gamma$.  
For each pair in \eqref{eq:UVs}, 
we have thus obtained an explicit inversion formula 
of the corresponding tropical RSK correspondence 
in terms of nonintersecting paths. 
The corresponding combinatorial formula for the 
inverse RSK correspondence is obtained simply 
by the standard procedure:
\begin{equation}
x^i_j=\tau^i_j-\tau^{i-1}_j-\tau^i_{j-1}
+\tau^{i-1}_{j-1},
\quad
\quad \tau^i_j=\max_{(\gamma_1,\ldots,\gamma_r)}\ 
(\wt(\gamma_1)+\cdots+\wt(\gamma_r)), 
\end{equation} 
where the weight of a path is defined as 
the sum of weights attached to the edges and the vertices; 
read the weights  
$\bar{u}^i_j$ and $\bar{v}^i_j$ 
in $\Gamma$ as $-u^i_j$ and $-v^i_j$ in the 
combinatorial setting. 

\par\medskip
In the case of the pair $(U,V)$, we can give a rectangular  
diagram as well, by deforming the diagram $\Gamma_{U,V}$. 
Note first that the diagram $\Gamma_{U,V}$ is equivalent to 
the following. 
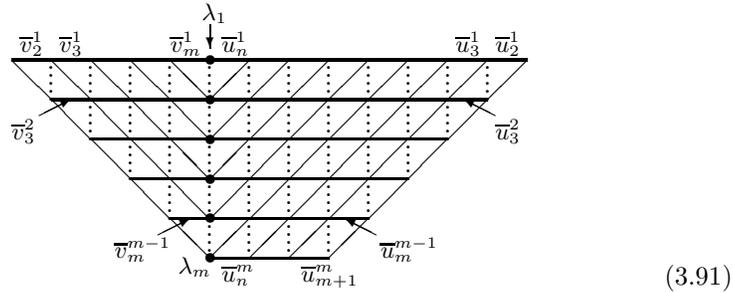
\begin{equation}
\begin{picture}(200,100)(0,-10)
\unitlength=1.5pt
{\thicklines
\put(50,0){\line(1,0){30}}
\put(50,10){\line(1,0){40}}
\put(50,20){\line(1,0){50}}
\put(50,30){\line(1,0){60}}
\put(50,40){\line(1,0){70}}
\put(50,50){\line(1,0){80}}
\put(50,10){\line(-1,0){10}}
\put(50,20){\line(-1,0){20}}
\put(50,30){\line(-1,0){30}}
\put(50,40){\line(-1,0){40}}
\put(50,50){\line(-1,0){50}}
}
\multiput(49,-0.3)(0,2){26}{$.$}
\multiput(59,-0.3)(0,2){26}{$.$}
\multiput(69,-0.3)(0,2){26}{$.$}
\multiput(79,-0.3)(0,2){26}{$.$}
\multiput(89,9.7)(0,2){21}{$.$}
\multiput(99,19.7)(0,2){16}{$.$}
\multiput(109,29.7)(0,2){11}{$.$}
\multiput(119,39.7)(0,2){6}{$.$}
\multiput(39,9.7)(0,2){21}{$.$}
\multiput(29,19.7)(0,2){16}{$.$}
\multiput(19,29.7)(0,2){11}{$.$}
\multiput(9,39.7)(0,2){6}{$.$}
\put(0,50){\line(1,-1){50}}
\put(10,50){\line(1,-1){40}}
\put(20,50){\line(1,-1){30}}
\put(30,50){\line(1,-1){20}}
\put(40,50){\line(1,-1){10}}
\put(60,50){\line(-1,-1){10}}
\put(70,50){\line(-1,-1){20}}
\put(80,50){\line(-1,-1){30}}
\put(90,50){\line(-1,-1){40}}
\put(100,50){\line(-1,-1){50}}
\put(110,50){\line(-1,-1){50}}
\put(120,50){\line(-1,-1){50}}
\put(130,50){\line(-1,-1){50}}
\put(122,52){\small$\overline{u}^1_2$}
\put(112,52){\small$\overline{u}^1_3$}
\put(53,52){\small$\overline{u}^1_n$}
\put(122,30){\small$\overline{u}^{2}_3$}
\put(122,36){\vector(-2,1){8}}
\put(93,0){\small$\overline{u}^{m-1}_m$}
\put(92,6){\vector(-2,1){8}}
\put(53,-6){\small$\overline{u}^m_n$}
\put(73,-6){\small$\overline{u}^m_{m+1}$}
\put(2,52){\small$\overline{v}^1_2$}
\put(12,52){\small$\overline{v}^1_3$}
\put(40,52){\small$\overline{v}^1_m$}
\put(-0,30){\small$\overline{v}^{2}_3$}
\put(7,36){\vector(2,1){8}}
\put(26,0){\small$\overline{v}^{m-1}_m$}
\put(37,6){\vector(2,1){8}}
\multiput(48.5,48.5)(0,-10){6}{\small$\bullet$}
\put(48,60){\small$\lambda_1$}
\put(50,59){\vector(0,-1){6}}
\put(42,-4){\small$\lambda_m$}
\end{picture}
\end{equation}
We deform this diagram to
\begin{equation}
\begin{picture}(200,100)(0,-10)
\unitlength=1.5pt
{\thicklines
\put(50,0){\line(1,0){30}}
\put(50,10){\line(1,0){40}}
\put(50,20){\line(1,0){50}}
\put(50,30){\line(1,0){60}}
\put(50,40){\line(1,0){70}}
\put(50,50){\line(1,0){80}}
\put(50,10){\line(-1,-1){10}}
\put(50,20){\line(-1,-1){20}}
\put(50,30){\line(-1,-1){30}}
\put(50,40){\line(-1,-1){40}}
\put(50,50){\line(-1,-1){50}}
}
\multiput(49,-0.3)(0,2){26}{$.$}
\multiput(59,-0.3)(0,2){26}{$.$}
\multiput(69,-0.3)(0,2){26}{$.$}
\multiput(79,-0.3)(0,2){26}{$.$}
\multiput(89,9.7)(0,2){21}{$.$}
\multiput(99,19.7)(0,2){16}{$.$}
\multiput(109,29.7)(0,2){11}{$.$}
\multiput(119,39.7)(0,2){6}{$.$}
\multiput(39,-0.3)(0,2){21}{$.$}
\multiput(29,-0.3)(0,2){16}{$.$}
\multiput(19,-0.3)(0,2){11}{$.$}
\multiput(9,-0.3)(0,2){6}{$.$}
\put(0,0){\line(1,0){50}}
\put(10,10){\line(1,0){40}}
\put(20,20){\line(1,0){30}}
\put(30,30){\line(1,0){20}}
\put(40,40){\line(1,0){10}}
\put(60,50){\line(-1,-1){10}}
\put(70,50){\line(-1,-1){20}}
\put(80,50){\line(-1,-1){30}}
\put(90,50){\line(-1,-1){40}}
\put(100,50){\line(-1,-1){50}}
\put(110,50){\line(-1,-1){50}}
\put(120,50){\line(-1,-1){50}}
\put(130,50){\line(-1,-1){50}}
\put(122,52){\small$\overline{u}^1_2$}
\put(112,52){\small$\overline{u}^1_3$}
\put(53,52){\small$\overline{u}^1_n$}
\put(122,30){\small$\overline{u}^{2}_3$}
\put(122,36){\vector(-2,1){8}}
\put(93,0){\small$\overline{u}^{m-1}_m$}
\put(92,6){\vector(-2,1){8}}
\put(54,-6){\small$\overline{u}^m_n$}
\put(73,-6){\small$\overline{u}^m_{m+1}$}
\put(-2,6){\small$\overline{v}^1_2$}
\put(8,16){\small$\overline{v}^1_3$}
\put(37,46){\small$\overline{v}^1_m$}
\put(13,-10){\small$\overline{v}^{2}_3$}
\put(15,-4){\vector(0,1){8}}
\multiput(48.5,48.5)(0,-10){6}{\small$\bullet$}
\put(44,53){\small$\lambda_1$}
\put(43,-6){\small$\lambda_m$}
\end{picture}
\end{equation}
and finally to the $m\times n$ rectangle: 
\begin{equation}\label{eq:GPsiUV}
\begin{picture}(210,110)(-20,0)
\unitlength=2pt
\multiput(0,0)(0,10){6}{\line(1,0){80}}
\multiput(0,0)(10,0){9}{\line(0,1){50}}
{\thicklines
\put(0,50){\line(1,0){80}}
\put(10,40){\line(1,0){70}}
\put(20,30){\line(1,0){60}}
\put(30,20){\line(1,0){50}}
\put(40,10){\line(1,0){40}}
\put(50,0){\line(1,0){30}}
\put(0,50){\line(0,-1){50}}
\put(10,40){\line(0,-1){40}}
\put(20,30){\line(0,-1){30}}
\put(30,20){\line(0,-1){20}}
\put(40,10){\line(0,-1){10}}
}
\multiput(-1.2,48.8)(10,-10){6}{\small$\bullet$}
\put(3,52){\small$\overline{u}^1_n$}
\put(13,52){\small$\overline{u}^1_{n-1}$}
\put(13,42){\small$\overline{u}^2_{n}$}
\put(63,52){\small$\overline{u}^1_3$}
\put(73,52){\small$\overline{u}^1_2$}
\put(73,42){\small$\overline{u}^2_3$}
\put(73,52){\small$\overline{u}^1_2$}
\put(73,2){\small$\overline{u}^m_{m+1}$}
\put(43,12){\small$\overline{u}^{m-1}_{n}$}
\put(53,2){\small$\overline{u}^m_{n}$}
\put(-6,43){\small$\overline{v}^1_{m}$}
\put(-8,33){\small$\overline{v}^1_{m-1}$}
\put(4,33){\small$\overline{v}^2_{m}$}
\put(34,3){\small$\overline{v}^{m-1}_{m}$}
\put(4,3){\small$\overline{v}^{2}_{3}$}
\put(-6,13){\small$\overline{v}^{1}_{3}$}
\put(-6,3){\small$\overline{v}^{1}_{2}$}
\put(-5,52){\small$\lambda_1$}
\put(5,42){\small$\lambda_2$}
\put(44,2){\small$\lambda_m$}
\multiput(7,0)(10,0){8}{\vector(1,0){0}}
\multiput(7,10)(10,0){8}{\vector(1,0){0}}
\multiput(7,20)(10,0){8}{\vector(1,0){0}}
\multiput(7,30)(10,0){8}{\vector(1,0){0}}
\multiput(7,40)(10,0){8}{\vector(1,0){0}}
\multiput(7,50)(10,0){8}{\vector(1,0){0}}
\multiput(0,7)(10,0){9}{\vector(0,1){0}}
\multiput(0,17)(10,0){9}{\vector(0,1){0}}
\multiput(0,27)(10,0){9}{\vector(0,1){0}}
\multiput(0,37)(10,0){9}{\vector(0,1){0}}
\multiput(0,47)(10,0){9}{\vector(0,1){0}}
\end{picture}
\end{equation}
By this rectangle, 
$\psi^i_j$ is 
expressed as the sum 
\begin{equation}
\psi^i_j=\sum_{\gamma: (i,1)\to\gamma(1,j)} \wt(\gamma)
\end{equation}
of weights defined as above, 
over all paths $\gamma: (i,1)\to(1,j)$. 
(This representation is similar to that by $Y=\pmatrix{y^i_j}_{i,j}$,
although the weights are defined in a different way.)
Hence we have 
\begin{equation}
x^i_j=\frac{\tau^i_j\ \tau^{i-1}_{j-1}}{\tau^{i-1}_j\tau^i_{j-1}},
\quad
\tau^i_j=
\sum_{(\gamma_1,\ldots,\gamma_r)} 
\wt(\gamma_1)\cdots\wt(\gamma_r),
\end{equation}
where $r=\min\br{i,j}$ and the summation is taken 
over all $r$-tuples of nonintersecting paths 
$\gamma_k: (m-i+k,1)\to(1,n-j+k)$ ($k=1,\ldots,r$)
in the $m\times n$ rectangle.
This inversion formula is essentially equivalent to the inverse 
\RSKs correspondence discussed in the previous subsection. 

\begin{remark}\rm 
As we have seen above,  
the RSK correspondence can be thought of as 
the Gauss (or $LR$) decomposition of 
{\em ultra-discretized} matrices with respect to the 
product defined by 
\begin{equation}
(XY)^i_j= \max_k\  ( X^i_k+Y^k_j ). 
\end{equation}
\end{remark}
\setcounter{equation}{0}
\section{Birational Weyl group actions} 

In this section, we introduce a subtraction-free birational affine Weyl
group action on the space of tropical transportation 
matrices. 
It induces an action of the symmetric group 
on the space of tropical tableaux through the 
tropical RSK correspondence.  
In this section, we work with the generic 
$m\times n$ matrix $X=\pmatrix{x^i_j}_{i,j}$, regarding 
$x^i_j$ as indeterminates.  

\subsection{Affine Weyl group action on the matrix space}
In what follows, we consider the following 
two (extended) affine Weyl groups 
$\widetilde{W}^m$ and $\widetilde{W}_n$ 
of type $A^{(1)}_{m-1}$ and $A^{(1)}_{n-1}$, 
respectively. 
We denote by 
\begin{equation}
\widetilde{W}^m=\pr{r_0,r_1,\ldots,r_{m-1},\omega}
\end{equation}
the group generated by the {\em simple reflections} 
$r_0,r_1,\ldots,r_{m-1}$ and the {\em diagram rotation} 
$\omega$ subject to the fundamental relations
\begin{equation}
\begin{array}{llll}\smallskip
r_i^2=1,\quad \cr\smallskip
r_ir_j=r_jr_i\quad&(j\not\equiv i, i\pm1\  \mod m),\cr\smallskip
r_ir_jr_i=r_jr_ir_j\quad&(j\equiv i\pm1\ \ \mod m),\cr\smallskip
\omega r_i=r_{i+1} \omega,
\end{array}
\end{equation}
where we understand the indices for $r_i$ as elements 
of $\BZ/m\BZ$. 
Notice that we have not imposed the relation $\omega^m=1$. 
This version of extended affine Weyl group is isomorphic to 
the semidirect product of the lattice $\BZ^m$ of rank $m$
({\em not} of $m-1$) and the symmetric group $\BS_m$ 
acting on it; the subgroup $\pr{r_1,\ldots,r_{m-1}}$ of 
$\widetilde{W}^m$ is identified with $\BS_m$ by 
mapping each $r_i$ to the adjacent transposition 
$\sigma_i=(i,i+1)$ $(i=1,\ldots,m-1)$. 
We define $\widetilde{W}_n=\pr{s_0,s_1,\ldots,s_{n-1},\pi}$ 
similarly to be the group generated by simple reflections 
$s_0,s_1,\ldots,s_{n-1}$ and the diagram rotation $\pi$\,:
\begin{equation}
\begin{array}{lll}\smallskip
s_i^2=1,\qquad 
s_is_j=s_js_i\quad(j\not\equiv i, i\pm1\  \mod n),\cr\smallskip
s_is_js_i=s_js_is_j\quad (j\equiv i\pm1\ \ \mod n),\quad
\pi s_i=s_{i+1} \pi. 
\end{array}
\end{equation}
The subgroup $\pr{s_1,\ldots,s_{n-1}}$ of $\widetilde{W}_n$ 
is identified with the symmetric group $\BS_n$. 

We now propose to realize these two affine Weyl groups 
as a group of automorphisms of the field of rational functions 
$\BK(x)$ in $mn$ variables $x=(x^i_j)_{i,j}$.  
With two extra parameters $p, q$, we take the field of 
rational functions $\BK=\BQ(p,q)$ in $(p,q)$ as the ground 
field.  
In our realization, 
the groups $\widetilde{W}^m$ and $\widetilde{W}_n$ 
concern the {\em nontrivial} permutation of rows 
and columns of the matrix $X=\pmatrix{x^i_j}_{i,j}$,
respectively.  
We first extend the indexing set 
$\br{1,\ldots,m}\times\br{1,\ldots,n}$ for the matrix 
$X=\pmatrix{x^i_j}_{i,j}$ to $\BZ\times\BZ$ by imposing 
the periodicity condition
\begin{equation}
x^{i+m}_j=q^{-1} x^i_j,\quad x^i_{j+n}=p^{-1} x^i_j \qquad(i,j\in\BZ). 
\end{equation}
We define the automorphism $r_k$ ($k\in\BZ/m\BZ$)
and $\omega$ of 
$\BK(x)$ by 
\begin{eqnarray}
&r_k(x^i_j)=p\,x^{i+1}_j\dfrac{P^i_j}{P^i_{j-1}},\quad
r_k(x^{i+1}_j)=p^{-1}x^{i}_j\dfrac{P^i_{j-1}}{P^i_{j}}\quad (i\equiv k \mod m),
\nonumber\\
&r_k(x^i_j)=x^i_j\quad(i\not\equiv k, k+1 \mod m),
\quad
\omega(x^{i}_j)=x^{i+1}_j
\end{eqnarray}
for $i,j\in\BZ$, where $P^i_j$ is the sum 
\begin{equation}
P^i_j=\sum_{k=1}^n 
x^{i+1}_{j+1}x^{i+1}_{j+2}
\cdots x^{i+1}_{j+k}
x^{i}_{j+k}x^{i}_{j+k+1}\cdots x^{i}_{j+n}
\end{equation} 
over all paths $\gamma:(i+1,j+1)\to(i,j+n)$ in the lattice 
$\BZ\times\BZ$. 
We define $s_l$ $(l\in\BZ/n\BZ)$
and $\pi$ by interchanging the roles 
of rows and columns, and of $p$ and $q$: 
\begin{eqnarray}
&s_l(x^i_j)=q\,x^{i}_{j+1}\dfrac{Q^i_j}{Q^{i-1}_{j}},\quad
s_l(x^i_{j+1})=q^{-1} x^{i}_j\dfrac{Q^{i-1}_{j}}{Q^i_{j}}
\quad (j\equiv l \mod n),
\nonumber\\
&s_l(x^i_j)=x^i_j\quad(j\not\equiv l, l+1 \mod n),
\quad
\pi(x^{i}_j)=x^{i}_{j+1}
\end{eqnarray}
for $i,j\in\BZ$, where
\begin{equation}
Q^i_j=\sum_{k=1}^m 
x^{i+1}_{j+1}x^{i+2}_{j+1}
\cdots x^{i+k}_{j+1}
x^{i+k}_{j}x^{i+k+1}_{j}\cdots x^{i+m}_{j}
\end{equation}
summed over all paths $\gamma:(i+1,j+1)\to(i+m,j)$. 
It is directly seen that these definitions are consistent 
with the periodicity conditions on $x^i_j$. 
Also, it is clear that $r_k$ and $s_l$ have rotational symmetry 
\begin{equation}
\begin{array}{llll}\smallskip
\omega \, r_k =r_{k+1} \omega,\quad & \pi \,r_k= r_{k} \pi
\quad&(k\in\BZ/m\BZ),\cr
\omega \, s_l =s_{l} \omega,\quad & \pi \,s_l= s_{l+1} \pi
\quad&(l\in\BZ/n\BZ),
\end{array}
\end{equation}
respectively. 

\begin{remark}\rm
The polynomials $P^i_j$ and $Q^i_j$ above are characterized 
by the recurrence relations
\begin{equation}
\begin{array}{ll}\smallskip
x^{i+1}_jP^i_j-P^i_{j-1} x^i_{j+n}
=x^{i+1}_{j}(x^{i+1}_{j+1}\cdots x^{i+1}_{j+n}-
x^i_j\cdots x^i_{j+n-1})x^{i+1}_{j+n},\cr
x^{i}_{j+1}Q^i_j-Q^{i-1}_{j} x^{i+m}_{j}
=x^{i}_{j+1}(x^{i+1}_{j+1}\cdots x^{i+m}_{j+1}-
x^i_j\cdots x^{i+m-1}_{j})x^{i+m}_{j},
\end{array}
\end{equation}
and the periodicity conditions $P^i_{j+n}=p^{-n-1}P^i_j$,
$Q^{i+m}_j=q^{-m-1}Q^i_j$.  
\end{remark}

\begin{theorem}\label{thm:BirAW}
The automorphisms $r_k$ $(k\in\BZ/m\BZ)$, $\omega$,
$s_l$ $(l\in\BZ/n\BZ)$, $\pi$ of $\ \BK(x)$ defined 
as above give a realization of the direct product 
$\widetilde{W}^m\times\widetilde{W}_n$ of two 
extended affine Weyl groups. 
In particular, the actions of 
$\widetilde{W}^m=\pr{r_0,\ldots,r_{m-1},\omega}$ 
and $\widetilde{W}_n=\pr{s_0,\ldots,s_{n-1},\pi}$ 
commute with each other. 
\end{theorem}

In the next two subsections, we give a proof of this theorem 
by using two characterizations of birational actions 
of $r_k$ and $s_l$.  

\begin{remark}\rm 
The realization of $\widetilde{W}^m\times \widetilde{W}_n$ 
mentioned above is the same as the one we gave in 
\cite{KNY2} ($p=q=1$),  
and \cite{KNY3}; 
the variables $x^i_j$ above correspond to $x_{ij}^{-1}$ 
in \cite{KNY3}.   
When $p=q=1$, 
it coincides with the birational realization of 
$\widetilde{W}^m\times \widetilde{W}_n$ 
constructed in \cite{K}, Theorem 4.12. 
\end{remark} 

\subsection{First characterization} 

By introducing the spectral parameter $z$, 
for an $n$-vector $\bx=(x_1,\ldots,x_n)$ with $x_i\ne 0$ given, 
we introduce the following two matrices: 
\begin{equation}
E(\bx;z)=\diag{\bx}+\Lambda(z),\quad
H(\bx;z)=(\diag{\overline{\bx}}-\Lambda(z))^{-1},
\end{equation}
where
\begin{equation}
\Lambda(z)=\sum_{k=1}^{n-1}E_{k,k+1}+z E_{n,1}. 
\end{equation}
Note that the definition of $H(\bx;z)$ makes sense since
\begin{equation}
\det(\diag{\overline{\bx}}-\Lambda(z))=
\overline{x}_1\cdots \overline{x}_n-z. 
\end{equation}
When $z=0$, these matrices reduce to 
$E(\bx)$ and $H(\bx)$ used in previous sections. 
Note also that $H(\bx;z)=DE(\overline{\bx};z)^{-1}D^{-1}$,
$D=\diag{(-1)^{i-1}}_{i=1}^n$.  
We remark that the entries of the matrix $H(\bx;z)$ are expressed explicitly as
\begin{equation}\label{eq:Hzentries}
H(\bx;z)^i_j=
\left\{\begin{array}{lll}\medskip
\dfrac{x_i x_{i+1}\cdots x_j}{1-x_1\cdots x_n z} \quad & (i\le j),\cr
\dfrac{x_1\cdots x_j x_{i}\cdots x_{n}z}{1-x_1\cdots x_n z} & (i> j). 
\end{array}\right.
\end{equation}

For two $n$-vectors $\bx=(x_1,\ldots,x_n)$, 
$\by=(y_1,\ldots,y_n)$ of indeterminates given, 
we consider the following matrix equation  
for unknown vectors $\bu=(u_1,\ldots,u_n)$, 
$\bv=(v_1,\ldots,v_n)$ such that $u_j\ne 0$, $v_j\ne 0$: 
\begin{equation}\label{eq:Hz2}
H(\by;z) H(\bx;pz) = H(\bv;z) H(\bu;pz),
\end{equation}
or equivalently, 
\begin{equation}\label{eq:Ez2}
E(\overline{\bx};pz)E(\overline{\by};z)=
E(\overline{\bu};pz)E(\overline{\bv};z). 
\end{equation}
As before we extend the indexing set for $x_j, y_j,\ldots$  
to $\BZ$ by setting $x_{j+n}=p^{-1} x_j$, $y_{j+n}=p^{-1}y_j,\ldots$. 
Then the matrix equation \eqref{eq:Ez2} is equivalent to 
the system of algebraic equations of 
{\em discrete Toda type}
\begin{equation}\label{eq:dToda}
x_jy_j=u_jv_j,\quad
\frac{1}{x_j}+\frac{1}{y_{j+1}}=\frac{1}{u_j}+\frac{1}{v_{j+1}}
\quad(j\in\BZ). 
\end{equation}
(For the discrete Toda equation, see Remark \ref{rem:dToda}.)
The next lemma is fundamental in the following 
argument.
\begin{lemma}\label{lem:dToda}
The matrix equation \eqref{eq:Hz2} has the following 
two solutions:\newline
\begin{equation}
\begin{array}{llll}\smallskip
(1)\quad& u_j=x_j,\quad &v_j=y_j\quad&(j=1,\ldots,n),\cr
(2)& u_j=p\,y_j\dfrac{P_j}{P_{j-1}},\quad &v_j=
p^{-1}x_j \dfrac{P_{j-1}}{P_{j}}
\quad&(j=1,\ldots,n),
\end{array}
\end{equation}
where 
\begin{equation}
P_j=\sum_{k=1}^n y_{j+1}\cdots y_{j+k}x_{j+k}\cdots x_{j+n}
\quad(j=0,1,\ldots,n). 
\end{equation}
\end{lemma}
\proof 
If $v_{j+1}=y_{j+1}$ for some $j\in\BZ$, form \eqref{eq:dToda} it follows 
that $u_{j}=x_{j}$, and $v_j=y_j$.  Hence 
we have $u_j=x_j$, $v_j=y_j$ for all $j\in \BZ$.  
Assuming that $v_{j}\ne y_{j}$ for any $j\in\BZ$, we introduce 
the variable $h_j$ ($j\in\BZ$) such that 
\begin{equation}
\frac{1}{v_{j+1}}=\frac{1}{y_{j+1}}+\frac{1}{h_j}\qquad(j\in \BZ),
\end{equation}
so that $h_{j+n}=p^{-1}h_j$.  Then by eliminating $u_j$ in 
\eqref{eq:dToda}, we obtain the recurrence relations
\begin{equation}\label{eq:rech}
\frac{h_j}{x_j}=1+\frac{h_{j-1}}{y_j},
\quad\mbox{i.e.,}\quad
h_j=x_j+\frac{x_j}{y_j} h_{j-1}\qquad(j\in\BZ)
\end{equation}
for $h_j$.  Hence we have
\begin{equation} 
\begin{array}{ll}\smallskip
h_{j+n}&=x_{j+n}+\dfrac{x_{j+n-1}x_{j+n}}{y_{j+n}}+
\cdots+\dfrac{x_{j+1}\cdots x_{j+n}}{y_{j+2}\cdots y_{j+n}}
+\dfrac{x_{j+1}\cdots x_{j+n}}{y_{j+1}\cdots y_{j+n}}h_j\cr
&=\dfrac{ P_j+x_{j+1}\cdots x_{j+n}h_{j}}{y_{j+1}\cdots y_{j+n}}.
\end{array}
\end{equation}
Since $h_{j+n}=p^{-1}h_j$, this equation determines $h_j$ as
\begin{equation}
h_j=\frac{P_j}{p^{-1}y_{j+1}\cdots y_{j+n}-x_{j+1}\cdots x_{j+n}}. 
\end{equation}
In fact, these $h_j$ satisfy the recurrence relations above, since 
\begin{equation}
y_jP_j-P_{j-1}x_{j+n}=
y_jy_{j+1}\cdots y_{j+n}x_{j+n}
-y_jx_jx_{j+1}\cdots x_{j+n},
\end{equation}
Hence we have
\begin{equation}
\frac{1}{v_{j}}=\frac{1}{y_{j}}+\frac{1}{h_{j-1}}
=\frac{h_j}{x_jh_{j-1}}=\frac{p\,P_{j}}{x_j P_{j-1}},
\end{equation}
and
\begin{equation}
\frac{1}{u_j}=\frac{1}{x_j}-\frac{1}{h_j}=\frac{h_{j-1}}{y_j h_j}
=\frac{P_{j-1}}{y_j p P_{j}}, 
\end{equation}
which gives the solution (2). 
\qed
\noindent
We remark that the two solutions above are characterized by 
the conditions
\begin{equation}
\begin{array}{lll}\smallskip
(1) & 
u_1\cdots u_n=x_1\cdots x_n,\quad &
v_1\cdots v_n=y_1\cdots y_n,\cr 
(2) & 
u_1\cdots u_n=p^{-1}y_1\cdots y_n,\quad &
v_1\cdots v_n= p\,x_1\cdots x_n,
\end{array}
\end{equation}
respectively. 
Note here that $P_n=p^{-n-1}P_0$. 
\par\medskip
Returning to the setting of the previous subsection, 
we consider the matrix $X=\pmatrix{x^i_j}_{i,j}$. 
We denote the row vectors and the column vectors of 
$X=\pmatrix{x^i_j}_{i,j}$ by 
$\bx^i=(x^i_1,\ldots,x^i_n)$ and 
$\bx_j=(x^1_j,\ldots,x^m_j)$, respectively. 
Then Lemma \ref{lem:dToda} implies 
\begin{equation}
H(\bx^{k+1};z)H(\bx^{k};pz)=H(r_{k}(\bx^{k+1});z)H(r_k(\bx^{k});pz),
\end{equation}
where we have used the notation 
$r_k(\bx)=(r_k(x_1),\ldots,r_k(x_n))$
for $\bx=(x_1,\ldots,x_n)$. 
Since $r_k(\bx^i)=\bx^i$ for $i\not\equiv k \mod m$, we have
\begin{eqnarray}
&&
H(\bx^{m};z)H(\bx^{m-1};pz)\cdots H(\bx^{1};p^{m-1}z)
\nonumber\\
&&=H(r_k(\bx^{m});z)H(r_k(\bx^{m-1});pz)\cdots H(r_k(\bx^{1});p^{m-1}z)
\end{eqnarray}
for $k=1,\dots,m-1$.  Namely, the product of matrices in the 
left-hand side is invariant under the action of $r_k$ ($k=1,\ldots,m-1$). 
Hence we see that 
\begin{eqnarray}\label{eq:Hzxxx}
&&H(\bx^{m};z)H(\bx^{m-1};pz)\cdots H(\bx^{1};p^{m-1}z)
\nonumber\\
&&=H(w(\bx^{m});z)H(w(\bx^{m-1});pz)\cdots H(w(\bx^{1});p^{m-1}z)
\end{eqnarray}
for any composition 
$w=r_{k_1}r_{k_2}\cdots r_{k_l}$ 
with $k_1,\ldots,k_l\in\br{1,\ldots,m-1}$. 
In the following, we set
\begin{equation}
H(X;z)=H(\bx^{m};z)H(\bx^{m-1};pz)\cdots H(\bx^{1};p^{m-1}z)
\end{equation}
and 
\begin{equation}
M(X;z)=E(\overline{\bx}^{1};p^{m-1}z)E(\overline{\bx}^{2};p^{m-1}z)
\cdots E(\overline{\bx}^{m};z)
\end{equation}
so that $H(X;z)=D M(X;z)^{-1} D^{-1}$. 
Then we have 
\begin{equation}
H(X;z)=H(w(X);z),\qquad M(X;z)=M(w(X);z)
\end{equation}
for any $w=r_{k_1}r_{k_2}\cdots r_{k_l}$ 
($k_1,\ldots,k_l\in\br{1,\ldots,m-1}$), 
where $w(X)=\pmatrix{w(x^i_j)}_{i,j}$ denotes the 
matrix obtained from $X$ by applying $w$ to its entries. 
\begin{proposition}\label{prop:RInv}
All the entries of the matrices
$H(X;z)$ and $M(X;z)$ are 
invariant under the action of $r_1,\ldots,r_{m-1}$. 
\end{proposition}

Considering $X=\pmatrix{x^i_j}_{i,j}$ as given, we now investigate 
in general 
the matrix equation $H(X;z)=H(Y;z)$ for an $m\times n$ unknown 
matrix $Y=(y^i_j)_{i,j}$, $y^i_j\ne 0$: 
\begin{eqnarray}\label{eq:HzHz}
&&H(\bx^m;z) H(\bx^{m-1};pz)\cdots H(\bx^1;p^{m-1}z)
\nonumber\\
&&=H(\by^m;z) H(\by^{m-1};pz)\cdots H(\by^1;p^{m-1}z). 
\end{eqnarray}
Note that this equation is equivalent to $M(X;z)=M(Y;z)$: 
\begin{eqnarray}\label{eq:EzEz}
&&E(\overline{\bx}^1;p^{m-1}z) E(\overline{\bx}^{2};p^{m-2}z)
\cdots E(\overline{\bx}^m;z)
\nonumber\\
&&=
E(\overline{\by}^1;p^{m-1}z) E(\overline{\by}^{2};p^{m-2}z)
\cdots E(\overline{\by}^m;z). 
\end{eqnarray}
Since 
$\det H(\bx;z)=(\overline{x}_1\cdots \overline{x}_n-z)^{-1}$, 
by comparing the determinants of the both sides of \eqref{eq:HzHz}, 
we see that, for any solution of \eqref{eq:HzHz}, there exists 
a unique permutation $\sigma\in\BS_m$ such that
\begin{equation}\label{eq:Hzsgm}
p^{m-i}y^i_1\cdots y^i_n
=p^{m-\sigma(i)}x^{\sigma(i)}_1\cdots x^{\sigma(i)}_n
\qquad(i=1,\ldots,m). 
\end{equation}
\begin{theorem}\label{thm:RChar}
For each permutation $\sigma\in\BS_m$, 
the matrix equation \eqref{eq:HzHz} has a unique solution 
satisfying the condition \eqref{eq:Hzsgm}.  
For any choice of expression 
$\sigma=\sigma_{k_1}\cdots\sigma_{k_l}$
of $\sigma$ as a product of adjacent transpositions 
$\sigma_k=(k,k+1)$ $(k=1,\ldots,m-1)$,  
the solution corresponding to $\sigma$ is given by
\begin{equation} 
y^i_j=w(x^i_j)\qquad (i=1,\ldots,m;\,j=1,\ldots,n),
\end{equation}
where $w=r_{k_1}\cdots r_{k_l}$. 
\end{theorem}
\proof  
Since $P^i_n=p^{-n-1}P^{i}_0$ for any $i$, 
we have
\begin{equation}
\begin{array}{lll}
\smallskip
r_k(x^k_1\ldots x^k_n)=p^{-1} x^{k+1}_1\ldots x^{k+1}_n,\quad
r_k(x^{k+1}_1\ldots x^{k+1}_n)=p\,x^{k}_1\ldots x^{k}_n \cr
r_k(x^i_1\ldots x^i_n)=x^i_1\ldots x^i_n
\qquad(i=1,\ldots,k-1,k+2,\ldots,n)
\end{array}
\end{equation}
for $k=1,\ldots,m-1$. 
Hence, 
\begin{equation}
r_k(x^{i}_1\ldots x^{i}_n)=p^{i-\sigma_k(i)}
x^{\sigma_{k}(i)}\ldots x^{\sigma_{k}(i)}_m
\qquad(i=1,\ldots,m). 
\end{equation}
This implies furthermore that, for any composition 
$w=r_{k_1}\ldots r_{k_l}$ of $r_k$ ($k=1,\ldots,m-1$), 
we have
\begin{equation}
w(x^{i}_1\ldots x^{i}_n) = p^{i-\sigma(i)}
x^{\sigma(i)}\ldots x^{\sigma(i)}_n 
\qquad(i=1,\ldots,m). 
\end{equation}
where $\sigma=\sigma_{k_1}\cdots\sigma_{k_l}$. 
Namely, $w(\bx^1),\ldots,w(\bx^m)$ 
give a solution of \eqref{eq:HzHz} satisfying the condition
\eqref{eq:Hzsgm}. 
In order to complete the proof of the theorem, 
we show that any solution $\by^1,\ldots,\by^m$ 
satisfying \eqref{eq:Hzsgm} must coincide with this 
solution.  
In the following we denote by 
$\xi_i=p^{-m+i}\,\overline{x}^i_1\cdots \overline{x}^i_n$ 
the pole of $H(\bx^i;p^{m-i}z)$.  
Consider the equality 
\begin{eqnarray}
&&H(w(\bx^m);z) H(w(\bx^{m-1});pz)\cdots H(w(\bx^1);p^{m-1}z)
\nonumber\\
&&=H(\by^m;z) H(\by^{m-1};pz)\cdots H(\by^1;p^{m-1}z), 
\end{eqnarray}
and multiply the both sides by $H(\by^m;z)^{-1}$ from the left 
to get
\begin{eqnarray}
&&H(\by^m;z)^{-1}H(w(\bx^m);z) H(w(\bx^{m-1});pz)
\cdots H(w(\bx^1);p^{m-1}z)
\nonumber\\
&&=H(\by^{m-1};pz)\cdots H(\by^1;p^{m-1}z). 
\end{eqnarray}
Since the right-hand side is regular
at $z=\xi_{\sigma(m)}=\overline{y}^m_1\cdots \overline{y}^m_n$,
the residue of the left-hand side at $z=\xi_{\sigma(m)}$ must 
vanish.  It implies 
\begin{equation}
\big(\diag{\overline{\by}^m}-\Lambda(\xi_{\sigma(m)})\big)
\,\Res_{z=\xi_{\sigma(m)}}(H(w(\bx^m);z)dz)=0
\end{equation}
since the matrices $H(w(\bx^i);\xi_{\sigma(m)})$ ($i=1,\ldots,m-1$) 
are all invertible. 
If we set $\widetilde{H}(\bx;z)
=(\overline{x}_1\cdots \overline{x}_n-z)H(\bx;z)$,
it is equivalent to 
\begin{equation}
(\diag{\overline{\by}^m}-\Lambda(\xi_{\sigma(m)}))
\widetilde{H}(w(\bx^m); \xi_{\sigma(m)})=0. 
\end{equation}
This equation determines $\by^m$ uniquely since 
$\widetilde{H}(w(\bx^m); \xi_{\sigma(m)})^i_j\ne 0$
for any $i,j$.  
Since $(\diag{w(\overline{\bx}^m)}-\Lambda(\xi_{\sigma(m)}))
\widetilde{H}(w(\bx^m); \xi_{\sigma(m)})=0$, 
we have $\by^m=w(\bx^m)$, and also 
\begin{equation}
H(w(\bx^{m-1});pz)\cdots H(w(\bx^1);p^{m-1}z)
=H(\by^{m-1};pz)\cdots H(\by^1;p^{m-1}z). 
\end{equation}
By repeating the same procedure, we finally 
obtain $\by^i=w(\bx^i)$ for all $i=1,\ldots,m$
as, desired.  
\qed
\begin{corollary}
Let $i_1,\ldots,i_k$ and $ j_1,\ldots,j_l$ be two sequences 
of elements of $\br{1,\ldots,m-1}$ such that 
\begin{equation}
\sigma_{i_1}\cdots \sigma_{i_k}=\sigma_{j_1}\cdots\sigma_{j_l}. 
\end{equation}
Then the automorphisms $r_1,\ldots,r_{m-1}$ satisfies the 
relation
\begin{equation}
r_{i_1}\cdots r_{i_k}=r_{j_1}\cdots r_{j_l}. 
\end{equation}
\end{corollary}
By this corollary and the rotational symmetry 
of $r_k$, we see that the automorphisms 
$r_0,r_1,\ldots,r_{m-1}, \omega$ 
satisfy the fundamental relations for the generators 
of $\widetilde{W}^m$.  The same statement is valid 
for $s_0,s_1,\ldots,s_{n-1}$, $\pi$ and 
$\widetilde{W}_n$ by the symmetry under the 
transposition of the matrix $X$. 

\begin{remark}\rm
Lemma \ref{lem:dToda} implies that 
the action of $r_k$ ($k=1,\ldots,m-1$) is characterized by 
the system of algebraic equations 
of discrete Toda type
\begin{equation}
x^i_j x^{i+1}_j= y_j^i y^{i+1}_j,
\quad
\frac{1}{x^i_j}+\frac{1}{x^{i+1}_{j+1}}
=\frac{1}{y^i_j}+\frac{1}{y^{i+1}_{j+1}}
\end{equation}
for $y^i_j=r_k(x^i_j)$ $(i,j\in\BZ)$
with periodicity 
condition $y^{i+m}_j=q^{-1}y^i_j$, $y^i_{j+n}=p^{-1} y^i_{j}$, 
and an extra constraint
\begin{equation}
y^i_{1}\cdots y^i_n=p^{i-\sigma_k(i)} 
x^{\sigma_k(i)}_1\cdots x^{\sigma_k(i)}_n\quad(i=1,\ldots,m). 
\end{equation}
\end{remark}
 
\subsection{Second characterization}

We give another characterization of $r_k$ and $s_l$,
and use it for proving the commutativity of the actions of 
$\pr{r_0,r_1,\ldots,r_{m-1},\omega}$ and 
$\pr{s_0,s_1,\ldots,s_{n-1},\pi}$. 

We define the $n\times n$ matrices 
$G_l(u;z)$, depending on a parameter $u$, by setting 
\begin{equation}
G_0(u;z)=1+ \frac{1}{u} E_{1,n}z^{-1},\quad
G_l(u;z)=1+\frac{1}{u}E_{l+1,l}\quad(l=1,\ldots,n-1). 
\end{equation}
For $l=1,\ldots,n-1$, we also use the notation 
$G_l(u)=G_l(u;z)$ since they do {\em not} depend on $z$. 
Fix an index $l=0,1,\ldots,n-1$, and consider the system 
of matrix equations
\begin{equation}\label{eq:GEEG}
G_l(g_{i-1};p z) E(\overline{\bx}^i;z)=E(\overline{\by}^i;z)G_l(g_i;z)
\quad (i=1,\ldots,m)
\end{equation}
for unknown variables $\by^i$ ($i=1,\ldots,m$) and $g_i$ ($i=0,1,\ldots,m$). 
\begin{theorem}\label{thm:SChar}
Under the periodicity condition $g_{m}=q^{-1}g_0$, 
the system of algebraic equations \eqref{eq:GEEG} has a unique 
solution. 
It is given explicitly as 
\begin{equation}
\begin{array}{ll}\smallskip
g_i=\dfrac{Q^i_l}
{q^{-1}x^{i+1}_{l+1}\cdots x^{i+m}_{l+1}-x^{i+1}_l\cdots x^{i+m}_l}
\quad (i=0,1,\ldots,m),\cr
y^i_j=s_l(x^i_j)\qquad(i=1,\ldots,m;\,j=1,\ldots,n). 
\end{array}
\end{equation}
\end{theorem}
\proof 
It is easily seen that the matrix equation \eqref{eq:GEEG} is 
equivalent to the recurrence relations
\begin{equation}\label{eq:rechi}
g_{i}=x^i_l+\frac{x^i_l}{x^{i}_{l+1}}g_{i-1}\qquad(i=1,\ldots,m), 
\end{equation}
together with
\begin{equation}\label{eq:xysss}
\begin{array}{ll}\smallskip
\dfrac{1}{y^i_l}=\dfrac{1}{x^i_l}-\dfrac{1}{g_i},
\quad
\dfrac{1}{y^i_{l+1}}=\dfrac{1}{x^i_{l+1}}+\dfrac{1}{g_{i-1}},\cr
y^i_j=x^i_j\qquad(j\not\equiv l,l+1 \mod n).
\end{array}
\end{equation} 
These are the same recurrence relations as 
we have discussed in Lemma \ref{lem:dToda}. 
As we already know, \eqref{eq:rechi}
determines $g_{i}$, and \eqref{eq:xysss} gives rise to the 
expressions we have used in defining $s_l$. 
\qed
\noindent
Note that the matrix equation \eqref{eq:GEEG} implies
\begin{equation}\label{eq:GMMG}
G_l(g_0;p^m z) M(X;z)=M(Y;z) G_l(q^{-1}g_0;z).  
\end{equation}
Hence, by Theorem \ref{thm:SChar}, we see that 
the action of $s_l$ on $M(X;z)$ is described by
\begin{equation}\label{eq:sMGMG}
M(s_l(X);z)=
G_l(g_0;p^m z) M(X;z)G_l(q^{-1}g_0;z)^{-1}. 
\end{equation} 
In terms of the matrix $H(X;z)$, this formula can be 
written as
\begin{equation}\label{eq:sonHXz}
H(s_l(X);z)=G_l(q^{-1}g_0;(-1)^n z)^{-1} H(X;z) G_l(g_0;(-1)^n z).
\end{equation}

We remark that the rational function $g_0$ can be determined 
only from $M(X;z)$.  It is an easy exercise to show
\begin{lemma} \label{lem:phialp}
Let $\mathfrak{b}$ be the space of all 
$n\times n $ matrix $M(z)$ with coefficients in $\BK(x)[z]$ 
such that $M(0)$ is upper triangular. 
For a matrix 
\begin{equation}
M(z)=M_0+M_1 z+\cdots +M_d z^d\in \mathfrak{b}
\end{equation}
given, set
\begin{equation}
\begin{array}{cl}\smallskip
\varepsilon_i=(M_0)^i_i,\quad & (i=1,\ldots,n),\cr
\varphi_0=(M_1)^n_1,\quad 
\varphi_i=(M_0)^i_{i+1}& (i=1,\ldots,n-1). 
\end{array}
\end{equation}
Then we have 
\begin{equation}
G_0(u;a z) M(z) G_0(q^{-1}u;z)^{-1}\in \mathfrak{b}
\ \  \Longleftrightarrow
\ \ 
(q^{-1}\varepsilon_n-a\,\varepsilon_{1})\,u= \varphi_0,
\end{equation}
and 
\begin{equation}
G_l(u) M(z) G_l(q^{-1}u)^{-1}\in \mathfrak{b}
\ \ 
\Longleftrightarrow
\ \ 
(q^{-1}\varepsilon_l-\,\varepsilon_{l+1})\,u= \varphi_l,
\end{equation}
for $l=1,\ldots,n-1$. 
In particular, the parameter $u$ is determined uniquely from 
from $M(z)$ if $\varepsilon_i$ and $\varphi_i$ are generic. 
\end{lemma}
This lemma implies that $g_0$ is expressed as 
a rational function of entries of $M(X;z)$. 
Hence, by Proposition \ref{prop:RInv}, 
we see that 
$g_0$ in invariant under the action of 
$r_1,\dots,r_{m-1}$.

We now prove the commutativity of the actions of 
$\pr{r_0,r_1,\ldots,r_{m-1}}$ and $\pr{s_0,s_1,\ldots,s_{n-1}}$. 
By the rotational symmetry of $r_k$, it suffices  
to prove $s_l w=w s_l$ ($l=0,1,\ldots,n-1$), 
assuming that $w\in\pr{r_1,\ldots,r_{m-1}}$.  
Applying $w$ to \eqref{eq:sMGMG}, we have
\begin{equation}\label{eq:GMMGws}
M(ws_l(X);z)=
G_l(w(g_0);p^m z) M(w(X);z)G_l(q^{-1}w(g_0);z)^{-1}. 
\end{equation} 
By Proposition \ref{prop:RInv}, we have $M(w(X);z)=M(X;z)$, 
and also, $w(g_0)=g_0$ as we remarked above. 
This implies that  $M(w s_l(X);z)=M(s_l(X);z)$. 
By applying $s_l$ again, we obtain
\begin{equation} 
M(s_lw s_l(X);z)=M(s_l^2(X);z)=M(X;z). 
\end{equation}
Note that 
$s_l(x^i_1\ldots x^i_n)=x^i_1\ldots x^i_n$ for any $i=1,\ldots,m$. 
Hence, for $Y=s_lws_l(X)$, we have
\begin{equation}
y^i_1\cdots y^i_n=
p^{i-\sigma(i)}x^{\sigma(i)}_1\ldots x^{\sigma(i)}_n\quad(i=1,\ldots,m),
\end{equation}
where $\sigma\in\BS_m$ is the permutation corresponding to $w$.  
Then, by Theorem \ref{thm:RChar}, we obtain
$Y=w(X)$.  This means that 
$s_lws_l(X)=w(X)$, namely, $s_lws_l=w$.
This completes the proof of Theorem \ref{thm:BirAW}. 

\par\medskip
Recall that the roles of $r_k$, $s_l$ are interchanged 
with each other by the transposition of the matrix 
$X=\pmatrix{x^i_j}_{i,j}$.  Accordingly, the two characterization 
we have discussed so far can be applied to both $r_k$ and $s_l$. 

\subsection{Passage to the tropical tableaux}

In what follows we set
$\BS^m=\pr{r_1,\ldots,r_{m-1}}$ and 
$\BS_n=\pr{s_1,\ldots,s_{n-1}}$. 

Let us consider the tropical \RSKs correspondence 
$X\mapsto (U,V)$ with the notation as in the previous section:
\begin{eqnarray}
& 
H(\bx^m)\cdots H(\bx^2)H(\bx^1)=
H_m(\bu^m) \cdots H_2(\bu^2)H_1(\bu^1)=H_U,\nonumber\\
&
H(\bx_n) \cdots H(\bx_2) H(\bx_1)=
H_m(\bv^m) \cdots H_2(\bv^2)
H_1(\bv^1)=H_{V}.
\end{eqnarray}
As before we assume that $m\le n$. 
We regard now the variables $u^i_j$ and $v^i_j$ ($i\le j$) 
as elements of $\BK(x)$.  
Note that by specializing the spectral parameter $z$ to zero, 
we have
\begin{equation}
H(X;0)=H(\bx^m)\cdots H(\bx^1)=H_U. 
\end{equation}
By Proposition \ref{prop:RInv}, 
we already know that $H(X;z)$, hence $H(X;0)$ is invariant under the 
action of the symmetric group $\BS^m=\pr{r_1,\ldots,r_{m-1}}$. 
Since the variables $u^i_j$ are determined uniquely 
from the matrix $H_U=H(X;0)$, we conclude that 
all $u^i_j$ are invariant under the action of 
$\BS^m=\pr{r_1,\ldots,r_{m-1}}$.  

We now consider the action of $\BS_n=\pr{s_1,\ldots,
s_{n-1}}$.  
For $l=1,\ldots,n-1$, 
from \eqref{eq:sonHXz}, we have
\begin{equation}
H(s_l(X);0)
=G_l(q^{-1}g_0)^{-1} H(X;0) G_l(g_0),
\end{equation}
hence
\begin{equation}\label{eq:slHU}
s_l(H_U)=G_l(q^{-1}g_0)^{-1} H_U \, G_l(g_0),
\end{equation}
where 
\begin{equation}
g_0=
\dfrac{\dsum{k=1}{m} x^1_{l+1}\cdots x^k_{l+1} x^k_{l}\cdots x^m_{l}}
{q^{-1}x^{1}_{l+1}\cdots x^{m}_{l+1}-x^{1}_l\cdots x^{m}_l}. 
\end{equation}
This formula is equivalent to
\begin{equation}\label{eq:slMU}
s_l(M_U)=G_l(g_0) M_U G_l(q^{-1}g_0),
\end{equation}
where 
\begin{equation}
M_U=D H_U^{-1} D^{-1}=
E_1(\overline{\bu}^1)\cdots E_m(\overline{\bu}^m). 
\end{equation}
By applying Lemma \ref{lem:phialp} to \eqref{eq:slMU},
we see that $g_0$ is expressed as follows in terms 
of the $u$-variables: 
\begin{equation}\label{eq:ginu}
g_0=
\left\{
\begin{array}{ll}\smallskip
\dfrac{\dsum{k=1}{l} u^1_{l+1}\cdots u^k_{l+1} u^k_{l}\cdots u^l_l}
{q^{-1}u^1_{l+1}\cdots u^{l+1}_{l+1}-u^1_l\cdots u^l_l}\quad
& (l=1,\ldots,m-1),\cr
\dfrac{\dsum{k=1}{m} u^1_{l+1}\cdots u^k_{l+1} u^k_{l}\cdots u^m_l}
{q^{-1}u^1_{l+1}\cdots u^{m}_{l+1}-u^1_l\cdots u^m_l}
& (l=m,\ldots,n-1). 
\end{array}
\right.
\end{equation}
Hence, formula \eqref{eq:slMU} as well as 
\eqref{eq:slHU} determines completely the action of $s_l$
on the $u$-variables. 

In order to describe the action of $s_l$ on the $u$-variables,  
for each $0\le i\le\min\br{l,m}$, 
we define
\begin{equation}\label{eq:defAil}
\begin{array}{lll}\smallskip
A^i_l=u^1_l\cdots u^i_l\dsum{k=i+1}{l} u^{i+1}_{l+1}
\cdots u^k_{l+1} u^k_{l}\cdots u^{l}_l
\cr
\quad
+ \,q^{-1}u^{i+1}_{l+1}\cdots u^{l+1}_{l+1}
\dsum{k=1}{i} u^1_{l+1}\cdots u^k_{l+1} 
u^k_l\cdots u^i_l
\quad&(1\le l\le m-1), \cr
A^i_l=u^1_l\cdots u^i_l\dsum{k=i+1}{m} u^{i+1}_{l+1}
\cdots u^k_{l+1} u^k_{l}\cdots u^{m}_l
\cr
\quad
+ \,q^{-1}u^{i+1}_{l+1}\cdots u^{m}_{l+1}
\dsum{k=1}{i} u^1_{l+1}\cdots u^k_{l+1} 
u^k_l\cdots u^i_l & (m\le l\le n-1). 
\end{array}
\end{equation}

\begin{theorem} \label{thm:sactsonU}
Under the tropical \RSKs correspondence 
$X\mapsto (U,V)$, 
the variables $u^i_j$ $(1\le i\le m;\ i\le j \le n)$ are
invariant with respect to the action of $\BS^m=\pr{r_1,\ldots,r_{m-1}}$. 
The action of $s_l$ $(l=1,\ldots,n-1)$ on 
$u^i_j$ is described as follows\,$:$ 
\begin{equation}\label{eq:slu}
\begin{array}{c}\smallskip
s_l(u^i_l)=u^i_{l+1} \, \dfrac{A^i_l}{A^{i-1}_l},
\quad
s_l(u^i_{l+1})=u^i_{l} \,\dfrac{A^{i-1}_l}{A^{i}_l},
\cr
s_l(u^i_j)=u^i_j\quad(j\ne l,l+1)
\end{array}
\end{equation}
for $1\le i\le\min\br{l,m}$ and 
$s_l(u^i_j)=u^i_j$ for $\min\br{l,m}+1\le i\le m$. 
\end{theorem} 
\proof 
Fixing the index $l=1,\ldots,n-1$, we consider the 
system of matrix equations
\begin{equation}\label{eq:auta}
G_l(a_{i-1})E_i(\overline{\bu}^i)=E_i(\overline{\bt}^i)
G_l(a_i)\quad(i=1,\ldots,m)
\end{equation}
for unknown variables $\bt^i=(1,\ldots,1,t^i_i,\ldots,t^i_n)$ 
($i=1,\ldots,m$) and $a_i$ ($i=0,1,\ldots,m$).  
We will construct below a solution of this system 
such that $a_m=q^{-1}a_0$, so that
\begin{equation}
G_l(a_0)M_U= M_T \, G_l(q^{-1}a_0); 
\end{equation}
this equation must imply $a_0=g_0$ and $T=s_l(U)$. 
The system of matrix equations \eqref{eq:auta} gives 
the recurrence relations
\begin{equation}\label{eq:reca}
\begin{array}{ll}
a_i=u^i_l+\dfrac{u^i_l}{u^i_{l+1}} a_{i-1}\qquad(1\le i\le l),\cr
a_{l+1}=\dfrac{1}{u^{l+1}_{l+1}}a_{l},
\quad
a_i=a_{i-1}\quad(l+2\le i\le m). 
\end{array}
\end{equation}
for $a_i$, and also 
\begin{equation}\label{eq:dTtu}
\frac{1}{t^i_{l}}=\frac{1}{u^i_l}-\frac{1}{a_{i}},
\quad
\frac{1}{t^i_{l+1}}=\frac{1}{u^i_{l+1}}+\frac{1}{a_{i-1}},\quad
t^i_j=u^i_j\quad(j\ne l,l+1). 
\end{equation}
for $1\le i\le l$ and $t^i_j=u^i_j$ for $l+1\le i\le m$. 
Under the condition  $a_m=q^{-1}a_0$, 
the recurrence relations \eqref{eq:reca} for $a_i$ 
are solved by  
\begin{equation}
a_i=\frac{A^i_l}
{q^{-1}u^1_{l+1}\cdots,u^{\min\br{l+1,m}}_{l+1}-u^1_l\cdots u^l_l}
\quad
(0\le i\le \min\br{l,m}). 
\end{equation}
Hence we obtain the expression for $t^i_j=s_l(u^i_j)$ 
as \eqref{eq:slu}.
\qed

By eliminating $a_i$ in \eqref{eq:reca}, \eqref{eq:dTtu}, 
we obtain
\begin{proposition}
The action of $s_l$ $(l=1,\ldots,n-1)$ on the tropical 
tableau $U=\pmatrix{u^i_j}_{i\le j}$ is characterized 
by the following system of algebraic equations 
of discrete Toda type for $t^i_j=s_l(u^i_j)$\,$:$ 
\begin{equation}
\begin{array}{llll}\smallskip
t^i_l t^i_{l+1}=u^i_{l}u^i_{l+1}\quad
(i=1,\ldots,l),\quad  &t^{l+1}_{l+1}=u^{l+1}_{l+1},
\cr\smallskip
\dfrac{1}{t^i_l}+\dfrac{1}{t^{i+1}_{l+1}}=
\dfrac{1}{u^i_l}+\dfrac{1}{u^{i+1}_{l+1}}\quad&(i=1,\ldots,l-1),\cr
\dfrac{1}{t^l_{l}}+\dfrac{q}{t^{l+1}_{l+1}t^1_{l+1}}
=\dfrac{1}{u^l_{l}}+\dfrac{q}{u^{l+1}_{l+1}u^1_{l+1}}
\end{array}
\end{equation}
for $l=1,\ldots,m-1$, and
\begin{equation}
\begin{array}{llll}\smallskip
t^i_l t^i_{l+1}=u^i_{l}u^i_{l+1}\quad
&(i=1,\ldots,m), 
\cr\smallskip
\dfrac{1}{t^i_l}+\dfrac{1}{t^{i+1}_{l+1}}=
\dfrac{1}{u^i_l}+\dfrac{1}{u^{i+1}_{l+1}}\quad&(i=1,\ldots,m-1),\cr
\dfrac{1}{t^m_{l}}+\dfrac{q}{t^1_{l+1}}
=\dfrac{1}{u^m_{l}}+\dfrac{q}{u^1_{l+1}}
\end{array}
\end{equation}
for $l=m,\ldots,n-1$, 
together with the constraint
\begin{eqnarray}
&&t^1_l\cdots t^{\min\br{l,m}}_l =u^1_{l+1}\cdots u^{\min\br{l+1,m}}_{l+1},
\nonumber\\
&&t^1_{l+1}\cdots t^{\min\br{l+1,m}}_{l+1} =
u^1_{l}\cdots u^{\min\br{l,m}}_{l}. 
\end{eqnarray}
\end{proposition}

The action of the tropical Sch\"utzenberger involution 
on the tropical tableau $U=\pmatrix{u^i_j}_{i\le j}$ 
plays the role of reversing the indices of the transformations 
$s_1,\ldots,s_{n-1}$ and 
interchanging $q$ and $q^{-1}$.   
Denoting by 
$\BK(u)$ the field of rational functions 
in the variables $u=(u^i_j)_{i,j}$, 
we define the involutive automorphism 
${\sf s}: \BK(u)\to \BK(u)$ by using the 
tropical Sch\"utzenberger involution of Theorem 
\ref{thm:tropSch}\,: 
\begin{equation}
{\sf s}(u^i_i)=\frac{\sigma^i_i}{\sigma^{i-1}_i},
\quad 
{\sf s}(u^i_j)=\frac{\sigma^i_j\ \sigma^{i-1}_{j-1}}
{\sigma^{i-1}_j\sigma^i_{j-1}}
\quad(i<j),
\end{equation}
where 
\begin{equation}
\sigma^i_j=\sum_{(\gamma_1,\ldots,\gamma_i)}\ 
u_{\gamma_1}\cdots u_{\gamma_i}
\end{equation}
is the sum of weights associated with $U$, 
over all $i$-tuples of nonintersecting paths 
$\gamma_k: (1,n-i+k)\to(\min\br{m,n-j+k},n-j+k)$ 
$(k=1,\ldots,i)$. 

\begin{theorem}
For each $l=1,\ldots, n-1$, 
let $s_l^q : \BK(u)\to\BK(u)$ the automorphisms 
defined as in Theorem $\ref{thm:sactsonU}$. 
Then we have  
${\sf s}\, s_l^q=s_{n-l}^{q^{-1}}\,{\sf s}$ 
for $(l=1,\ldots,n-1)$.
\end{theorem}
\proof
The tropical tableau ${\sf s}(U)=\pmatrix{{\sf s}(u^i_j)}_{i\le j}$
is characterized by the condition 
\begin{equation}
H_{{\sf s}(U)}=\theta(H_U)=J_n H_U^\trp J_n,
\end{equation} 
or equivalently, by 
\begin{equation}
M_{{\sf s}(U)}=\theta(M_U)=J_n M_U^\trp J_n. 
\end{equation} 
Hence, by applying $s_l^q$ to this equality, 
we have
\begin{equation}\label{eq:Mss}
M_{s_l^q\,{\sf s}(U)}=\theta(M_{s_l^q(U)}). 
\end{equation}
Recall that the action of $s_l$ ($l=1,\ldots,n-1$) is 
characterized by 
\begin{equation}
M_{s_l^q(U)}=G_l(a_0)\ M_{U}\ G_l(q^{-1}a_0)^{-1}. 
\end{equation}
Since $\theta(G_l(a))=G_{n-l}(-a)=G_{n-l}(a)^{-1}$, we 
obtain
\begin{eqnarray}
&&
\theta(M_{s_l^q(U)})
=G_{n-l}(q^{-1}a_0)\ \theta(M_{U})\ G_{n-l}(a_0)^{-1}
\nonumber\\
&& 
\phantom{\theta(M_{s_l^q(U)})}
=G_{n-l}(q^{-1}a_0)\ M_{{\sf s}(U)}\ G_{n-l}(a_0)^{-1}
\end{eqnarray}
By combining this with \eqref{eq:Mss}, 
we obtain
\begin{equation}
M_{s_l^q{\sf s}(U)}
=G_{m-l}(q^{-1}a_0)\ M_{{\sf s}(U)}\ G_{n-l}(a_0)^{-1}. 
\end{equation}
By applying ${\sf s}$ again, we have 
\begin{eqnarray}
&& M_{{\sf s}\,s_l^q{\sf s}(U)}
=G_{m-l}(q^{-1}{\sf s}(a_0))\ M_{s^2(U)}\ G_{n-l}({\sf s}(a_0))^{-1}
\nonumber\\
&& \phantom{M_{{\sf s}s_l{\sf s}(U)}}
=G_{m-l}(q^{-1}{\sf s}(a_0))\ M_{U}\ G_{n-l}({\sf s}(a_0))^{-1}
\nonumber\\
&& \phantom{M_{{\sf s}s_l{\sf s}(U)}}
=M_{s_l^{q^{-1}}(U)}. 
\end{eqnarray}
The last equality is a consequence of Lemma 
\ref{lem:phialp}.  
This implies ${\sf s}\,s_l^q{\sf s}(u^i_j)=s_l^{q^{-1}}(u^i_j)$ for all 
$i\le j$, as desired.  
\qed

\subsection{Combinatorial formulas for the Weyl group action}

By the standard procedure of Section 1.3, we can derive the 
piecewise linear action of 
the direct product $\widetilde{W}^m\times \widetilde{W}_n$ 
of affine Weyl groups
on the space of transportation matrices $X$.  
Also, via the \RSKs correspondence, we obtain 
the piecewise linear action of $\BS_n=\pr{s_1,\ldots,s_{n-1}}$ 
(resp. $\BS^m=\pr{r_1,\ldots,r_{m-1}}$ ) 
on the space of tableaux 
$U=\pmatrix{u^i_j}_{i\le j}$ (resp. $V=\pmatrix{v^i_j}_{i\le j}$).  
\par\medskip

Consider the space $\Mat_{m,n}(\BR)$ of real $m\times n$ 
matrices $X=\pmatrix{x^i_j}_{i,j}$, 
regarding $x=(x^i_j)_{i,j}$ as the canonical coordinates.  
For each multi-index $\alpha=(\alpha^{i}_j)_{i,j}\in\BN^{mn}$ 
and $a,b\in \BN$, we define the linear function 
$\ell_{\alpha,a,b}(x)$ on $\Mat_{m,n}(\BR)$ by 
\begin{equation}
\ell_{\alpha,a,b}(x)=\sum_{i,j} \alpha^i_j x^i_j+a p+b q,
\end{equation}
where $p,q$ are parameters.  
Note that 
$\ell_{\alpha,a,b}(x)=M(x^\alpha p^a q^b)$ 
in the notation of Section 1.3.
We denote by $\CF_X$ the set of all piecewise linear 
functions $f=f(x)$ in the form 
\begin{equation}
f(x) = \max\br{\ell_{\alpha,a,b}(x)\mid (\alpha,a,b)\in A}
-\max\br{\ell_{\beta,c,d}(x)\mid (\beta,c,d)\in B}, 
\end{equation}
where $A$, $B$  are nonempty finite sets of triples 
$(\alpha,a,b)$ of $\alpha\in\BN^{mn}$ and $a,b\in\BN$. 
Since $\CF_X=M(\BQ(x,p,q)_{>0})$, 
$\CF_X$ is closed under the addition, the subtraction 
and ``$\max$''; 
it is also closed under ``$\min$'' since 
$\min\br{f,g}=f+g-\max\br{f,g}$. 
We say that an isomorphism 
$w: \CF_X\to \CF_X$ of $\BZ$-modules 
is {\em combinatorial} 
if $w(\max\br{f,g})=\max\br{w(f),w(g)}$ 
for any $f,g\in \CF_X$.  

We first extend the indexing set for $x^i_j$ 
by setting 
\begin{equation}
x^{i+m}_j=x^i_j-q,\quad x^i_{j+n}=x^i_j-p \qquad(i,j\in\BZ). 
\end{equation}
We define the action of $r_k$ ($k=0,1,\ldots,m$), 
$\omega$ and $s_l$ $(l=0,1,\ldots,n)$, $\pi$ on the 
variables $x^i_j$ as follows: 
\begin{eqnarray}
&
\ \begin{array}{l}\smallskip
r_k(x^i_j)=x^{i+1}_j+{P^i_j}-{P^i_{j-1}}+p\cr
r_k(x^{i+1}_j)=x^{i}_j+{P^i_{j-1}}-{P^i_{j}}-p
\end{array}
\quad (i\equiv k \mod m),
\nonumber\\
&\ \ \quad r_k(x^i_j)=x^i_j\quad(i\not\equiv k, k+1 \mod m),
\quad
\omega(x^{i}_j)=x^{i+1}_j,
\nonumber\\
&
\begin{array}{l}\smallskip
s_l(x^i_j)=x^{i}_{j+1}+{Q^i_j}-{Q^{i-1}_{j}}+q\cr
s_l(x^i_{j+1})=x^{i}_j+{Q^{i-1}_{j}}-{Q^i_{j}}-q
\end{array}
\quad (j\equiv l \mod n),
\nonumber\\
&\ \ \  s_l(x^i_j)=x^i_j\quad(j\not\equiv l, l+1 \mod n),
\quad
\pi(x^{i}_j)=x^{i}_{j+1}, 
\end{eqnarray}
for $i,j\in\BZ$, where 
\begin{eqnarray}
&&P^i_j=\max_{1\le k\le n}\big(\ 
\sum_{a=1}^k x^{i+1}_{j+a}
+\sum_{a=k}^n x^{i}_{j+a}\ \big),
\nonumber\\
&&
Q^i_j=\max_{1\le k\le m} \big(\ 
\sum_{a=1}^k
x^{i+a}_{j+1}
+\sum_{a=k}^m
x^{i+a}_{j}\ \big). 
\end{eqnarray} 
These formulas can also be written in terms of ``$\min$'';
for instance, 
\begin{equation}
\ \begin{array}{l}\smallskip
r_k(x^i_j)=x^{i+1}_j-{R^i_j}+{R^i_{j-1}}+p\cr
r_k(x^{i+1}_j)=x^{i}_j-{R^i_{j-1}}+{R^i_{j}}-p
\end{array}
\quad (i\equiv k \mod m),
\end{equation}
where 
\begin{equation}
R^i_j=\min_{1\le k\le n}\big(\ 
\sum_{a=1}^{k-1} x^{i}_{j+a}
+\sum_{a=k+1}^n x^{i+1}_{j+a}\ \big).
\end{equation}

\begin{theorem}
Define the mappings 
$r_k$ $(k\in\BZ/m\BZ)$, 
$\omega$ and $s_l$ $(l\in\BZ/n\BZ)$, $\pi$ 
from the set of variables $x^i_j$ to $\CF_X$ 
as above. 
Then each of them extends uniquely to 
a combinatorial isomorphism $\CF_X\to\CF_X$. 
Furthermore, they give a realization of 
the direct product $\widetilde{W}^m\times\widetilde{W}_n$ 
of affine Weyl groups as a group 
of combinatorial isomorphisms of $\CF_X$. 
\end{theorem}

\begin{proposition}\label{prop:uTodaR}
By using the action of $r_k$ $(k=1,\ldots,m-1)$,
set $y^i_j=r_k(x^i_j)$ for all $i,j\in \BZ$.  
Then we obtain a solution to the ultra-discrete equation 
of Toda type 
\begin{equation}
x^i_j+x^{i+1}_j= y_j^i+y^{i+1}_j,
\quad
\min\br{x^i_j,x^{i+1}_{j+1}}
=\min\br{y^i_j, y^{i+1}_{j+1}}
\end{equation}
with periodicity 
condition $y^{i+m}_j=y^i_j-q$, $y^i_{j+n}=y^i_{j}-p$, 
satisfying the constraint 
\begin{equation}
y^i_{1}+\cdots +y^i_n=
x^{\sigma_k(i)}_1+\cdots+x^{\sigma_k(i)}_n
+(i-\sigma_k(i))p
\quad(i=1,\ldots,m). 
\end{equation}
\end{proposition}

\begin{remark} \rm
Let $B=\bigcup_{l=0}^\infty B_l$ the crystal basis of the 
symmetric tensor representation 
$S(V)=\bigoplus_{l=0}^\infty S_l(V)$ of $\mathfrak{gl}(n)$ 
associated with the vector representation $V=\BC^n$.  
Then $B$ is identified with the set of $n$-vectors 
$\bx=(x_1,\ldots,x_n)$ of nonnegative integers.
The crystal basis 
$B^{\otimes m}=B\otimes \cdots \otimes B$ 
($m$ times) 
for the $m$-th tensor product $S(V)^{\otimes m}$ is parametrized 
by $\BN^{mn}$.  
We identify the matrix $X=\pmatrix{x^i_j}_{i,j}$ with the 
coordinates of $B^{\otimes m}=\BN^{mn}$, 
regarding $\bx^i=(x^1_1,\ldots,x^i_n)$ as corresponding 
to the $i$-th component.   
When $p=q=0$, 
the actions of $r_k$ $(k=1,\ldots,m-1)$ 
and $s_l$ ($l=1,\ldots,n-1$) 
on the 
variables $x^i_j$ 
coincide with the combinatorial $R$-matrix 
acting on the $k$-th and $(k+1)$-st components 
of $B^{\otimes m}$,
and Kashiwara's Weyl group actions, 
respectively (see \cite{HHIKTT}, \cite{Y}). 
\end{remark}

Assuming that $m\le n$, we consider 
the variables $u^i_j$ 
($1\le i\le m;\ i\le j\le n$) 
and $v^i_j$ ($1\le i\le j\le m$), 
associated with the column strict tableaux $U$ and $V$,
through the \RSKs correspondence 
$X\mapsto(U,V)$.  
Recall that each $u^i_j$ (resp. $v^i_j$) denotes the 
number of $j$'s in the $i$-th row of $U$ (resp. $V$).  
Then, by the explicit piecewise linear formulas 
described in Theorem \ref{thm:CIRSKs}, 
$u^i_j$ and $v^i_j$ are regarded as elements 
in $\CF_X$. 
We thus 
obtain a combinatorial action 
of $\widetilde{W}^m\times\widetilde{W}_n$ 
on the variables $u^i_j$ and $v^i_j$. 
We describe below the action 
of the subgroups 
$\BS^m=\pr{r_1,\ldots,r_{m-1}}$ 
and 
$\BS_n=\pr{s_1,\ldots,s_{n-1}}$ 
on $u^i_j$; 
their action on $v^i_j$ is 
given by an obvious modification. 

In view of \eqref{eq:defAil}, we define $A^i_l$ 
($0\le i\le\min\br{l,m}$)
for the combinatorial version by
\begin{equation}\label{eq:defAilC}
\begin{array}{lll}\smallskip
A^i_l=\max\big\{\dmax{i+1\le k\le l}(
u^1_l+\cdots+u^i_l+u^{i+1}_{l+1}+
\cdots +u^k_{l+1}+u^k_{l}\cdots +u^{l}_l
),
\cr
\qquad
\dmax{1\le k\le i}(
u^{i+1}_{l+1}+\cdots+u^{l+1}_{l+1}
+u^1_{l+1}+\cdots+u^k_{l+1}+
u^k_l+\cdots+u^i_l -q)\big\}
\end{array}
\end{equation}
for $1\le l\le m-1$, and by 
\begin{equation}
\begin{array}{lll}\smallskip
A^i_l=\max\big\{\dmax{i+1\le k\le m}(
u^1_l+\cdots+u^i_l+u^{i+1}_{l+1}+
\cdots +u^k_{l+1}+u^k_{l}\cdots +u^{m}_l
),
\cr
\qquad
\dmax{1\le k\le i}(
u^{i+1}_{l+1}+\cdots+u^{m}_{l+1}
+u^1_{l+1}+\cdots+u^k_{l+1}+
u^k_l+\cdots+u^i_l -q)\big\}
\end{array}
\end{equation}
for $m\le l\le n-1$. 
Then, from Theorem \ref{thm:sactsonU} we obtain
\begin{theorem} 
The variables $u^i_j$ are invariant under the action of 
the symmetric group $\BS^m=\pr{r_1,\ldots,r_{m-1}}$ 
induced via the \RSKs correspondence.  
The action of $s_l$ $(l=1,\ldots,n-1)$ is given 
explicitly as follows: 
\begin{equation}\label{eq:sluC}
\begin{array}{c}\medskip
s_l(u^i_l)=u^i_{l+1} +{A^i_l}-{A^{i-1}_l},
\quad
s_l(u^i_{l+1})=u^i_{l}+{A^{i-1}_l}-{A^{i}_l},
\cr
s_l(u^i_j)=u^i_j\quad(j\ne l,l+1)
\end{array} 
\end{equation}
for $1\le i\le\min\br{l,m}$ and $s_l(u^i_j)=u^i_j$ for 
$\min\br{l,m}+1\le i\le m$.
\end{theorem} 

Note that $v^i_j$ are invariant under the action 
of $\BS_n=\br{s_1,\ldots,s_{n-1}}$, and that 
$r_k$ ($k=1,\ldots,m-1$) act on $v^i_j$ by 
explicit piecewise linear formulas 
similar to those described above. 

\begin{proposition}
By using the action of $s_l$ $(l=1,\ldots,n-1)$,
set $t^i_j=s_l(u^i_j)$ for $1\le i\le m$, $i\le j\le n$.  
Then we obtain a solution to 
the ultra-discrete equation of Toda type
\begin{equation}
\begin{array}{llll}\smallskip
t^i_l+t^i_{l+1}=u^i_{l}+u^i_{l+1}\quad
(i=1,\ldots,l),\qquad  t^{l+1}_{l+1}=u^{l+1}_{l+1},
\cr\smallskip
\min\br{t^i_l, t^{i+1}_{l+1}}=
\min\br{u^i_l, u^{i+1}_{l+1}}\qquad(i=1,\ldots,l-1),\cr
\min\br{t^l_{l},t^{l+1}_{l+1}+t^1_{l+1}-q}
=\min\br{u^l_{l},u^{l+1}_{l+1}+u^1_{l+1}-q}
\end{array}
\end{equation}
for $l=1,\ldots,m-1$, and
\begin{equation}
\begin{array}{llll}\smallskip
t^i_l + t^i_{l+1}=u^i_{l}+ u^i_{l+1}\quad &
(i=1,\ldots,m),
\cr\smallskip
\min\br{t^i_l, t^{i+1}_{l+1}}=
\min\br{u^i_l,u^{i+1}_{l+1}}\quad&(i=1,\ldots,m-1),\cr
\multicolumn{2}{c}{
\min\br{t^m_{l},{t^1_{l+1}-q}}
=\min\br{u^m_{l},u^1_{l+1}-q}}
\end{array}
\end{equation}
for $l=m,\ldots,n-1$, satisfying the constraint 
\begin{eqnarray}
&&t^1_l+\cdots+t^{\min\br{l,m}}_l =
u^1_{l+1}+\cdots +u^{\min\br{l+1,m}}_{l+1},
\nonumber\\
&&t^1_{l+1}+\cdots +t^{\min\br{l+1,m}}_{l+1} =
u^1_{l}+\cdots +u^{\min\br{l,m}}_{l}. 
\end{eqnarray}
\end{proposition}

\begin{remark}\rm
The variables $u^i_j$ are identified with 
the coordinates of crystal bases for general 
finite dimensional 
irreducible representations 
of $\mathfrak{gl}(n)$
as in \cite{KN}.   
Then, by using an argument as in \cite{HHIKTT}, 
it can be shown that 
the combinatorial action of $\BS_n=\pr{s_1,\ldots,s_{n-1}}$ 
on $u^i_j$ with $q=0$ 
provides with the description of 
Kashiwara's Weyl group actions 
on the crystal bases.  
This symmetric group action on the set 
of column strict tableaux thus coincides 
with the one introduced earlier 
by A.~Lascoux and M.P.\ Sch\"utzenberger 
\cite{LS}(see also \cite{LLT}). 
The corresponding piecewise linear action is discussed 
in \cite{KB}. 
\end{remark}


\end{document}